\documentclass[prd,aps,showpacs,onecolumn,12pt]{revtex4}
\usepackage{amssymb}
\usepackage{amsmath}
\usepackage{graphicx}
\usepackage{bm}

\input{tcilatex}
\begin{document}

\title{Diagonal Representation for a Generic Matrix Valued Quantum
Hamiltonian}
\author{Pierre Gosselin$^{1}$ and Herv\'{e} Mohrbach$^{2}$}

\begin{abstract}
A general method to derive the diagonal representation for a generic matrix
valued quantum Hamiltonian is proposed. In this approach new mathematical
objects like non-commuting operators evolving with the Planck constant
promoted as a running variable are introduced. This method leads to a formal
compact expression for the diagonal Hamiltonian which can be expanded in a
power series of the Planck constant. In particular, we provide an explicit
expression for the diagonal representation of a generic Hamiltonian to the
second order in the Planck constant. This last result is applied, as a
physical illustration, to Dirac electrons and neutrinos in external fields.
\end{abstract}

\affiliation{Institut Fourier, UMR 5582 CNRS-UJF UFR de Math\'{e}matiques, Universit\'{e}
Grenoble I, BP74, 38402 Saint Martin d'H\`{e}res, Cedex, France }
\affiliation{Laboratoire de Physique Mol\'{e}culaire et des Collisions, ICPMB-FR CNRS
2843, Universit\'{e} Paul Verlaine-Metz, 57078 Metz Cedex 3, France}
\maketitle

\section{Introduction}

The spacetime evolution of some important quantum systems is governed by a
multicomponent Schr\"{o}dinger like equation whose Hamiltonian is a matrix
valued operators. In many cases, these systems are too complicated to be
solved directly using the Schr\"{o}dinger equation and some simplifications
are required. For systems displaying a separation of scales in terms of slow
and fast degrees of freedom a major simplification is to treat quantally the
fast ones while the slow ones can be approximated classically or
semiclassically. For a slow cyclic variation of the classical part, Berry
has shown that the wave function of the quantum part gets an additional
geometric phase factor \cite{BERRY1}. It has also been understood, in the
context of the Born-Oppenheimer theory of molecules \cite{MEAD}\cite{MOODY}%
\cite{ZYGELMAN} that the geometric phases induced by the quantum part of the
system provide a correction, reacting on the classical part with geometric
Lorentz and electric forces \cite{BERRY3}\cite{STERN}. Going beyond this
first correction has proved extraordinarily difficult due to the intricate
entanglement of noncommuting operators \cite{LITTLEJOHN}.

In fact, many physical systems share\ a very similar mathematical structure
with molecular systems. For instance, the translational movement of a Dirac
electron in a slowly varying external field can be separated from the spin
evolution and therefore considered semiclassically.

In this paper, we adress the question of the dynamics governing the slow
part of an arbitrary quantum system and present a formalism for calculating
the Hamiltonian of this slow part to all orders in the Planck's constant. Of
course, we do not expect the convergence of the series in general, since for
most quantum systems the limit $\hbar \rightarrow 0$ is singular, which
means that an exact quantum behavior can not be approximated even by an
infinite expansion.

The philosophy behind the approach developed here, consists in a mapping of
the initial quantum system to a classical one which can be easily
diagonalized and then to return to the full quantum system. The method
requires the introduction of new mathematical objects like non-commuting
operators evolving with the Planck constant promoted as a running variable.
This leads to a differential calculus on a non-commutative space which shows
some similarities with the stochastic calculus. This innovation is the clue
to derive the formal exact diagonal representation for any arbitrary matrix
valued Hamiltonian.

The results presented here extend the work presented in \cite{SERIESPIERRE},
where a differential equation with respect to $\hbar $ was proposed for the
required diagonal Hamiltonian. The semiclassical limit, which is often
enough to get physical insight for the problem considered, was obtained by a
straightforward integration of this differential equation and a general
semiclassical formula for an arbitrary diagonal Hamiltonian in terms of
covariant operators and commutators between Berry connections was obtained.
This formalism allowed to consider several straightforward applications.
These ones included the deflection of light by the interaction between its
polarization state and external inhomogeneities, such as a varying
refractive index or a gravitational field \cite{PIERRE1}; Dirac electrons
and Bloch electrons in crystals \cite{SEMIDIAG}, leading to new effective
geometrical forces and a clarification of the Peierls substitution \cite%
{PIERREBLOCH} as well as to a generalization of the Bohr-Sommerfeld
quantification rule \cite{PIERRESOMMERFELD}. In principle, one could solve
the differential equation to the desired order in a series of $\hbar ,$ but
although this was done for simple examples to order $\hbar ^{2}$ (see \cite%
{SERIESPIERRE}), we could not derive such an expression for an arbitrary
Hamiltonian.

The main advantage of the new method presented in this paper is that being
not based on a differential equation, it directly leads to a formal exact
diagonal representation for any arbitrary matrix valued Hamiltonian, which,
as proven in this paper, is an exact solution of the differential equation
of \cite{SERIESPIERRE}. It appears also, as one could expect, that the
expansion in series of $\hbar $ for the exact diagonal Hamiltonian derived
here is much more easier to obtain than by the successive integration of the
differential equation of \cite{SERIESPIERRE}. In particular, we provide an
explicit formula for the diagonal Hamiltonian to order $\hbar ^{2}$ that can
be directly applied in principle to any physical systems.

In this paper, for the sake of clarity and to keep the length reasonable, we
will consider only two simple but still physically relevant examples. More
applications can be found in \cite{INTERACTINGBLOCH}, where interacting
Bloch electrons are considered, and in \cite{BornPierre} for a
generalization of the Born-Oppenheimer approximation.

It is worth mentioning that some other diagonalization procedures exist,
each of them having its range of validity, advantages and defects. The
paradigmic example is provided by the Foldy-Wouthuysen (FW) representation
of the Dirac Hamiltonian for relativistic particles interacting with an
external electromagnetic field. In this representation the positive and
negative energy states are separately represented and the non-relativistic
Pauli-Hamiltonian is obtained \cite{FOLDY}. Actually even if several exact
FW transformations have been found for some definite classes of potentials 
\cite{ERIKSEN}\cite{NIKITIN}\cite{SILENKO1}, the diagonalization is a
difficult mathematical problem requiring approximations, essentially a
perturbation expansion in weak fields. A generalisation of the FW
transformation developed by Blount and also based on an expansion in weak
electromagnetic fields, is in principle applicable to Bloch electrons and
Dirac electrons \cite{Blount}. But this method is limited by construction to
weak external perturbations and provides a formal series expansion which
leads to cumbersome calculations for practical applications. Recently a
variant of the Foldy Wouthuysen transformation valid for strong fields and
based also on an expansion in $\hbar $ of the Dirac Hamiltonian was
presented \cite{SILENKO}.

Among other approaches, Weigert and Littlejohn developed a systematic method
to diagonalize general quantum Hamiltonian in a series expansion in $\hbar $ 
\cite{LITTLEJOHN}. It leads also to a formal series expansion written in
terms of symbols of operators which also makes the method complicated for
practical applications. In particular this method does not provide explicit
formulas, but instead is an algorithm to compute case by case the series
order by order. To our knowledge, this approach was only used for
Born-Oppenheimer molecular systems. Note also, that a different method based
on a adiabatic perturbation theory was proposed in \cite{TEUFFEL} (and
references therein) and applied to the Dirac electron and electrons in a
cristal at the first order.

Our approach, although similar in spirit to these methods, is technically
very different and allows us for the first time to derive the expression of
the diagonal representation at order $\hbar ^{2}$ of an arbitrary quantum
Hamiltonian. This efficiency could be explained by the fact that, unlike
other approaches, no Weyl calculus and Moyal Algebra are required. Another
interesting feature of our approach is that it confirms the fundamental role
played by Berry curvatures since it leads to an effective diagonal
Hamiltonian with Berry phase corrections as well as noncommutative (Berry
connections dependent) coordinates and momentum covariant operators as in
previous approaches \cite{SERIESPIERRE}\cite{SEMIDIAG} (see also \cite{ALAIN}
for Dirac electrons in an electric field and \cite{BLIOKH1} for the
extension to the full electromagnetic field). The resulting generic
equations of motion are also corrected by Berry curvatures terms.

The paper is organized as follows. In the next section we develop the
differential calculus in noncommutative space. We then derive, in section 3,
the diagonalization procedure and show the link with the differential
equation in section 4. Section 5 is devoted to the discussion of the
dynamical coordinate and momentum operators. In section 6, we give the
general diagonal energy operator formula to the second order in $\hbar $ and
in section 7, Dirac electron in an external electric field and neutrinos in
a static symmetric gravitational field are considered as an illustration of
the general formalism. Last section is for the conclusion.

\section{Preliminaries:}

\subsection{Differential calculus on noncommutative space}

Consider a quantum mechanical system whose state space is a tensor product $%
L^{2}\left( \mathcal{R}^{3}\right) \otimes V$\ with $V$\ some internal
space. In other words, the Hamiltonian can be written as a matrix $H\left( 
\mathbf{P,R}\right) $\ of size $\dim V$\ whose elements are operators
depending on a couple of canonical operators $\left[ R^{i},P^{j}\right]
=i\hbar \delta _{ij}$, the archetype example being the Dirac Hamiltonian
with $V=C^{4}$. Our goal is to derive the formal diagonal representation of
this matrix valued quantum Hamiltonian. By diagonal Hamiltonian it is meant
an effective in-band Hamiltonian which has a matrix representation with
block-diagonal matrix elements associated with energy band subspaces. As in
paper \cite{SERIESPIERRE} the procedure we propose for the removal of the
interband matrix elements needs the use of some unconventional mathematics
compared to the usual formalism of quantum mechanics. Since the Planck
constant is here considered as a variable, operators do not satisfy the
usual rules of quantum mechanics. Therefore before directly embarking on the
diagonalization procedure we first introduce the\ required mathematical
tools.\textbf{\ }For that purpose we start with some definitions and
notations.

\subsubsection{Running coordinates and momentum operators}

To begin with, we introduce a formal space of non commuting infinitesimal
operators $dX_{\alpha }^{i}\equiv \left\{ dR_{\alpha }^{i},dP_{\alpha
}^{i}\right\} $ $\ i=1,2,3$ indexed by a continuous parameter $\alpha $,
that satisfy the following infinitesimal Heisenberg algebra (with a reversed
sign) : 
\begin{equation}
\left[ dR_{\alpha }^{i},dP_{\alpha \prime }^{j}\right] =-id\alpha \delta
_{\alpha ,\alpha \prime }\delta _{ij}\ \ \ \ \text{and\ }\ \ \ \ \ \left[
dR_{\alpha }^{i},dR_{\alpha \prime }^{j}\right] =\left[ dP_{\alpha
}^{i},dP_{\alpha \prime }^{j}\right] =0.  \label{commdaplpha}
\end{equation}
Then, we define a set of running coordinate and momentum operators by
witting the following formal sums : 
\begin{equation}
R_{\alpha }^{i}=R^{i}-\int_{\alpha }^{\hbar }dR_{\lambda }^{i}\text{, \ \
and \ \ }P_{\alpha }^{i}=P^{i}-\int_{\alpha }^{\hbar }dP_{\lambda }^{i}
\label{RRR}
\end{equation}
with the convention $dR_{\alpha }^{i}=R_{\alpha }^{i}-R_{\alpha -d\alpha
}^{i}$ and $dP_{\alpha }^{i}=P_{\alpha }^{i}-P_{\alpha -d\alpha }^{i}.$ This
\textquotedblright downward\textquotedblright\ choice of the differential
element notably implies the commutation rules $\left[ R_{\alpha
}^{i},dP_{\alpha }^{j}\right] =\left[ dR_{\alpha }^{i},P_{\alpha }^{j}\right]
=0$ which turns out to be absolutely necessary to develop later on a
differential calculus on this noncommutative space.

For $\alpha =\hbar $ we recover the usual canonical operators $R^{i}\equiv
R_{\hbar }^{i}$and $P^{i}\equiv P_{\hbar }^{i}$ which satisfy the canonical
Heisenberg algebra $\left[ R^{i},P^{j}\right] =i\hbar \delta _{ij}$, whereas
the running operators satisfy 
\begin{equation}
\left[ R_{\alpha }^{i},P_{\alpha }^{j}\right] =i\alpha \delta _{ij}\text{ \
\ \ \ and \ \ \ \ \ }\left[ R_{\alpha }^{i},R_{\alpha }^{j}\right] =\left[
P_{\alpha }^{i},P_{\alpha }^{j}\right] =0.  \label{commalpha}
\end{equation}
Note that in this paper we will never consider the algebra of the operators $%
R_{\alpha }^{i}$ and $P_{\alpha \prime }^{j}$ for $\alpha \neq \alpha
^{\prime }$, which from the definition Eq. $\left( \ref{RRR}\right) $ is
clearly not a Heisenberg one.

The variables $d\mathbf{X}_{\alpha }$ have to be understood as fictitious
variables that make the link between quantal operators $\left( \alpha =\hbar
\right) $ and classical variables $\left( \alpha =0\right) $. As will be
explicit later on, their role is to transport our quantum system to a formal
classical one that can be in general easily diagonalized, and then back from
the formal classical one to the required\textbf{\ }quantized system. By
writing $d\mathbf{X}_{\alpha }=\sqrt{d\alpha /\hbar }\mathbf{\hat{X}}%
_{\alpha }$ with $\mathbf{\hat{X}}_{\alpha }$ a normalized canonical
operator we see that the infinitesimal quantities $d\mathbf{X}_{\alpha }$
are actually of order $\sqrt{d\alpha }.$ However, having this in mind, we
will never use the $\mathbf{\hat{X}}_{\alpha }$ notation and always work
with $d\mathbf{X}_{\alpha }.$

\subsubsection{Differential algebra}

For the sequel, we need to define the differential $dF\left( \mathbf{X}%
_{\alpha },\alpha \right) $ of an arbitrary function $F\left( \mathbf{X}%
_{\alpha },\alpha \right) $ where $X_{\alpha }^{i}\equiv \left\{ R_{\alpha
}^{i},P_{\alpha }^{i}\right\} $. For this purpose, consider the operators $%
R_{\alpha }^{i}-\int_{\alpha }^{\hbar }dR_{\lambda }^{i}$, and $P_{\alpha
}^{i}=P^{i}-\int_{\alpha }^{\hbar }dP_{\lambda }^{i}$ as acting on a space $%
W=\left( V\otimes L^{2}\left( \mathcal{R}^{3}\right) \right) \otimes \left(
\otimes _{\alpha <\hbar }L^{2}\left( \mathcal{R}^{3}\right) _{\alpha
}\right) $ which is the tensor product of $V$ and an infinite number of
copies of $L^{2}\left( \mathcal{R}^{3}\right) $. The tensor product $%
V\otimes L^{2}\left( \mathcal{R}^{3}\right) $ and the space $L^{2}\left( 
\mathcal{R}^{3}\right) _{\alpha }$ refer respectively to the spaces on which
the canonical operators $\left( R^{i},P^{i}\right) $ and the differential
operators $dX_{\alpha }^{i}$ act.

Now to be consistent with our convention for the\ differential element $%
dX_{\alpha }^{i}$, the differential of the function $F\left( \mathbf{X}%
_{\alpha },\alpha \right) $\ in defined in an unusual backward manner as $%
dF\left( \mathbf{X}_{\alpha },\alpha \right) \equiv F\left( \mathbf{X}%
_{\alpha },\alpha \right) -F\left( \mathbf{X}_{\alpha -d\alpha },\alpha
-d\alpha \right) .$\ This construction is essential as it permits the
commutation between the non-differential and the differential elements $%
dX_{\alpha }^{i}$. The differential will be written as a second order
expansion plus some neglected terms: 
\begin{align}
dF\left( \mathbf{X}_{\alpha },\alpha \right) & =\nabla _{R_{i}}FdR_{\alpha
}^{i}+\nabla _{P_{i}}FdP_{\alpha }^{i}-\frac{1}{2}\left( \nabla
_{R_{i}}\nabla _{R_{j}}F\right) dR_{\alpha }^{i}dR_{\alpha }^{j}-\frac{1}{2}%
\left( \nabla _{P_{i}}\nabla _{P_{j}}F\right) dP_{\alpha }^{i}dP_{\alpha
}^{j}  \notag \\
& -\frac{1}{2}\left( \nabla _{R_{i}}\nabla _{P^{j}}F\right) \left(
dR_{\alpha }^{i}dP_{\alpha }^{j}+dP_{\alpha }^{j}dR_{\alpha }^{i}\right)
+\left\langle F\left( \mathbf{X}_{\alpha },\alpha \right) \right\rangle
d\alpha +\frac{\partial F}{\partial \alpha }d\alpha  \notag \\
& +\text{terms of order }3\text{,}  \label{dhbare}
\end{align}
where all expressions in the r.h.s. are evaluated at\textbf{\ }$\left( 
\mathbf{X}_{\alpha },\alpha \right) $. Note that here, we have kept the
terms of order square in $dX_{\alpha }^{i}$ since they are of order $d\alpha 
$ and thus contribute to the differential, whereas higher orders can safely
be disregarded as they are negligible when the integration over $\alpha $\
is considered. These second order terms have been organized in a certain
form that will be practical for us later. There is nothing to say about the
usual $dR_{\alpha }^{i}dR_{\alpha }^{j}$ or $dP_{\alpha }^{i}dP_{\alpha
}^{j} $ terms. The crossed terms involving products of $dP_{\alpha }^{i}$
and $dR_{\alpha }^{j}$ have to be taken with care since these two
infinitesimal elements do not commute with each other. For this purpose we
have decomposed the second order terms involving products such as $%
dP_{\alpha }^{i}dR_{\alpha }^{j}$ and $dR_{\alpha }^{j}dP_{\alpha }^{i}$\ in
a symmetric part proportional to $\left( dR_{\alpha }^{i}dP_{\alpha
}^{j}+dP_{\alpha }^{j}dR_{\alpha }^{i}\right) $ giving the contribution%
\textbf{\ }$-\frac{1}{2}\left( \nabla _{R_{i}}\nabla _{P^{j}}F\right) \left(
dR_{\alpha }^{i}dP_{\alpha }^{j}+dP_{\alpha }^{j}dR_{\alpha }^{i}\right) $,
and an antisymmetric part proportional to $\left( dR_{\alpha }^{i}dP_{\alpha
}^{j}-dP_{\alpha }^{j}dR_{\alpha }^{i}\right) =-i\delta ^{ij}d\alpha $
corresponding to the bracket $\left\langle F\left( \mathbf{X}_{\alpha
},\alpha \right) \right\rangle d\alpha $ that we now explain in details. The
notation $\left\langle F\left( \mathbf{X}_{\alpha },\alpha \right)
\right\rangle $ (which in \cite{SERIESPIERRE} was corresponding to the
operation $-\frac{i}{2}Asym\nabla _{R_{i}}\nabla _{P^{i}}F\left( \mathbf{X}%
_{\alpha },\alpha \right) $) is defined as a specific procedure on a series
expansion of $F$ in the variables $R_{\alpha }^{i}$, $P_{\alpha }^{i}$ in
the following way : let $F$ be a sum of monomials of the kind $M_{1}\left( 
\mathbf{R}_{\alpha }\right) M_{2}\left( \mathbf{P}_{\alpha }\right)
M_{3}\left( \mathbf{R}_{\alpha }\right) ....$ the $M_{i}$ being arbitrary
monomials in $R_{\alpha }$ or $P_{\alpha }$ alternatively. The operator $%
\nabla _{R_{i}}\nabla _{P^{j}}$ acts on such an expression by deriving all
combinations of one monomial in $\mathbf{R}_{\alpha }$ and one monomial in $%
\mathbf{P}_{\alpha }$. For each of these combinations, insert a $dR_{\alpha
}^{i}$ at the place where the derivative $\nabla _{R_{i}}$ is acting and in
a same manner a $dP_{\alpha }^{j}$ at the place where the derivative $\nabla
_{P^{j}}$ is acting. This leads to an expression with two kind of terms, one
kind being proportional to the $dR_{\alpha }^{i}dP_{\alpha }^{j},$ and the
second proportional to $dP_{\alpha }^{j}dR_{\alpha }^{i}.$ Then rewrite this
expression in terms of $dR_{\alpha }^{i}dP_{\alpha }^{j}+dP_{\alpha
}^{j}dR_{\alpha }^{i}$ and $dR_{\alpha }^{i}dP_{\alpha }^{j}-dP_{\alpha
}^{j}dR_{\alpha }^{i}=-i\delta ^{ij}d\alpha $. Then $\left\langle F\left( 
\mathbf{X}_{\alpha },\alpha \right) \right\rangle $ is defined as minus the
sum over $i$ and $j$ of all the terms proportional to $d\alpha $ in the
computation procedure just considered. This definition implies a procedure
which is clearly dependent of the symmetrization chosen for the expansion of 
$F$.

To make the definition of $\left\langle F\left( \mathbf{X}_{\alpha },\alpha
\right) \right\rangle $\ clearer, consider some important practical
examples. If the function $F$\ has the following form $F=\frac{1}{2}\left(
A\left( \mathbf{R}_{\alpha }\right) B\left( \mathbf{P}_{\alpha }\right)
+B\left( \mathbf{P}_{\alpha }\right) A\left( \mathbf{R}_{\alpha }\right)
\right) $\ which corresponds to a frequent choice of symmetrization in $%
R_{\alpha }$\ and $P_{\alpha }$, then $\left\langle F\left( \mathbf{X}%
_{\alpha },\alpha \right) \right\rangle =\frac{i}{4}\dsum\nolimits_{i}\left[
\nabla _{R_{i}}A\left( \mathbf{R}_{\alpha }\right) ,\nabla _{P^{i}}B\left( 
\mathbf{P}_{\alpha }\right) \right] $. Another choice of symmetrization
leads in general to a different result.

For instance, if we rewrite the same function $F$ in a fully symmetrized
form in $R_{\alpha }$\ and $P_{\alpha }$\ (that is invariant by all
permutations in $R_{\alpha }$\ and $P_{\alpha }$) which is also often used,
we now have a different result since $\left\langle F\left( \mathbf{X}%
_{\alpha },\alpha \right) \right\rangle =0$.

Nevertheless, this dependence of $\left\langle F\left( \mathbf{X}_{\alpha
},\alpha \right) \right\rangle $ in the symmetrization choice is not
astonishing at all. \ Actually changing the symmetrization of a function $%
F\left( \mathbf{X}_{\alpha },\alpha \right) $ introduces some explicit terms
in $\alpha $ which changes also the term $\partial _{\alpha }Fd\alpha $
present in the differential Eq. $\left( \ref{dhbare}\right) $. As a
consequence, neither the partial derivative with respect to $\alpha $, nor
the bracket are invariant by a change of form. But, what is invariant is the
sum $\partial _{\alpha }F+\left\langle F\right\rangle $.

We show this assertion by first giving a useful formula for the differential
of a product of functions : 
\begin{eqnarray}
d\left( F\left( \mathbf{X}_{\alpha },\alpha \right) G\left( \mathbf{X}%
_{\alpha },\alpha \right) \right) &=&dFG+FdG-\nabla _{R_{i}}F\nabla
_{R_{j}}GdR_{\alpha }^{i}dR_{\alpha }^{j}-\nabla _{P_{i}}F\nabla
_{P_{j}}GdP_{\alpha }^{i}dP_{\alpha }^{j}  \notag \\
&&-\left( \nabla _{R_{i}}F\nabla _{P^{j}}G\right) dR_{\alpha }^{i}dP_{\alpha
}^{j}-\left( \nabla _{P^{j}}F\nabla _{R_{i}}G\right) dP_{\alpha
}^{j}dR_{\alpha }^{i}  \notag \\
&=&dFG+FdG-\nabla _{R_{i}}F\nabla _{R_{j}}GdR_{\alpha }^{i}dR_{\alpha
}^{j}-\nabla _{P_{i}}F\nabla _{P_{j}}GdP_{\alpha }^{i}dP_{\alpha }^{j} 
\notag \\
&&-\frac{1}{2}\left( \nabla _{R_{i}}F\nabla _{P^{j}}G+\nabla _{P^{j}}F\nabla
_{R_{i}}G\right) \left( dR_{\alpha }^{i}dP_{\alpha }^{j}+dP_{\alpha
}^{j}dR_{\alpha }^{i}\right)  \notag \\
&&-\frac{i}{2}\left( \nabla _{P_{i}}F\nabla _{R_{i}}G-\nabla _{R_{i}}F\nabla
_{P_{i}}G\right) d\alpha  \label{dFG}
\end{eqnarray}

Now, the invariance of $\partial _{\alpha }F+\left\langle F\right\rangle $
is the consequence of two facts. The first one is that the definition of the
differential $dF\left( \mathbf{X}_{\alpha },\alpha \right) $ is independent
of the way the function $F\left( \mathbf{X}_{\alpha },\alpha \right) $ has
been symmetrized. Actually, considering $F\left( \mathbf{X}_{\alpha },\alpha
\right) $ as a monomial, a change of symmetrization amounts to move
successively powers of the momentum on the right or the left of the position
variable. Recursively, one just has to check that moving only one power of
the momentum does not change the differential. Considering thus a monomial
of the form $F\left( \mathbf{X}_{\alpha },\alpha \right) =M_{1}P_{\alpha
}^{i}R_{\alpha }^{i}M_{2}$ with $M_{1}$ and $M_{2}$ arbitrary, move the
momentum to rewrite $F\left( \mathbf{X}_{\alpha },\alpha \right)
=M_{1}\left( R_{\alpha }^{i}P_{\alpha }^{i}-i\alpha \right) M_{2}$. Our
formula for a product of differential, allows to assert that the
differentials of the two terms $M_{1}P_{\alpha }^{i}R_{\alpha }^{i}M_{2}$
and $M_{1}\left( R_{\alpha }^{i}P_{\alpha }^{i}-i\alpha \right) M_{2}$\
differ only by the contributions $M_{1}d\left( P_{\alpha }^{i}R_{\alpha
}^{i}\right) M_{2}$ \ and $M_{1}d\left( R_{\alpha }^{i}P_{\alpha
}^{i}-i\alpha \right) M_{2}$. As a consequence, the differential $dF\left( 
\mathbf{X}_{\alpha },\alpha \right) $ is independent of the choice of
symmetrization if and only if $d\left( P_{\alpha }^{i}R_{\alpha }^{i}\right)
=d\left( R_{\alpha }^{i}P_{\alpha }^{i}-i\alpha \right) $. Given that 
\begin{eqnarray*}
d\left( P_{\alpha }^{i}R_{\alpha }^{i}\right) &=&dP_{\alpha }^{i}R_{\alpha
}^{i}+P_{\alpha }^{i}dR_{\alpha }^{i}-\frac{1}{2}\left( dR_{\alpha
}^{i}dP_{\alpha }^{j}+dP_{\alpha }^{j}dR_{\alpha }^{i}\right) -\frac{i}{2}%
d\alpha \\
d\left( R_{\alpha }^{i}P_{\alpha }^{i}-i\alpha \right) &=&R_{\alpha
}^{i}dP_{\alpha }^{i}+dR_{\alpha }^{i}P_{\alpha }^{i}-\frac{1}{2}\left(
dR_{\alpha }^{i}dP_{\alpha }^{j}+dP_{\alpha }^{j}dR_{\alpha }^{i}\right) +%
\frac{i}{2}d\alpha -id\alpha
\end{eqnarray*}
and that the $dP_{\alpha }^{i}$,$dR_{\alpha }^{i}$\ commute with $R_{\alpha
}^{i}$, $P_{\alpha }^{i}$\ the equality of the two previous lines holds as
well as the assertion of symmetrization independence.

This result joined to the second fact that in the series expansion of $%
dF\left( \mathbf{X}_{\alpha },\alpha \right) $, the variables $dX_{\alpha
}^{i}$, $dX_{\alpha }^{i}dX_{\alpha }^{j}$, $d\alpha $, are independent, and
thus that the coefficients of the expansion of $dF\left( \mathbf{X}_{\alpha
},\alpha \right) $ are uniquely defined, leads directly to the announced
result that $\partial _{\alpha }F+\left\langle F\right\rangle $ which is the
coefficient of $d\alpha $ in the expansion of $dF\left( \mathbf{X}_{\alpha
},\alpha \right) $ is independent of any symmetrization choice. Therefore it
is convenient to introduce a symmetrization invariant derivative $D_{\alpha
} $ as given by 
\begin{equation}
D_{\alpha }F\left( \mathbf{X}_{\alpha },\alpha \right) =\partial _{\alpha
}F\left( \mathbf{X}_{\alpha },\alpha \right) +\left\langle F\left( \mathbf{X}%
_{\alpha },\alpha \right) \right\rangle
\end{equation}

We end up this paragraph by introducing two formulas that will be convenient
later. First, we still more compactify our notations for the differential by
writing : 
\begin{align}
dF\left( \mathbf{X}_{\alpha },\alpha \right) & =\sum_{i=1}^{6}\nabla
_{X_{\alpha }^{i}}F\left( \mathbf{X}_{\alpha },\alpha \right) dX_{\alpha
}^{i}-\frac{1}{4}\sum_{i,j=1}^{6}\nabla _{X_{\alpha }^{i}}\nabla _{X_{\alpha
}^{j}}F\left( \mathbf{X}_{\alpha },\alpha \right) \left( dX_{\alpha
}^{i}dX_{\alpha }^{j}+dX_{\alpha }^{i}dX_{\alpha }^{j}\right)  \notag \\
& +D_{\alpha }F\left( \mathbf{X}_{\alpha },\alpha \right) d\alpha  \label{dF}
\end{align}
with $i$,$j=1..6$. We also assume that $X_{\alpha }^{i}\equiv R_{\alpha
}^{i} $ for $i=1,2,3$ and $X_{\alpha }^{i}\equiv P_{\alpha }^{i}$ for $%
i=4,5,6$. In fact, as we will show in the next sections the quantity of real
importance for us in Eq. $\left( \ref{dF}\right) $ is the term proportional
to $d\alpha $. Note that in this notation, our previous formula for a
product of differential Eq. $\left( \ref{dFG}\right) $ takes a compact form
: 
\begin{equation}
d\left( F\left( \mathbf{X}_{\alpha },\alpha \right) G\left( \mathbf{X}%
_{\alpha },\alpha \right) \right) =dFG+FdG-\sum_{i,j=1}^{6}\nabla
_{X_{\alpha }^{i}}\left( F\right) \nabla _{X_{\alpha }^{j}}\left( G\right)
dX_{\alpha }^{i}dX_{\alpha }^{j}
\end{equation}
where the last term could of course be developed as before in symmetric and
antisymmetric parts.

Second, we also give the bracket formula $\left\langle .\right\rangle $ for
a product of two functions. Using the procedure defined previously (see also
ref. \cite{SERIESPIERRE}) one obtains the following expression 
\begin{equation}
<F\left( \mathbf{X}_{\alpha },\alpha \right) G\left( \mathbf{X}_{\alpha
},\alpha \right) >=\left\langle F\right\rangle G+F\left\langle
G\right\rangle -\frac{i}{2}\nabla _{P_{i}}F\nabla _{R_{i}}G+\frac{i}{2}%
\nabla _{R_{i}}F\nabla _{P_{i}}G.  \label{BFG}
\end{equation}
This formula shows that the bracket operation for a product $F\left( \mathbf{%
X}_{\alpha },\alpha \right) G\left( \mathbf{X}_{\alpha },\alpha \right) $
can also be seen as a sort of deformation of the Poisson bracket including
some \textquotedblright internal\textquotedblright\ contributions $%
\left\langle F\left( \mathbf{X}_{\alpha },\alpha \right) \right\rangle $ and 
$\left\langle G\left( \mathbf{X}_{\alpha },\alpha \right) \right\rangle $.

Let us remark ultimately that the term $\left\langle F\left( \mathbf{X}%
_{\alpha },\alpha \right) \right\rangle d\alpha $ in Eq. $\left( \ref{dhbare}%
\right) $ which is of the second order in the derivatives $\nabla _{R_{i}}$
and $\nabla _{P^{i}}$ is very reminiscent of the bracket introduced in the
stochastic calculus. Given the non commutativity of the operators at stake,
one should in fact rather expect our formalism to be close to the quantum
stochastic calculus \cite{PARTA}. However, it does not seem at first sight
that our objects fit in such a framework which deals rather with a formalism
of creation, annihilation and conservation operators. A full comparison is
out of the scope of our paper.

\subsubsection{Integration}

Now, we will prove that the formula Eq. $\left( \ref{dhbare}\right) $\
allows to express a function $F\left( \mathbf{X}_{\hbar },\hbar \right) $\
which depends on the physical quantum operators $X_{\hbar }$\ defined at the
quantum scale $\alpha =\hbar ,$\ as the integration of a differential.

\textbf{Proposition} 
\begin{equation}
F\left( \mathbf{X}_{\hbar },\hbar \right) =F(\mathbf{X}_{0},0)+\int_{0}^{%
\hbar }dF\left( \mathbf{X}_{\alpha },\alpha \right) .  \label{FXII}
\end{equation}

Before proving Eq. (\ref{FXII}), remark that this expression has the
following immediate generalization : 
\begin{equation*}
F\left( \mathbf{X}_{\hbar },\hbar \right) =F(\mathbf{X}_{\alpha },\alpha
)+\int_{\alpha }^{\hbar }dF\left( \mathbf{X}_{\beta },\beta \right)
\end{equation*}

\textbf{Proof}\textit{. }

The proof of formula Eq. (\ref{FXII}) is by recursion. Expanding $F\left( 
\mathbf{X}_{\alpha },\alpha \right) $ as sum of monomials in the canonical
variable, it is enough to prove our assertion for a monomial of a certain
degree in the $R_{\hbar }^{i}$, $P_{\hbar }^{i}$. For monomials of degree
one, that is linear expressions in the $R_{\hbar }^{i}$, $P_{\hbar }^{i}$,
the result is trivial given our definitions. Now, assume that the result is
true for all monomial $M$ of, say, bidegree $m$, $n$ in the $R_{\hbar }^{i}$%
, $P_{\hbar }^{i}$. We will show the result for a monomial whose degree has
increased by one in one variable. Such a monomial can be written $X_{\hbar
}^{i}M$ or $MX_{\hbar }^{i}$, with $M$ of bidegree $m$, $n$ (recall that $%
X_{\hbar }^{i}=\left( R_{\hbar }^{i},P_{\hbar }^{i}\right) $). We will
concentrate on the first possibility, the proof being obviously the same for
the other case. We compute directly $X_{0}^{i}M(\mathbf{X}%
_{0},0)+\int_{0}^{\hbar }dX_{\alpha }^{i}M\left( \mathbf{X}_{\alpha },\alpha
\right) $.

Given our formula for the differential of a product as well as the
recurrence hypothesis which states that the integral formula is true for $M$
we have : 
\begin{eqnarray*}
X_{0}^{i}M(\mathbf{X}_{0},0)+\int_{0}^{\hbar }d\left( X_{\alpha }^{i}M\left( 
\mathbf{X}_{\alpha },\alpha \right) \right) &=&\left( X_{\hbar
}^{i}-\int_{0}^{\hbar }dX_{\alpha }^{i}\right) \left( M(X_{\hbar }^{i},\hbar
)-\int_{0}^{\hbar }dM\left( \mathbf{X}_{\alpha },\alpha \right) \right) \\
&&+\int_{0}^{\hbar }dX_{\alpha }^{i}M\left( \mathbf{X}_{\alpha },\alpha
\right) +\int_{0}^{\hbar }X_{\alpha }^{i}dM\left( \mathbf{X}_{\alpha
},\alpha \right) \\
&&-\sum_{j}\int_{0}^{\hbar }dX_{\alpha }^{i}dX_{\alpha }^{j}\nabla
_{X_{\alpha }^{j}}M\left( \mathbf{X}_{\alpha },\alpha \right)
\end{eqnarray*}
In the last line we have chosen (for the sake of simplicity) not to separate
the product $dX_{\alpha }^{i}dX_{\alpha }^{j}$ into symmetric and
antisymmetric part, but to keep the second order terms in a compact form.
Note again the minus sign arising in front of this term due to our choice of
definition for the differential. As a consequence one has : 
\begin{eqnarray*}
&&X_{0}^{i}M(\mathbf{X}_{0},0)+\int_{0}^{\hbar }d\left( X_{\alpha
}^{i}M\left( \mathbf{X}_{\alpha },\alpha \right) \right) \\
&=&X_{\hbar }^{i}M(X_{\hbar }^{i},\hbar )-\int_{0}^{\hbar }dX_{\alpha
}^{i}M(X_{\hbar }^{i},\hbar )-X_{\hbar }^{i}\int_{0}^{\hbar }dM\left( 
\mathbf{X}_{\alpha },\alpha \right) +\int_{0}^{\hbar }dX_{\alpha
}^{i}\int_{0}^{\hbar }dM\left( \mathbf{X}_{\alpha },\alpha \right) \\
&&+\int_{0}^{\hbar }dX_{\alpha }^{i}M\left( \mathbf{X}_{\alpha },\alpha
\right) +\int_{0}^{\hbar }X_{\alpha }^{i}dM\left( \mathbf{X}_{\alpha
},\alpha \right) -\sum_{j}\int_{0}^{\hbar }dX_{\alpha }^{i}dX_{\alpha
}^{j}\nabla _{X_{\alpha }^{j}}M\left( \mathbf{X}_{\alpha },\alpha \right)
\end{eqnarray*}
Now, due again to the recursion hypothesis, we rewrite $A=\int_{0}^{\hbar
}dX_{\alpha }^{i}M\left( \mathbf{X}_{\alpha },\alpha \right)
+\int_{0}^{\hbar }X_{\alpha }^{i}dM\left( \mathbf{X}_{\alpha },\alpha
\right) $ in the following manner : 
\begin{eqnarray*}
A &=&\int_{0}^{\hbar }dX_{\alpha }^{i}\left[ M(X_{\hbar }^{i},\hbar
)-\int_{\alpha }^{\hbar }dM\left( \mathbf{X}_{\alpha ^{\prime }},\alpha
^{\prime }\right) \right] +\int_{0}^{\hbar }\left( X_{\hbar
}^{i}-\int_{\alpha }^{\hbar }dX_{\alpha ^{\prime }}^{i}\right) dM\left( 
\mathbf{X}_{\alpha },\alpha \right) \\
&=&\int_{0}^{\hbar }dX_{\alpha }^{i}M(X_{\hbar }^{i},\hbar )+X_{\hbar
}^{i}\int_{0}^{\hbar }dM\left( \mathbf{X}_{\alpha },\alpha \right)
-\int_{0}^{\hbar }dX_{\alpha }^{i}\int_{\alpha }^{\hbar }dM\left( \mathbf{X}%
_{\alpha ^{\prime }},\alpha ^{\prime }\right) -\int_{0}^{\hbar }\int_{\alpha
}^{\hbar }dX_{\alpha ^{\prime }}^{i}dM\left( \mathbf{X}_{\alpha },\alpha
\right)
\end{eqnarray*}
Now, we have to take care about the meaning of the two last integrals.
Actually, given our definition of the differential, in the first integral $%
\int_{0}^{\hbar }dX_{\alpha }^{i}\left[ \int_{\alpha }^{\hbar }dM\left( 
\mathbf{X}_{\alpha ^{\prime }},\alpha ^{\prime }\right) \right] $, one has
to consider that $\alpha ^{\prime }>\alpha $, since our differentials are
pointing downward, and as a consequence when $\alpha ^{\prime }$ is getting
closer to $\alpha $, the \textquotedblright last\textquotedblright\ element
of integration is $dM\left( \mathbf{X}_{\alpha +d\alpha },\alpha +d\alpha
\right) =M\left( \mathbf{X}_{\alpha +d\alpha },\alpha +d\alpha \right)
-M\left( \mathbf{X}_{\alpha },\alpha \right) $. In the integral $%
\int_{0}^{\hbar }\int_{\alpha }^{\hbar }dX_{\alpha ^{\prime }}^{i}dM\left( 
\mathbf{X}_{\alpha },\alpha \right) $ one has rather to consider $\alpha
^{\prime }<\alpha $. As a consequence, one has : 
\begin{eqnarray*}
&&\int \int_{\alpha ^{\prime }<\alpha <\hbar }dX_{\alpha }^{i}dM\left( 
\mathbf{X}_{\alpha ^{\prime }},\alpha ^{\prime }\right) +\int \int_{\alpha
<\alpha ^{^{\prime }}<\hbar }dX_{\alpha }^{i}dM\left( \mathbf{X}_{\alpha
^{\prime }},\alpha ^{\prime }\right) \\
&=&\int_{0}^{\hbar }dX_{\alpha }^{i}\int_{0}^{\hbar }dM\left( \mathbf{X}%
_{\alpha },\alpha \right) -\int_{0}^{\hbar }dX_{\alpha }^{i}dM\left( \mathbf{%
X}_{\alpha },\alpha \right)
\end{eqnarray*}
the last term in the right cancelling the diagonal contribution that should
not appear given our considerations just above. Note that this contribution
would be negligible for an ordinary integral. As a consequence, we obtain : 
\begin{eqnarray*}
A &=&\int_{0}^{\hbar }dX_{\alpha }^{i}M(X_{\hbar }^{i},\hbar )+X_{\hbar
}^{i}\int_{0}^{\hbar }dM\left( \mathbf{X}_{\alpha },\alpha \right)
-\iint\limits_{\alpha ^{\prime }<\alpha <\hbar }dX_{\alpha }^{i}dM\left( 
\mathbf{X}_{\alpha ^{\prime }},\alpha ^{\prime }\right)
-\iint\limits_{\alpha <\alpha ^{^{\prime }}<\hbar }dX_{\alpha }^{i}dM\left( 
\mathbf{X}_{\alpha ^{\prime }},\alpha ^{\prime }\right) \\
&=&\int_{0}^{\hbar }dX_{\alpha }^{i}M(X_{\hbar }^{i},\hbar )+X_{\hbar
}^{i}\int_{0}^{\hbar }dM\left( \mathbf{X}_{\alpha },\alpha \right) -\left(
\int_{0}^{\hbar }dX_{\alpha }^{i}\int_{0}^{\hbar }dM\left( \mathbf{X}%
_{\alpha },\alpha \right) -\int_{0}^{\hbar }dX_{\alpha }^{i}dM\left( \mathbf{%
X}_{\alpha },\alpha \right) \right)
\end{eqnarray*}
Gathering all the terms leads thus ultimately to : 
\begin{eqnarray*}
X_{0}^{i}M(\mathbf{X}_{0},0)+\int_{0}^{\hbar }d\left( X_{\alpha }^{i}M\left( 
\mathbf{X}_{\alpha },\alpha \right) \right) &=&X_{\hbar }^{i}M(X_{\hbar
}^{i},\hbar ) \\
&&+\int_{0}^{\hbar }dX_{\alpha }^{i}dM\left( \mathbf{X}_{\alpha },\alpha
\right) -\sum_{j}\int_{0}^{\hbar }dX_{\alpha }^{i}dX_{\alpha }^{j}\nabla
_{X_{\alpha }^{j}}M\left( \mathbf{X}_{\alpha },\alpha \right)
\end{eqnarray*}
Now, the integral on the diagonal $\int_{0}^{\hbar }dX_{\alpha }^{i}dM\left( 
\mathbf{X}_{\alpha },\alpha \right) $ reduces to $\sum_{j}\int_{0}^{\hbar
}dX_{\alpha }^{i}dX_{\alpha }^{j}\nabla _{X_{\alpha }^{j}}M\left( \mathbf{X}%
_{\alpha },\alpha \right) $. Actually as explained before in the definition
of the differential, the second order terms in $dM\left( \mathbf{X}_{\alpha
},\alpha \right) $, multiplied by $dX_{\alpha }^{i}$ contribute to the order 
$d\alpha ^{\frac{3}{2}}$ which yields a zero contribution while summing over 
$\alpha $. We thus end up with : 
\begin{equation}
X_{0}^{i}M(\mathbf{X}_{0},0)+\int_{0}^{\hbar }d\left( X_{\alpha }^{i}M\left( 
\mathbf{X}_{\alpha },\alpha \right) \right) =X_{\hbar }^{i}M(X_{\hbar
}^{i},\hbar )
\end{equation}
which proves Eq. $\left( \ref{FXII}\right) $ $\blacksquare $.

Now, we interpret Eq. $\left( \ref{FXII}\right) $ as follows. Starting from
the classical regime the integration over $\alpha $\ leads in continuous
manner to the fully quantum regime. This original manner to achieve a sort
of quantification is very reminiscent of the renormalization group technics
in which the integration of the high energy modes leads to an effective low
energy theory with all quantum fluctuations included.

Eq. $\left( \ref{FXII}\right) $ allows us to rewrite formally a function $%
F\left( \mathbf{X}_{\hbar },\hbar \right) $ depending on the physical
variables in terms of the same function evaluated at the classical variables 
$(\mathbf{X}_{0},0)$ plus an integral of a differential. This introduce
apparently a useless complexity since the expansion of the right hand side
of Eq. $\left( \ref{FXII}\right) $ around $\mathbf{X}_{\hbar }$ involve a
infinite series of the infinitesimal increment of the $d\mathbf{X}_{\alpha
}. $ By construction, this expansion implies trivially that all terms of the
series exactly vanishes except obviously the initial term which is equal to
the function $F\left( \mathbf{X}_{\hbar },\hbar \right) .$ Although trivial,
this decomposition will prove useful later.

Note again that the integral in Eq. $\left( \ref{FXII}\right) $ has to be
understood as being computed downward as seen in the definition of ($\mathbf{%
R}_{\alpha },\mathbf{P}_{\alpha })$ : the starting point is at $\hbar $ and
the differentials are pointed downward. For example $\mathbf{\nabla }_{%
\mathbf{R}_{\alpha }}F.d\mathbf{R}_{\alpha }=-\mathbf{\nabla }_{\mathbf{R}%
_{\alpha }}F\left( \mathbf{X}_{\alpha },\alpha \right) \left( \mathbf{R}%
_{\alpha -d\alpha }-\mathbf{R}_{\alpha }\right) $. However, this equation
has to be taken with some care. Actually the sum over the terms proportional
to $d\mathbf{X}_{\alpha }$ being a sum of terms of magnitude $\sqrt{d\alpha }
$, it converges only if the sum is discretized, and this will be implicitly
assumed in this paper. Defining properly the continuous limit is out of the
scope of this paper.

\subsection{Expectation operator}

Having now a differential set up, we aim at defining a linear conditional
expectation operator $\mathcal{E}\left( .\right) .$ We first set the
following formula : 
\begin{eqnarray}
\mathcal{E}\left( Gd\alpha \right) &=&\mathcal{E}\left( G\right) d\alpha 
\text{, whatever }G  \label{epsi1} \\
\mathcal{E}\left( F\left( \mathbf{X}_{\hbar },\hbar \right) \right)
&=&F\left( \mathbf{X}_{\hbar },\hbar \right) \text{ and }\mathcal{E}\left(
F\left( \mathbf{X}_{\alpha },\alpha \right) \right) =F\left( \mathbf{X}%
_{\hbar },\hbar \right) -\int_{\alpha }^{\hbar }\mathcal{E}\left( dF\left( 
\mathbf{X}_{\alpha },\alpha \right) \right)
\end{eqnarray}
The first equality will allow to define recursively expectations of
integrals of a function $F\left( \mathbf{X}_{\alpha },\alpha \right) $ with
a deterministic increment $d\alpha .$ The second one will make the
expectation conditional by fixing the starting point $\mathbf{X}_{\hbar }$
of the path $\mathbf{X}_{\alpha }$. The third equation allows to define the
expectation of an arbitrary function $F\left( \mathbf{X}_{\alpha },\alpha
\right) $ with the help of its initial value $F\left( \mathbf{X}_{\hbar
},\hbar \right) $ and the integral of the expectation of a differential
whose rule of computation is given below. Note that, due to the required
linearity of the expectation operator, the expectation has to commute with
the integration. This property is implied by the third equality which yields 
$\mathcal{E}\int_{\alpha }^{\hbar }\left( dF\left( \mathbf{X}_{\alpha
},\alpha \right) \right) =\int_{\alpha }^{\hbar }\mathcal{E}\left( dF\left( 
\mathbf{X}_{\alpha },\alpha \right) \right) $. By analogy with the
stochastic calculus, $\mathbf{X}_{\alpha }$ will be seen as a random path
whose infinitesimal increments $d\mathbf{X}_{\alpha }$ will have zero
expectation. However, again as in stochastic calculus, the expectation of a
quadratic term like $dX_{\alpha }^{i}dX_{\alpha }^{j}$ can not be set
consistently to vanish since it can be written in terms of $\mathcal{E}%
\left( dR_{\alpha }^{i}dP_{\alpha }^{j}+dP_{\alpha }^{j}dR_{\alpha
}^{i}\right) $ and $\mathcal{E}\left( dR_{\alpha }^{i}dP_{\alpha
}^{j}-dP_{\alpha }^{j}dR_{\alpha }^{i}\right) =-i\delta ^{ij}d\alpha ,$ and
the last term can not vanish due to the first definition in Eq.$\left( \ref%
{epsi1}\right) $, and only the symmetric part $\mathcal{E}\left( dR_{\alpha
}^{i}dP_{\alpha }^{j}+dP_{\alpha }^{j}dR_{\alpha }^{i}\right) $ can be
chosen to vanish.

Consequently we set the expectation rules for products at the same
\textquotedblright time\textquotedblright\ $\alpha $ 
\begin{equation}
\mathcal{E}\left( \prod_{i=1}^{n}dR_{\alpha _{i}}\right) =\mathcal{E}\left(
\prod_{i=1}^{n}dP_{\alpha _{i}}\right) =0\text{, \ \ \ }\mathcal{E}\left(
\prod_{i=1}^{n}\left( dX_{\alpha _{i}}^{k_{i}}dX_{\alpha
_{i}}^{l_{i}}+dX_{\alpha _{i}}^{l_{i}}dX_{\alpha _{i}}^{k_{i}}\right)
\right) =0  \label{epsi2}
\end{equation}
and for different times 
\begin{equation}
\mathcal{E}\left( \prod_{i=1}^{n}\prod_{j=1}^{p}dX_{\alpha
_{i}}^{k_{i}}dX_{\alpha _{j}}^{l_{j}}\right) =0\ \ \ \text{for}\ \ \ \alpha
_{i}\neq \alpha _{j}  \label{epsi3}
\end{equation}
We need also the independence of the increments $d\mathbf{X}_{\alpha }$ with
respect to a function evaluated at the corresponding $\mathbf{X}_{\alpha }$
or all the \textquotedblright previous ones\textquotedblright\ $\mathbf{X}%
_{\alpha ^{\prime }}$, $\alpha ^{\prime }\geq \alpha $, which is formulated
as 
\begin{equation}
\mathcal{E}\left( F\left( \mathbf{X}_{\alpha ^{\prime }}\right) d\mathbf{Z}_{%
\mathbf{\alpha }}\right) =\mathcal{E}\left( F\left( \mathbf{X}_{\alpha
^{\prime }}\right) \right) \mathcal{E}\left( d\mathbf{Z}_{\mathbf{\alpha }%
}\right) =0\text{ for }\alpha ^{\prime }\geq \alpha \text{ }  \label{indep}
\end{equation}
where $\mathbf{\alpha =}\left( \alpha _{1},...,\alpha _{p}\right) $ is an
arbitrary number of values all different and all lower or equal to $\alpha
^{\prime }$. $d\mathbf{Z}_{\mathbf{\alpha }}$ is condensed notation for a
product $\prod_{i=1}^{p}dZ_{\alpha _{i}}$ where the $dZ_{\alpha _{i}}$ can
be $dX_{\alpha _{i}}^{k_{i}}$ or $\left( dX_{\alpha _{i}}^{k_{i}}dX_{\alpha
_{i}}^{l_{i}}+dX_{\alpha _{i}}^{l_{i}}dX_{\alpha _{i}}^{k_{i}}\right) $ with 
$k_{i}=1..6.$

We could impose Eq. $\left( \ref{indep}\right) $ directly, but this property
is in fact the direct consequence of a single condition that we will thus
enforce which is the independence of the increments $d\mathbf{X}_{\alpha }$
with respect to the initial value $\mathbf{X}_{\hbar }$ of the path : 
\begin{equation}
\mathcal{E}\left( F\left( \mathbf{X}_{\hbar }\right) d\mathbf{Z}_{\mathbf{%
\alpha }}\right) =\mathcal{E}\left( F\left( \mathbf{X}_{\hbar }\right)
\right) \mathcal{E}\left( d\mathbf{Z}_{\mathbf{\alpha }}\right) =0\text{ }
\label{indepinit}
\end{equation}
That Eq. $\left( \ref{indep}\right) $ follows from this single condition is
a direct recursive computation. Actually, Eq. $\left( \ref{indep}\right) $
is trivially checked for $F\left( \mathbf{X}_{\alpha ^{\prime }}\right) $ a
polynomial of degree $0$ in $\left( \mathbf{X}_{\alpha ^{\prime }},\alpha
^{\prime }\right) $ since $\mathcal{E}\left( ad\mathbf{X}_{\alpha }\right) =a%
\mathcal{E}\left( d\mathbf{X}_{\alpha }\right) =0$ for $a$ constant. If Eq. $%
\left( \ref{indep}\right) $ is true for a polynomial of degree $N$ in the
variables $\left( \mathbf{X}_{\alpha ^{\prime }},\alpha ^{\prime }\right) $,
then consider $F\left( \mathbf{X}_{\alpha ^{\prime }},\alpha ^{\prime
}\right) $ to be of degree $N+1$. As a consequence, 
\begin{eqnarray*}
\mathcal{E}\left( F\left( \mathbf{X}_{\alpha ^{\prime }}\right) d\mathbf{Z}_{%
\mathbf{\alpha }}\right) &=&\mathcal{E}\left( \left( F\left( \mathbf{X}%
_{\hbar }\right) -\int_{\alpha ^{\prime }}^{\hbar }dF\left( \mathbf{X}%
_{\beta },\beta \right) \right) d\mathbf{Z}_{\mathbf{\alpha }}\right) \\
&=&\mathcal{E}\left( F\left( \mathbf{X}_{\hbar }\right) d\mathbf{Z}_{\alpha
}\right) -\mathcal{E}\left( \int_{\alpha ^{\prime }}^{\hbar }dF\left( 
\mathbf{X}_{\beta },\beta \right) d\mathbf{Z}_{\mathbf{\alpha }}\right) \\
&=&-\mathcal{E}\left( \int_{\alpha ^{\prime }}^{\hbar }dF\left( \mathbf{X}%
_{\beta },\beta \right) d\mathbf{Z}_{\mathbf{\alpha }}\right)
\end{eqnarray*}
the last equality being a consequence of $\left( \ref{indepinit}\right) $.
Now, since 
\begin{eqnarray*}
\mathcal{E}\left( dF\left( \mathbf{X}_{\beta },\beta \right) \right) &=&%
\mathcal{E}\left( D_{\beta }F\left( \mathbf{X}_{\beta },\beta \right) d\beta
+\sum_{i=1}^{6}\nabla _{X_{\beta }^{i}}F\left( \mathbf{X}_{\beta },\beta
\right) dX_{\beta }^{i}\right. \\
&&\left. -\frac{1}{4}\sum_{i,j=1}^{6}\nabla _{X_{\beta }^{i}}\nabla
_{X_{\beta }^{j}}F\left( \mathbf{X}_{\beta },\beta \right) \left( dX_{\beta
}^{i}dX_{\beta }^{j}+dX_{\beta }^{i}dX_{\beta }^{j}\right) \right)
\end{eqnarray*}
all terms in the integral are of degree lower or equal to $N$, so that the
recurrence applies (recall that since the integration is downward, in the
integral over $\beta $, $\beta >\alpha ^{\prime }\geq \mathbf{\alpha }$ and
thus in the set $\left( \beta ,\mathbf{\alpha }\right) $ all elements are
different) and $\mathcal{E}\left( dF\left( \mathbf{X}_{\beta },\beta \right)
\right) d\mathbf{Z}_{\mathbf{\alpha }}=0$. Thus Eq. $\left( \ref{indep}%
\right) $ is true for polynomials. Since all along the paper we consider
functions $F\left( \mathbf{X}_{\alpha ^{\prime }},\alpha ^{\prime }\right) $
that are converging series in the variables $\left( \mathbf{X}_{\alpha
^{\prime }},\alpha ^{\prime }\right) $ (seen as classical commuting real
variables), the density of the polynomial in that space of functions ends
the proof.

A first consequence of the definitions and Eq. $\left( \ref{indep}\right) $
is 
\begin{align*}
\mathcal{E}\left( dF\left( \mathbf{X}_{\alpha },\alpha \right) \right) & =%
\mathcal{E}\left( \partial _{\alpha }F\left( \mathbf{X}_{\alpha },\alpha
\right) +\left\langle F\left( \mathbf{X}_{\alpha },\alpha \right)
\right\rangle \right) d\alpha \\
& =\mathcal{E}\left( D_{\alpha }F\left( \mathbf{X}_{\alpha },\alpha \right)
\right) d\alpha
\end{align*}
This formula allows to compute the integral of $\mathcal{E}\left( dF\left( 
\mathbf{X}_{\alpha },\alpha \right) \right) $ set previously, as an
\textquotedblright ordinary\textquotedblright\ integral since the
integration element is only $d\alpha $.

The main interest of these definitions will show up later on when a function 
$F\left( \mathbf{X}_{\hbar },\hbar \right) $ will be written as an series
expansion of iterated integrals along the path $\mathbf{X}_{\alpha }.$
Indeed, as already said before, the expansion of the two terms in the right
hand side of Eq. $\left( \ref{FXII}\right) $ around $\mathbf{X}_{\hbar },$
implies a mechanism of cancellation of all contributions proportional to
products of $dX_{\alpha _{i}}^{k_{i}}$. The expectation operator is built to
cancel separately the contributions of products of $dX_{\alpha _{i}}^{k_{i}}$
at different times and symmetric products $\left( dX_{\alpha
_{i}}^{k_{i}}dX_{\alpha _{i}}^{l_{i}}+dX_{\alpha _{i}}^{l_{i}}dX_{\alpha
_{i}}^{k_{i}}\right) $ at the same time, coming from the two terms of the
right hand side of Eq. $\left( \ref{FXII}\right) $. The rest of the
contributions form the right hand side of Eq. $\left( \ref{FXII}\right) $
proportional to $d\alpha $ cancel exactly. As a consequence of this, an
arbitrary function can be written

\begin{align*}
F_{\hbar }\left( \mathbf{X}_{\hbar }\right) & =\mathcal{E}\left( F_{\hbar
}\left( \mathbf{X}_{\hbar }\right) \right) =\mathcal{E}\left( F(\mathbf{X}%
_{0},0)+\int_{0}^{\hbar }dF\left( \mathbf{X}_{\alpha },\alpha \right) \right)
\\
& =\mathcal{E}\left( F(\mathbf{X}_{0},0)\right) +\mathcal{E}\left(
\int_{0}^{\hbar }D_{\alpha }F\left( \mathbf{X}_{\alpha },\alpha \right)
\right)
\end{align*}
Once again we have apparently introduced some useless complexity, since the
function $F_{\hbar }\left( \mathbf{X}_{\hbar }\right) $ has been replaced by 
$\mathcal{E}\left( F(\mathbf{X}_{0},0)\right) $ which cannot be computed
directly, plus the integral $\mathcal{E}\left( \int_{0}^{\hbar }D_{\alpha
}\left( \mathbf{X}_{\alpha },\alpha \right) d\alpha \right) $. Irrelevant
sums of infinitesimal increments have been introduced in both terms and thus
ultimately have to cancel . However, this formulation presents the advantage
of replacing $\mathbf{X}_{\hbar }$ by $\mathbf{X}_{0}$ in the first
contribution which will simplify some of our computations later. The
integration over $d\alpha $ will allow to start a recursive expansion in the
Planck constant. But to do so, we first need to introduce an other operator
connecting a function evaluated at $\mathbf{X}_{0}$ to the same function
evaluated at $\mathbf{X}_{\hbar }$. This is the role of the EBS operation
defined below that realizes a kind of connection between the spaces of
operators at two different values of $\alpha $.

A second consequence of our definitions is the factorization of the initial
value in the expectation : 
\begin{equation*}
\mathcal{E}\left( F\left( \mathbf{X}_{\hbar },\hbar \right) G\left( \mathbf{X%
}_{\alpha },\alpha \right) \right) =F\left( \mathbf{X}_{\hbar }\right) 
\mathcal{E}\left( G\left( \mathbf{X}_{\alpha },\alpha \right) \right) \text{%
, }\alpha \leq \hbar
\end{equation*}
This property is usually trivial in probability, but here has to be derived
given we have started with expectations of infinitesimal increments. It
comes from iterating the differentiation of $G\left( \mathbf{X}_{\alpha
}\right) $ similar to the chaotic expansion in stochastic calculus.
Actually, we can write : 
\begin{equation*}
\mathcal{E}\left( F\left( \mathbf{X}_{\hbar },\hbar \right) G\left( \mathbf{X%
}_{\alpha },\alpha \right) \right) =F\left( \mathbf{X}_{\hbar },\hbar
\right) G\left( \mathbf{X}_{\hbar },\hbar \right) -\mathcal{E}\left( F\left( 
\mathbf{X}_{\hbar },\hbar \right) \int_{\alpha }^{\hbar }dG\left( \mathbf{X}%
_{\alpha },\alpha \right) d\alpha \right)
\end{equation*}
given Eq. $\left( \ref{indep}\right) $, this reduces to : 
\begin{eqnarray*}
\mathcal{E}\left( F\left( \mathbf{X}_{\hbar },\hbar \right) G\left( \mathbf{X%
}_{\alpha },\alpha \right) \right) &=&F\left( \mathbf{X}_{\hbar },\hbar
\right) G\left( \mathbf{X}_{\hbar },\hbar \right) -\mathcal{E}\left( F\left( 
\mathbf{X}_{\hbar },\hbar \right) \int_{\alpha }^{\hbar }D_{\alpha }G\left( 
\mathbf{X}_{\alpha },\alpha \right) d\alpha \right) \\
&=&F\left( \mathbf{X}_{\hbar },\hbar \right) G\left( \mathbf{X}_{\hbar
},\hbar \right) -\int_{\alpha }^{\hbar }\mathcal{E}\left( F\left( \mathbf{X}%
_{\hbar },\hbar \right) D_{\alpha }G\left( \mathbf{X}_{\alpha },\alpha
\right) \right) d\alpha
\end{eqnarray*}
and once again the recurrence works recursively for polynomials. If $G\left( 
\mathbf{X}_{\alpha },\alpha \right) $ is of degree $N$ in $\left( \mathbf{X}%
_{\alpha },\alpha \right) $, then $D_{\alpha }G\left( \mathbf{X}_{\alpha
},\alpha \right) $ is of degree $N-1$, and thus the property can be assumed
to be true for $D_{\alpha }G\left( \mathbf{X}_{\alpha },\alpha \right) $, so
that : 
\begin{eqnarray*}
\mathcal{E}\left( F\left( \mathbf{X}_{\hbar },\hbar \right) G\left( \mathbf{X%
}_{\alpha },\alpha \right) \right) &=&F\left( \mathbf{X}_{\hbar },\hbar
\right) G\left( \mathbf{X}_{\hbar },\hbar \right) -F\left( \mathbf{X}_{\hbar
},\hbar \right) \mathcal{E}\left( \int_{\alpha }^{\hbar }D_{\alpha }G\left( 
\mathbf{X}_{\alpha },\alpha \right) d\alpha \right) \\
&=&F\left( \mathbf{X}_{\hbar },\hbar \right) \mathcal{E}\left( G\left( 
\mathbf{X}_{\alpha },\alpha \right) \right)
\end{eqnarray*}
as needed.

We close this paragraph by a general remark about the nature of the
expectation operation. Despite the formal similarities with the stochastic
calculus, the expectation defined above is really different, since the
expectation is in fact a function of operators evaluated at $\mathbf{X}%
_{\hbar }$, not a number. This has some consequences concerning the
cancellation of certain types of expectations that have no counterpart in
usual probability theory. But to derive such consequences, we first need to
define some more tools.

\subsection{EBS operator}

We now introduce a way to relate two expectations of the same function
evaluated at two different variables namely $\mathcal{E}\left( F\left( 
\mathbf{X}_{\alpha },\alpha \right) \right) $ to $\mathcal{E}\left( F\left( 
\mathbf{X}_{\alpha \prime },\alpha \right) \right) $ but with constant
explicit dependence in $\alpha $. Given our previous definitions Eq. $\left( %
\ref{dF}\right) $, one has 
\begin{align}
F\left( \mathbf{X}_{\alpha -d\alpha },\alpha -d\alpha \right) -F\left( 
\mathbf{X}_{\alpha },\alpha -d\alpha \right) & =-\left\langle F\left( 
\mathbf{X}_{\alpha },\alpha -d\alpha \right) \right\rangle d\alpha
-\sum_{i=1}^{6}\nabla _{X_{\alpha }^{i}}F\left( \mathbf{X}_{\alpha },\alpha
-d\alpha \right) dX_{\alpha }^{i}  \notag \\
& +\frac{1}{4}\sum_{i,j=1}^{6}\nabla _{X_{\alpha }^{j}}\nabla _{X_{\alpha
}^{i}}F\left( \mathbf{X}_{\alpha },\alpha -d\alpha \right) \left( dX_{\alpha
}^{i}dX_{\alpha }^{j}+dX_{\alpha }^{i}dX_{\alpha }^{j}\right)
\end{align}

The absence of the $\partial /\partial \alpha $ term implies that this
expression really depends on the choice of the symmetrization of the
function $F\left( \mathbf{X}_{\alpha },\alpha \right) $. As a consequence
considering the expectation, one has : 
\begin{equation}
\mathcal{E}\left( F\left( \mathbf{X}_{\alpha -d\alpha },\alpha -d\alpha
\right) \right) =\mathcal{E}\left( \left( 1-\left\langle .\right\rangle
d\alpha \right) F\left( \mathbf{X}_{\alpha },\alpha -d\alpha \right) \right)
\label{PP}
\end{equation}
where $\left\langle .\right\rangle $ is the bracket operator defined
previously. At this point we have to remark that, since the $\left\langle
.\right\rangle $ operation depends on the symmetrization procedure, the
right hand side of Eq.$\left( \ref{PP}\right) $ seems to depend on the
choice of symmetrization in the variables $\mathbf{X}_{\alpha }$ in $F\left( 
\mathbf{X}_{\alpha },\alpha -d\alpha \right) $. However, and this is the
most important for us, the operation that sends $F\left( \mathbf{X}_{\alpha
-d\alpha },\alpha -d\alpha \right) $ to $\left( 1-\left\langle
.\right\rangle d\alpha \right) F\left( \mathbf{X}_{\alpha },\alpha -d\alpha
\right) $ by first replacing $\mathbf{X}_{\alpha -d\alpha }$ by $\mathbf{X}%
_{\alpha }$ in $F$ and then applying $\left( 1-\left\langle .\right\rangle
d\alpha \right) $, is independent at the first order in $d\alpha $ of any
choice of symmetrization for the operators in the series expansion of $%
F\left( \mathbf{X}_{\alpha -d\alpha },\alpha -d\alpha \right) $. This is a
trivial fact since $\left( \left( 1-\left\langle .\right\rangle d\alpha
\right) F\left( \mathbf{X}_{\alpha },\alpha -d\alpha \right) \right)
=F\left( \mathbf{X}_{\alpha },\alpha \right) -\left( \frac{\partial }{%
\partial \alpha }+\left\langle .\right\rangle d\alpha \right) F\left( 
\mathbf{X}_{\alpha },\alpha \right) $, and this expression is independent of
the symmetrization choice as explained before. This operation, that will be
the only one relevant for us, is thus a consistent and independent of any
symmetrization. The apparent trouble with the right hand side in Eq.\ $%
\left( \ref{PP}\right) $ comes only from the fact that changing the order of
operators $\mathbf{X}_{\alpha -d\alpha }$ is not the same operation as
changing the order of operators $\mathbf{X}_{\alpha }$ due to different
commutation relations. To avoid any confusion a subscript $\alpha -d\alpha $
should be added to $F$ to remind that the symmetrization has to be
considered for the variables $\mathbf{X}_{\alpha -d\alpha }$, but it would
complicate the notations in a useless way for the present paper.

The meaning of our operator $1-\left\langle .\right\rangle d\alpha $ being
now clarified, the above differential equation Eq. $\left( \ref{PP}\right) $
can be integrated to get the following relation

\textbf{Proposition} 
\begin{equation}
\mathcal{E}\left( F\left( \mathbf{X}_{\alpha _{2}},\alpha _{2}\right)
\right) =\mathcal{E}\left( S_{\mathbf{X}_{\alpha _{1}}}\left[ T\exp \left(
-\int_{\alpha _{2}}^{\alpha _{1}}\left[ S_{\mathbf{X}_{\alpha
_{1}}}\left\langle .\right\rangle _{\alpha }S_{\mathbf{X}_{\alpha }}\right]
d\alpha \right) F\left( \mathbf{X}_{\alpha _{2}},\alpha _{2}\right) \right]
\right)  \label{CXION}
\end{equation}
\textit{valid for }$\alpha _{1}>\alpha _{2}.$

The bracket $\left\langle .\right\rangle _{\alpha }$ reminds that all
commutators between canonical variables have to be computed with values $%
i\alpha $. The operator $S_{\mathbf{X}_{\alpha }}$ is a shift operator that
sets the dynamical variables to $\mathbf{X}_{\alpha }$. that is $S_{\mathbf{X%
}_{\alpha }}F\left( \mathbf{X}_{\beta }\right) =F\left( \mathbf{X}_{\alpha
}\right) $. Its action does not depend on the relative values of $\alpha $
and $\beta $ and $S_{\mathbf{X}_{\alpha }}S_{\mathbf{X}_{\beta }}=S_{\mathbf{%
X}_{\alpha }}$ whatever the values of $\alpha $ and $\beta $. In addition, $%
T $ is the notation for the ordered exponential, that is : 
\begin{align*}
Te^{-\int_{\alpha _{2}}^{\alpha _{1}}S_{\mathbf{X}_{\alpha
_{1}}}\left\langle .\right\rangle _{\alpha }S_{\mathbf{X}_{\alpha }}d\alpha
}& =\sum_{n=0}^{\infty }\int\limits_{\alpha _{2}<\beta _{n}<..\beta
_{1}<\alpha _{1}}\left[ -S_{\mathbf{X}_{\alpha _{1}}}\left\langle
.\right\rangle _{\beta _{n}}S_{\mathbf{X}_{\beta _{n}}}\right] ..\left[ -S_{%
\mathbf{X}_{\alpha _{1}}}\left\langle .\right\rangle _{\beta _{1}}S_{\mathbf{%
X}_{\beta _{1}}}\right] d\beta _{1}..d\beta _{n} \\
& =\sum_{n=0}^{\infty }\int\limits_{\alpha _{2}<\beta _{n}<..\beta
_{1}<\alpha _{1}}\left[ -S_{\mathbf{X}_{\alpha _{1}}}\left\langle
.\right\rangle _{\beta _{n}}S_{\mathbf{X}_{\beta _{n}}}\right] ..\left[
-\left\langle .\right\rangle _{\beta _{1}}S_{\mathbf{X}_{\beta _{1}}}\right]
d\beta _{1}..d\beta _{n}
\end{align*}
Once again, the result is independent of the choice of initial
symmetrization chosen in the variables $\mathbf{X}_{\alpha _{2}}$. The
introduction of the shift $S_{\mathbf{X}_{\alpha }}$ is crucial and induced
by our previous considerations since the bracket operation has to be
accompanied by a shift of the variable all along the integration process, so
that ultimately $\mathbf{X}_{\alpha _{2}}$ is replaced by $\mathbf{X}%
_{\alpha _{1}}$.

\textbf{Proof.\ }

The proof of Eq. $\left( \ref{CXION}\right) $ is as follows. Define the
function : 
\begin{equation*}
G\left( \mathbf{X}_{\alpha _{1}},\alpha _{1}\right) =S_{\mathbf{X}_{\alpha
_{1}}}\left[ T\exp \left( -\int_{\alpha _{2}}^{\alpha _{1}}\left[ S_{\mathbf{%
X}_{\alpha _{1}}}\left\langle .\right\rangle _{\alpha }S_{\mathbf{X}_{\alpha
}}\right] d\alpha \right) F\left( \mathbf{X}_{\alpha _{2}},\alpha
_{2}\right) \right]
\end{equation*}
The dependence in the parameter $\alpha _{2}$ has been skipped for the sake
of convenience. Differentiating $G\left( \mathbf{X}_{\alpha _{1}},\alpha
_{1}\right) $ with respect to $\alpha _{1}$ yields : 
\begin{eqnarray*}
dG\left( \mathbf{X}_{\alpha _{1}},\alpha _{1}\right) &=&G\left( \mathbf{X}%
_{\alpha _{1}},\alpha _{1}\right) -G\left( \mathbf{X}_{\alpha _{1}-d\alpha
},\alpha _{1}-d\alpha \right) \\
&=&S_{\mathbf{X}_{\alpha _{1}}}\left( Te^{-\int_{\alpha _{2}}^{\alpha _{1}}%
\left[ S_{\mathbf{X}_{\alpha _{1}}}\left\langle .\right\rangle _{\alpha }S_{%
\mathbf{X}_{\alpha }}\right] d\alpha F\left( \mathbf{X}_{\alpha _{2}},\alpha
_{2}\right) }\right) -G\left( \mathbf{X}_{\alpha _{1}-d\alpha },\alpha
_{1}-d\alpha \right) \\
&=&S_{\mathbf{X}_{\alpha _{1}}}\left( 1-\left\langle .\right\rangle _{\alpha
_{1}-d\alpha }d\alpha \right) S_{\mathbf{X}_{\alpha _{1}-d\alpha
_{1}}}\left( Te^{-\int_{\alpha _{2}}^{\alpha _{1}-d\alpha _{1}}\left[ S_{%
\mathbf{X}_{\alpha _{1}}}\left\langle .\right\rangle _{\alpha }S_{\mathbf{X}%
_{\alpha }}\right] d\alpha }F\left( \mathbf{X}_{\alpha _{2}},\alpha
_{2}\right) \right) \\
&&-G\left( \mathbf{X}_{\alpha _{1}-d\alpha },\alpha _{1}-d\alpha \right) \\
&=&\left( 1-\left\langle .\right\rangle _{\alpha _{1}}d\alpha \right) S_{%
\mathbf{X}_{\alpha _{1}}}S_{\mathbf{X}_{\alpha _{1}-d\alpha _{1}}}\left[
Te^{-\int_{\alpha _{2}}^{\alpha _{1}-d\alpha _{1}}\left[ S_{\mathbf{X}%
_{\alpha _{1}}}\left\langle .\right\rangle _{\alpha }S_{\mathbf{X}_{\alpha }}%
\right] d\alpha }F\left( \mathbf{X}_{\alpha _{2}},\alpha _{2}\right) \right]
\\
&&-G\left( \mathbf{X}_{\alpha _{1}-d\alpha },\alpha _{1}-d\alpha \right)
\end{eqnarray*}
where the last equality is taken at the lowest order in $d\alpha $.
Considering the expectations of the quantities involved, and using Eq. $%
\left( \ref{PP}\right) $, we are thus ultimately left with : 
\begin{eqnarray*}
\mathcal{E}dG\left( \mathbf{X}_{\alpha _{1}},\alpha _{1}\right) &=&\mathcal{E%
}\left( \left( 1-\left\langle .\right\rangle _{\alpha _{1}}d\alpha \right)
S_{\mathbf{X}_{\alpha _{1}}}G\left( \mathbf{X}_{\alpha _{1}-d\alpha },\alpha
_{1}-d\alpha \right) -G\left( \mathbf{X}_{\alpha _{1}-d\alpha },\alpha
_{1}-d\alpha \right) \right) \\
&=&\mathcal{E}\left( \left( 1-\left\langle .\right\rangle _{\alpha
_{1}}d\alpha \right) G\left( \mathbf{X}_{\alpha _{1}},\alpha _{1}-d\alpha
\right) -G\left( \mathbf{X}_{\alpha _{1}-d\alpha },\alpha _{1}-d\alpha
\right) \right) =0
\end{eqnarray*}
and as a consequence, using the properties of the expectation operator : 
\begin{equation*}
\mathcal{E}\left( G\left( \mathbf{X}_{\alpha _{1}},\alpha _{1}\right)
-G\left( \mathbf{X}_{\alpha _{2}},\alpha _{2}\right) \right) =\mathcal{E}%
\int_{\alpha _{1}}^{\alpha _{2}}dG\left( \mathbf{X}_{\alpha _{1}},\alpha
_{1}\right) =\int_{\alpha _{1}}^{\alpha _{2}}\mathcal{E}dG\left( \mathbf{X}%
_{\alpha _{1}},\alpha _{1}\right) =0
\end{equation*}
The fact that $G\left( \mathbf{X}_{\alpha _{2}},\alpha _{2}\right) =F\left( 
\mathbf{X}_{\alpha _{2}},\alpha _{2}\right) $ ends the proof $\blacksquare $.

To gain some space in the sequel, we will define an abbreviation for the
previous operation. We will denote ultimately : 
\begin{equation}
\exp \left( -\left\langle .\right\rangle _{\alpha _{2}\rightarrow \alpha
_{1}}^{S}\right) \equiv S_{\mathbf{X}_{\alpha _{1}}}T\exp \left(
-\int_{\alpha _{2}}^{\alpha _{1}}S_{\mathbf{X}_{\alpha _{1}}}\left\langle
.\right\rangle _{\alpha }S_{\mathbf{X}_{\alpha }}d\alpha \right)  \label{EBS}
\end{equation}
as the \textbf{exponentiated bracket plus shift}, called EBS operation. As a
consequence, Eq. $\left( \ref{CXION}\right) $ rewrites as : 
\begin{equation}
\mathcal{E}\left( F\left( \mathbf{X}_{\alpha _{2}},\alpha _{2}\right)
\right) =\mathcal{E}\left( \exp \left( -\left\langle .\right\rangle _{\alpha
_{2}\rightarrow \alpha _{1}}^{S}\right) F\left( \mathbf{X}_{\alpha
_{2}},\alpha _{2}\right) \right) \text{ for }\alpha _{1}>\alpha _{2}
\label{CXION2}
\end{equation}
This last formula can be understood intuitively as follows. since the EBS
operation changes the function (by the action of the bracket defined above)
but also changes progressively the variables from $\mathbf{X}_{\alpha _{2}}$
to $\mathbf{X}_{\alpha _{1}}$(through the shift operator) both the bracket
operation and the shift of variable compensate each over to produce the
equality with the left hand side.

Let us ultimately insist on the fact that the expression in the right hand
side $\exp \left( -\left\langle .\right\rangle _{\alpha _{2}\rightarrow
\alpha _{1}}^{S}\right) F\left( \mathbf{X}_{\alpha _{2}},\alpha _{2}\right) $
is a function of $(\mathbf{X}_{\alpha _{1}},\alpha _{1})$ due to the action
of the shift and is therefore a different function than $F\left( \mathbf{X}%
_{\alpha _{2}},\alpha _{2}\right) .$ Only the expectations of both
expressions of Eq. (\ref{CXION2}) are equal.

In the particular case where $\alpha _{1}=\hbar $, since the right hand side
of Eq. (\ref{CXION2}) depends only on $(\mathbf{X}_{\hbar },\hbar )$ the
expection can be safely removed so that 
\begin{equation}
\mathcal{E}\left( F\left( \mathbf{X}_{\alpha _{2}},\alpha _{2}\right)
\right) =\exp \left( -\left\langle .\right\rangle _{\alpha _{2}\rightarrow
\hbar }^{S}\right) F\left( \mathbf{X}_{\alpha _{2}},\alpha _{2}\right)
\label{CXION2SIMPLE}
\end{equation}
\bigskip We end up this section by giving a full generalization of the
previous formula to a product of $n$ functions and show the

\textbf{Proposition} 
\begin{eqnarray}
&&\mathcal{E}\left( H_{1}\left( \mathbf{X}_{\alpha _{1}},\alpha _{1}\right)
....H_{n}\left( \mathbf{X}_{\alpha _{n}},\alpha _{n}\right) \right)  \notag
\\
&=&\mathcal{E}\left( e^{-\left\langle .\right\rangle _{\alpha
_{1}\rightarrow \alpha }^{S}}H_{1}\left( \mathbf{X}_{\alpha _{1}},\alpha
_{1}\right) e^{-\left\langle .\right\rangle _{\alpha _{2}\rightarrow \alpha
_{1}}^{S}}H_{2}\left( \mathbf{X}_{\alpha _{2}},\alpha _{2}\right)
...e^{-\left\langle .\right\rangle _{\alpha _{n}\rightarrow \alpha
_{n-1}}^{S}}H_{n}\left( \mathbf{X}_{\alpha _{n}},\alpha _{n}\right) \right)
\label{CXION3}
\end{eqnarray}
\textit{for }$\alpha _{n}<...<\alpha _{1}<\alpha $. \textit{This expression
for }$\alpha =\hbar $\textit{\ becomes } 
\begin{eqnarray}
&&\mathcal{E}\left( H_{1}\left( \mathbf{X}_{\alpha _{1}},\alpha _{1}\right)
....H_{n}\left( \mathbf{X}_{\alpha _{n}},\alpha _{n}\right) \right)  \notag
\\
&=&\mathcal{E}\left( e^{-\left\langle .\right\rangle _{\alpha
_{1}\rightarrow \hbar }^{S}}H_{1}\left( \mathbf{X}_{\alpha _{1}},\alpha
_{1}\right) e^{-\left\langle .\right\rangle _{\alpha _{2}\rightarrow \alpha
_{1}}^{S}}H_{2}\left( \mathbf{X}_{\alpha _{2}},\alpha _{2}\right)
...e^{-\left\langle .\right\rangle _{\alpha _{n}\rightarrow \alpha
_{n-1}}^{S}}H_{n}\left( \mathbf{X}_{\alpha _{n}},\alpha _{n}\right) \right)
\label{CXION4}
\end{eqnarray}

Note that the order of the functions in the product is irrelevant, and the
formula is also true if some functions are permuted.

\textbf{Proof.} We start the proof with the following lemma:

\textbf{Lemma}

Let $F_{1}\left( \mathbf{X}_{\alpha _{1}},\alpha _{1}\right) $, $F_{2}\left( 
\mathbf{X}_{\alpha _{2}},\alpha _{1}\right) $ be two arbitrary functions
with $\alpha _{2}<\alpha _{1}$. We state that : 
\begin{equation}
\mathcal{E}\left( dF_{2}\left( \mathbf{X}_{\alpha _{2}},\alpha _{2}\right)
\right) =0\Rightarrow \mathcal{E}\left( F_{1}\left( \mathbf{X}_{\alpha
_{1}},\alpha _{1}\right) dF_{2}\left( \mathbf{X}_{\alpha _{2}},\alpha
_{2}\right) \right) =0  \label{epsilondf2}
\end{equation}

This result has no counterpart in usual probability, and, as said in the
previous section, this is the consequence of our definition of the
expectation which sends an operator to an operator, not to a number.

\textbf{Proof of the Lemma}\textit{. }

We first show that, for $\alpha <\hbar $ and every function $F\left( \mathbf{%
X}_{\alpha },\alpha \right) $ that can be expanded in a series of $\left( 
\mathbf{X}_{\alpha },\alpha \right) $ converging \textquotedblright in the
classical sense, that is when the $\left( \mathbf{X}_{\alpha },\alpha
\right) $ are seen as commuting real variables, $\mathcal{E}\left( F\left( 
\mathbf{X}_{\alpha },\alpha \right) \right) =0$ implies that $F\left( 
\mathbf{X}_{\alpha },\alpha \right) =0$. This is trivially true for a
polynomial of degree $0$ or $1$ in $\mathbf{X}_{\alpha }$ since $\mathbf{X}%
_{\alpha }=\mathbf{X}_{\hbar }-\int_{\alpha }^{\hbar }d\mathbf{X}_{\beta }$
and $\mathcal{E}\mathbf{X}_{\alpha }=\mathbf{X}_{\hbar }-\int_{\alpha
}^{\hbar }\mathcal{E}d\mathbf{X}_{\beta }=\mathbf{X}_{\hbar }$ as derived
from Eq. $\left( \ref{indep}\right) $. As a consequence $\mathcal{E}\left( a%
\mathbf{X}_{\alpha }+b\right) =\left( a\mathbf{X}_{\hbar }+b\right) $ which
is null only if $a=b=0$.

Now, assume that $\mathcal{E}\left( F\left( \mathbf{X}_{\alpha },\alpha
\right) \right) =0\Rightarrow F\left( \mathbf{X}_{\alpha },\alpha \right) =0$
for all polynomial of degree lower or equal to $N$ in the variables $\mathbf{%
X}_{\alpha }$. Consider then for $F\left( \mathbf{X}_{\alpha },\alpha
\right) $ a polynomial of degree $N+1$ (the coefficients depend on $\alpha $%
). Then using the EBS operator, we can write : 
\begin{equation*}
\mathcal{E}\left( F\left( \mathbf{X}_{\alpha },\alpha \right) \right) =%
\mathcal{E}\left( \exp \left( -\left\langle .\right\rangle _{\alpha
\rightarrow \alpha _{1}}^{S}\right) F\left( \mathbf{X}_{\alpha },\alpha
\right) \right)
\end{equation*}
given the bracket operation is defined through a derivative of second order,
we thus have : 
\begin{equation*}
\mathcal{E}\left( F\left( \mathbf{X}_{\alpha },\alpha \right) \right)
=F\left( \mathbf{X}_{\hbar },\alpha \right) +\text{Polynomial of degree }N-1%
\text{ in }\mathbf{X}_{\hbar }
\end{equation*}
If this is $0$, then necessarily the monomials of degree $N+1$ in $F\left( 
\mathbf{X}_{\hbar },\alpha \right) $ have to be null. As a consequence $%
F\left( \mathbf{X}_{\hbar },\alpha \right) $ is of degree $N$ and thus null
by hypothesis. As a consequence, the lemma to be shown is true for every
polynomial, and by a density argument for all the kind of series considered.

In a second step, notice that a direct consequence of the above proposition
is that 
\begin{equation*}
\mathcal{E}\left( dF\left( \mathbf{X}_{\alpha },\alpha \right) \right)
=0\Rightarrow D_{\alpha }F\left( \mathbf{X}_{\alpha },\alpha \right) =0
\end{equation*}
actually, $\mathcal{E}\left( dF\left( \mathbf{X}_{\alpha },\alpha \right)
\right) =\mathcal{E}\left( D_{\alpha }F\left( \mathbf{X}_{\alpha },\alpha
\right) \right) d\alpha =0$, which implies that $D_{\alpha }F\left( \mathbf{X%
}_{\alpha },\alpha \right) =0$.

We can then prove the required proposition. Actually, by construction of our
expectation operator, since $\alpha _{2}<\alpha _{1}$, 
\begin{eqnarray*}
&&\mathcal{E}\left( F_{1}\left( \mathbf{X}_{\alpha _{1}},\alpha _{1}\right)
dF_{2}\left( \mathbf{X}_{\alpha _{2}},\alpha _{2}\right) \right) \\
&=&\mathcal{E}\left( F_{1}\left( \mathbf{X}_{\alpha _{1}},\alpha _{1}\right)
\left( D_{\alpha _{2}}F_{2}\left( \mathbf{X}_{\alpha _{2}},\alpha
_{2}\right) d\alpha _{2}+\nabla _{X_{\alpha _{2}}^{i}}F_{2}\left( \mathbf{X}%
_{\alpha _{2}},\alpha _{2}\right) dX_{\alpha _{2}}^{i}\right) \right) \\
&&+\mathcal{E}\left( F_{1}\left( \mathbf{X}_{\alpha _{1}},\alpha _{1}\right)
\left( -\frac{1}{4}\nabla _{X_{\alpha _{2}}^{j}}\nabla _{X_{\alpha
_{2}}^{i}}F_{2}\left( \mathbf{X}_{\alpha _{2}},\alpha _{2}\right) dX_{\alpha
_{2}}^{i}dX_{\alpha _{2}}^{j}+dX_{\alpha _{2}}^{j}dX_{\alpha
_{2}}^{i}\right) \right) \\
&=&\mathcal{E}\left( F_{1}\left( \mathbf{X}_{\alpha _{1}},\alpha _{1}\right)
D_{\alpha _{2}}F_{2}\left( \mathbf{X}_{\alpha _{2}},\alpha _{2}\right)
\right) d\alpha _{2} \\
&=&0
\end{eqnarray*}
the second equality is a consequence of Eq. $\left( \ref{indep}\right) $.
This last result shows the lemma $\blacksquare $.

The proof of the proposition is now in two steps. We first consider a
product of two functions and show : 
\begin{eqnarray}
\mathcal{E}\left( H_{1}\left( \mathbf{X}_{\alpha _{1}},\alpha _{1}\right)
H_{2}\left( \mathbf{X}_{\alpha _{2}},\alpha _{2}\right) \right) &=&\mathcal{E%
}\left( e^{-\left\langle .\right\rangle _{\alpha _{1}\rightarrow \alpha
}^{S}}H_{1}\left( \mathbf{X}_{\alpha _{1}},\alpha _{1}\right)
e^{-\left\langle .\right\rangle _{\alpha _{2}\rightarrow \alpha
_{1}}^{S}}H_{2}\left( \mathbf{X}_{\alpha _{2}},\alpha _{2}\right) \right) 
\text{ }  \notag \\
\text{for }\alpha _{1} &>&\alpha _{2}
\end{eqnarray}
To do so, let us start again with $G\left( \mathbf{X}_{\alpha _{\gamma
}},\gamma \right) =\exp \left( -\left\langle .\right\rangle _{\alpha
_{2}\rightarrow \gamma }^{S}\right) H_{2}\left( \mathbf{X}_{\alpha
_{2}},\alpha _{2}\right) $, $\gamma $ is an arbitrary parameter varying
between $\alpha _{1}$ and $\alpha _{2}$. The dependence in $\alpha _{2}$ in $%
G\left( \mathbf{X}_{\alpha _{\gamma }},\gamma \right) $ is forgotten here
for the sake of simplicity. By differentiation with respect to $\gamma $ we
have, as before, at the lowest order in $d\alpha $ : 
\begin{equation*}
\mathcal{E}\left( dG\left( \mathbf{X}_{\alpha _{\gamma }},\gamma \right)
\right) =\mathcal{E}\left( H_{2}\left( \mathbf{X}_{\alpha _{\gamma }},\gamma
\right) -H_{2}\left( \mathbf{X}_{\alpha _{\gamma -d\gamma }},\gamma -d\gamma
\right) \right) =0
\end{equation*}
Now, multiply by $H_{1}\left( \mathbf{X}_{\alpha _{1}},\alpha _{1}\right) $
and take the expectation. Since $\gamma <\alpha _{1}$, one has using Eq. $%
\left( \ref{epsilondf2}\right) $ : 
\begin{equation*}
\mathcal{E}\left( H_{1}\left( \mathbf{X}_{\alpha _{1}},\alpha _{1}\right)
dG\left( \mathbf{X}_{\alpha _{\gamma }},\gamma \right) \right) =0
\end{equation*}
the integration of this last relation (recall that the integral and the
expectation commute), yields : 
\begin{equation*}
\mathcal{E}\left( H_{1}\left( \mathbf{X}_{\alpha _{1}},\alpha _{1}\right)
\int_{\alpha _{2}}^{\alpha _{1}}dG\left( \mathbf{X}_{\alpha },\alpha \right)
\right) =0
\end{equation*}
that is : 
\begin{equation*}
\mathcal{E}\left( H_{1}\left( \mathbf{X}_{\alpha _{1}},\alpha _{1}\right)
e^{-\left\langle .\right\rangle _{\alpha _{2}\rightarrow \alpha
_{1}}^{S}}H_{2}\left( \mathbf{X}_{\alpha _{2}},\alpha _{2}\right) \right) =%
\mathcal{E}\left( H_{1}\left( \mathbf{X}_{\alpha _{1}},\alpha _{1}\right)
H_{2}\left( \mathbf{X}_{\alpha _{2}},\alpha _{2}\right) \right)
\end{equation*}
Then, use Eq. $\left( \ref{CXION2}\right) $ with $F\left( \mathbf{X}_{\alpha
_{1}},\alpha _{1}\right) =H_{1}\left( \mathbf{X}_{\alpha _{1}},\alpha
_{1}\right) \exp \left( -\left\langle .\right\rangle _{\alpha
_{2}\rightarrow \alpha _{1}}^{S}\right) H_{2}\left( \mathbf{X}_{\alpha
_{2}},\alpha _{2}\right) $ between $\alpha _{1}$ and $\hbar $\ to get : 
\begin{eqnarray*}
\mathcal{E}\left( H_{1}\left( \mathbf{X}_{\alpha _{1}},\alpha _{1}\right)
H_{2}\left( \mathbf{X}_{\alpha _{2}},\alpha _{2}\right) \right) &=&\mathcal{E%
}\left( H_{1}\left( \mathbf{X}_{\alpha _{1}},\alpha _{1}\right)
e^{-\left\langle .\right\rangle _{\alpha _{2}\rightarrow \alpha
_{1}}^{S}}H_{2}\left( \mathbf{X}_{\alpha _{2}},\alpha _{2}\right) \right) \\
&=&\mathcal{E}\left( e^{-\left\langle .\right\rangle _{\alpha
_{1}\rightarrow \alpha }^{S}}H_{1}\left( \mathbf{X}_{\alpha _{1}},\alpha
_{1}\right) e^{-\left\langle .\right\rangle _{\alpha _{2}\rightarrow \alpha
_{1}}^{S}}H_{2}\left( \mathbf{X}_{\alpha _{2}},\alpha _{2}\right) \right)
\end{eqnarray*}
which is the required result. In a second step, the generalization to a
product of $n$ arbitrary functions is shown recursively by starting from the
right and replacing $H_{1}\left( \mathbf{X}_{\alpha _{n}},\alpha _{n}\right) 
$ by $\exp \left( -\left\langle .\right\rangle _{\alpha _{n}\rightarrow
\alpha _{n-1}}^{S}\right) H_{n}\left( \mathbf{X}_{\alpha _{n}},\alpha
_{n}\right) $, then to do the same with $H_{1}\left( \mathbf{X}_{\alpha
_{n-1}},\alpha _{n-1}\right) \exp \left( -\left\langle .\right\rangle
_{\alpha _{n}\rightarrow \alpha _{n-1}}^{S}\right) H_{n}\left( \mathbf{X}%
_{\alpha _{n}},\alpha _{n}\right) $ and so on $\blacksquare $.

We can now exploit the mathematical construction developed in this section
to consider the formal diagonalization of an arbitrary matrix valued quantum
Hamiltonian.

\section{The diagonalization procedure}

We now consider a generic matrix valued quantum Hamiltonian $H\left( \mathbf{%
R},\mathbf{P}\right) $ where $\mathbf{R}$ and $\mathbf{P}$ are the usual
canonical coordinate and momentum operators satisfying the canonical
Heisenberg algebra. Our goal is the find an unitary transformation $U$ such
that $UHU^{+}$ is a diagonal matrix valued operator $\varepsilon\left( 
\mathbf{R,P}\right) $ (block diagonal for the Dirac Hamiltonian). This is in
general an excessively difficult mathematical problem. For this reason we
consider this problem by dividing it in several steps.

\subsection{The Hamiltonian}

First we introduce the unitary matrix\textbf{\ }$U_{\alpha }\left( \mathbf{X}%
_{\alpha }\right) \equiv U\left( \mathbf{X}_{\alpha },\alpha \right) $%
\textbf{\ }which diagonalizes the Hamiltonian $H\left( \mathbf{X}_{\alpha
}\right) $ where the canonical variables $\mathbf{X}_{\hbar }\equiv \mathbf{X%
}$ have been replaced by the running ones $\mathbf{X}_{\alpha }$, so that we
can write 
\begin{equation}
U_{\alpha }\left( \mathbf{X}_{\alpha }\right) H\left( \mathbf{X}_{\alpha
}\right) U_{\alpha }^{+}\left( \mathbf{X}_{\alpha }\right) =\varepsilon
_{\alpha }\left( \mathbf{X}_{\alpha }\right)
\end{equation}
with $\varepsilon _{\alpha }\left( \mathbf{X}_{\alpha }\right) \equiv
\varepsilon _{\alpha }\left( \mathbf{X}_{\alpha },\alpha \right) $. With the
help of the previous identities Eqs. $\left( \ref{FXII}\right) \left( \ref%
{epsi1}\right) \left( \ref{epsi2}\right) $, we can compute $\varepsilon
_{\hbar }\left( \mathbf{X}_{\hbar }\right) \equiv \varepsilon \left( \mathbf{%
X}\right) $. Indeed as $\varepsilon \left( \mathbf{X}\right) =\mathcal{E}%
\left( \varepsilon \left( \mathbf{X}\right) \right) $ we can write 
\begin{equation}
\varepsilon \left( \mathbf{X}\right) =\mathcal{E}\left( \varepsilon
_{0}\left( \mathbf{X}_{0}\right) +\int_{0}^{\hbar }d\varepsilon _{\alpha
_{1}}\left( \mathbf{X}_{\alpha _{1}}\right) \right) .  \label{deltaepsilon}
\end{equation}
Clearly $\varepsilon _{0}\left( \mathbf{X}_{0}\right) $\ corresponds to the
diagonal representation of the original Hamiltonian $H(\mathbf{X})$\ where
the canonical operators $\mathbf{X}$\ have been replaced by the classical
variables $\mathbf{X}_{0}.$\ In practice, it is usually quite easy to
diagonalize the Hamiltonian when the operators are commuting. This is the
essence of the method. Starting with classical variables we can recursively
introduce more and more \textquotedblright quantification\textquotedblright\
through the running parameter $\alpha $\ until we get the full quantum
Hamiltonian. This procedure is now described in the following.

The quantity $\mathcal{E}\left( d\varepsilon _{\alpha _{1}}\right) =\mathcal{%
E}\left( \left( \frac{\partial }{\partial \alpha _{1}}+\left\langle
.\right\rangle \right) \varepsilon _{\alpha _{1}}\left( \mathbf{X}_{\alpha
_{1}}\right) \right) d\alpha _{1}$ can be straightforwardly computed by
using the rule previously given for the bracket of a product Eq. $\left( \ref%
{BFG}\right) $. Indeed we find 
\begin{align}
\left( \frac{\partial }{\partial \alpha _{1}}+\left\langle .\right\rangle
\right) \varepsilon _{\alpha _{1}}\left( \mathbf{X}_{\alpha _{1}}\right) & =%
\frac{\partial }{\partial \alpha _{1}}U_{\alpha _{1}}U_{\alpha
_{1}}^{+}\varepsilon _{\alpha _{1}}+\varepsilon _{\alpha _{1}}U_{\alpha _{1}}%
\frac{\partial }{\partial \alpha _{1}}U_{\alpha _{1}}^{+}+U_{\alpha _{1}}%
\frac{\partial }{\partial \alpha _{1}}H\left( \mathbf{X}_{\alpha
_{1}}\right) U_{\alpha _{1}}^{+}  \notag \\
& +\left\langle U_{\alpha _{1}}\right\rangle U_{\alpha _{1}}^{+}\varepsilon
_{\alpha _{1}}+U_{\alpha _{1}}\left\langle H\left( \mathbf{X}_{\alpha
_{1}}\right) \right\rangle U_{\alpha _{1}}^{+}+\varepsilon _{\alpha
_{1}}U_{\alpha _{1}}\left\langle U_{\alpha _{1}}^{+}\right\rangle  \notag \\
& -\frac{i}{2}\left( \nabla _{P_{i}}U_{\alpha _{1}}\nabla _{R_{i}}H\left( 
\mathbf{X}_{\alpha _{1}}\right) U_{\alpha _{1}}^{+}-\nabla _{R_{i}}U_{\alpha
_{1}}\nabla _{P_{i}}H\left( \mathbf{X}_{\alpha _{1}}\right) U_{\alpha
_{1}}^{+}\right)  \notag \\
& -\frac{i}{2}\left( U_{\alpha _{1}}\nabla _{P_{i}}H\left( \mathbf{X}%
_{\alpha _{1}}\right) \nabla _{R_{i}}U_{\alpha _{1}}^{+}-U_{\alpha
_{1}}\nabla _{R_{i}}H\left( \mathbf{X}_{\alpha _{1}}\right) \nabla
_{P_{i}}U_{\alpha _{1}}^{+}\right)  \notag \\
& -\frac{i}{2}\left( \nabla _{P_{i}}U_{\alpha _{1}}H\left( \mathbf{X}%
_{\alpha _{1}}\right) \nabla _{R_{i}}U_{\alpha _{1}}^{+}-\nabla
_{R_{i}}U_{\alpha _{1}}H\left( \mathbf{X}_{\alpha _{1}}\right) \nabla
_{P_{i}}U_{\alpha _{1}}^{+}\right)  \label{diagoanc}
\end{align}
As in \cite{SERIESPIERRE} we now introduce the notations $\mathcal{A}%
_{\alpha _{1}}^{R_{l}}=iU_{\alpha _{1}}\left( \mathbf{X}_{\alpha
_{1}}\right) \nabla _{P_{i}}U_{\alpha _{1}}^{+}\left( \mathbf{X}_{\alpha
_{1}}\right) $ and $\mathcal{A}_{\alpha _{1}}^{P_{l}}=-iU_{\alpha
_{1}}\left( \mathbf{X}_{\alpha _{1}}\right) \nabla _{R_{i}}U_{\alpha
_{1}}^{+}\left( \mathbf{X}_{\alpha _{1}}\right) $ so that Eq. $\left( \ref%
{diagoanc}\right) $ can be written in a more useful form in terms of
quantities of physical interest. Actually, replacing $H\left( \mathbf{X}%
_{\alpha _{1}}\right) $ by $U_{\alpha _{1}}^{+}\varepsilon _{\alpha
_{1}}U_{\alpha _{1}}$ everywhere, allows to write 
\begin{eqnarray*}
&&-\frac{i}{2}\left( \nabla _{P_{i}}U_{\alpha _{1}}\nabla _{R_{i}}H\left( 
\mathbf{X}_{\alpha _{1}}\right) U_{\alpha _{1}}^{+}-\nabla _{R_{i}}U_{\alpha
_{1}}\nabla _{P_{i}}H\left( \mathbf{X}_{\alpha _{1}}\right) U_{\alpha
_{1}}^{+}\right) \\
&&-\frac{i}{2}\left( U_{\alpha _{1}}\nabla _{P_{i}}H\left( \mathbf{X}%
_{\alpha _{1}}\right) \nabla _{R_{i}}U_{\alpha _{1}}^{+}-U_{\alpha
_{1}}\nabla _{R_{i}}H\left( \mathbf{X}_{\alpha _{1}}\right) \nabla
_{P_{i}}U_{\alpha _{1}}^{+}\right) \\
&=&\frac{1}{2}\mathcal{A}_{\alpha _{1}}^{R_{l}}\nabla _{R_{l}}\varepsilon
_{\alpha _{1}}+\nabla _{R_{l}}\varepsilon _{\alpha _{1}}\mathcal{A}_{\alpha
_{1}}^{R_{l}}+\mathcal{A}_{\alpha _{1}}^{P_{l}}\nabla _{P_{l}}\varepsilon
_{\alpha _{1}}+\nabla _{P_{l}}\varepsilon _{\alpha _{1}}\mathcal{A}_{\alpha
_{1}}^{P_{l}} \\
&&+\frac{i}{2}\left[ \mathcal{A}_{\alpha _{1}}^{R_{l}},\mathcal{A}_{\alpha
_{1}}^{P_{l}}\right] \varepsilon _{\alpha _{1}}+\frac{i}{2}\varepsilon
_{\alpha _{1}}\left[ \mathcal{A}_{\alpha _{1}}^{R_{l}},\mathcal{A}_{\alpha
_{1}}^{P_{l}}\right] -i\mathcal{A}_{\alpha _{1}}^{R_{l}}\varepsilon _{\alpha
_{1}}\mathcal{A}_{\alpha _{1}}^{P_{l}}+i\mathcal{A}_{\alpha
_{1}}^{P_{l}}\varepsilon _{\alpha _{1}}\mathcal{A}_{\alpha _{1}}^{R_{l}}
\end{eqnarray*}
and 
\begin{eqnarray*}
&&-\frac{i}{2}\left( \nabla _{P_{i}}U_{\alpha _{1}}H\left( \mathbf{X}%
_{\alpha _{1}}\right) \nabla _{R_{i}}U_{\alpha _{1}}^{+}-\nabla
_{R_{i}}U_{\alpha _{1}}H\left( \mathbf{X}_{\alpha _{1}}\right) \nabla
_{P_{i}}U_{\alpha _{1}}^{+}\right) \\
&=&\frac{i}{2}\mathcal{A}_{\alpha _{1}}^{R_{l}}\varepsilon _{\alpha _{1}}%
\mathcal{A}_{\alpha _{1}}^{P_{l}}-\frac{i}{2}\mathcal{A}_{\alpha
_{1}}^{P_{l}}\varepsilon _{\alpha _{1}}\mathcal{A}_{\alpha _{1}}^{R_{l}}
\end{eqnarray*}
so that ultimately, one has: 
\begin{align}
\left( \frac{\partial }{\partial \alpha _{1}}+\left\langle .\right\rangle
\right) \varepsilon _{\alpha _{1}}\left( \mathbf{X}_{\alpha _{1}}\right) &
=\left( \left( \frac{\partial }{\partial \alpha _{1}}+\left\langle
.\right\rangle \right) U_{\alpha _{1}}\right) U_{\alpha _{1}}^{+}\varepsilon
_{\alpha _{1}}+\varepsilon _{\alpha _{1}}U_{\alpha _{1}}\left( \frac{%
\partial }{\partial \alpha _{1}}+\left\langle .\right\rangle \right)
U_{\alpha _{1}}^{+}  \notag \\
& +U_{\alpha _{1}}\left( \frac{\partial }{\partial \alpha _{1}}+\left\langle
.\right\rangle \right) H\left( \mathbf{X}_{\alpha _{1}}\right) U_{\alpha
_{1}}^{+}  \notag \\
& +\frac{1}{2}\mathcal{A}_{\alpha _{1}}^{R_{l}}\nabla _{R_{l}}\varepsilon
_{\alpha _{1}}+\nabla _{R_{l}}\varepsilon _{\alpha _{1}}\mathcal{A}_{\alpha
_{1}}^{R_{l}}+\mathcal{A}_{\alpha _{1}}^{P_{l}}\nabla _{P_{l}}\varepsilon
_{\alpha _{1}}+\nabla _{P_{l}}\varepsilon _{\alpha _{1}}\mathcal{A}_{\alpha
_{1}}^{P_{l}}  \notag \\
& +\frac{i}{2}\left\{ \mathcal{A}_{\alpha _{1}}^{P_{l}}\varepsilon _{\alpha
_{1}}\mathcal{A}_{\alpha _{1}}^{R_{l}}-\mathcal{A}_{\alpha
_{1}}^{R_{l}}\varepsilon _{\alpha _{1}}\mathcal{A}_{\alpha
_{1}}^{P_{l}}+\varepsilon _{\alpha _{1}}\left[ \mathcal{A}_{\alpha
_{1}}^{R_{l}},\mathcal{A}_{\alpha _{1}}^{P_{l}}\right] +\left[ \mathcal{A}%
_{\alpha _{1}}^{R_{l}},\mathcal{A}_{\alpha _{1}}^{P_{l}}\right] \varepsilon
_{\alpha _{1}}\right\}  \label{diago}
\end{align}

We note that by construction $d\varepsilon _{\alpha _{1}}$ is a diagonal
matrix and we obviously have the following identities

\begin{align}
\left( \frac{\partial }{\partial \alpha _{1}}+\left\langle .\right\rangle
\right) \varepsilon _{\alpha _{1}}\left( \mathbf{X}_{\alpha _{1}}\right) & =%
\mathcal{P}_{+}\left( \text{R.H.S. of Eq. \ref{diago}}\right)  \label{pplus}
\\
0& =\mathcal{P}_{-}\left( \text{R.H.S. of Eq. \ref{diago}}\right)
\label{pmoins}
\end{align}
where $\mathcal{P}_{+}$ and $\mathcal{P}_{-}$ are the projection on the
diagonal and off the diagonal respectively. 
\begin{equation}
\left( \frac{\partial }{\partial \alpha _{1}}+\left\langle .\right\rangle
\right) \varepsilon _{\alpha _{1}}\left( \mathbf{X}_{\alpha _{1}}\right)
=O_{\alpha _{1}}.\varepsilon _{\alpha _{1}}\left( \mathbf{X}_{\alpha
_{1}}\right) +\mathcal{P}_{+}\left\{ U_{\alpha _{1}}\left( D_{\alpha }\left( 
\mathbf{X}_{\alpha _{1}},\alpha _{1}\right) H\left( \mathbf{X}_{\alpha
_{1}}\right) \right) U_{\alpha _{1}}^{+}\right\}  \label{O}
\end{equation}
with $O_{\alpha _{1}}$ is given by the projected \textquotedblright linear
part\textquotedblright\ of the r.h.s. expression of Eq. $\left( \ref{diago}%
\right) $ :

\begin{align}
O_{\alpha _{1}}\varepsilon _{\alpha _{1}}\left( \mathbf{X}_{\alpha
_{1}}\right) & =\mathcal{P}_{+}\left( \left( D_{\alpha }\left( \mathbf{X}%
_{\alpha _{1}},\alpha _{1}\right) U_{\alpha _{1}}\right) U_{\alpha
_{1}}^{+}\varepsilon _{\alpha _{1}}+\varepsilon _{\alpha _{1}}U_{\alpha
_{1}}D_{\alpha }\left( \mathbf{X}_{\alpha _{1}},\alpha _{1}\right) U_{\alpha
_{1}}^{+}\right)  \notag \\
& +\mathcal{P}_{+}\left\{ \frac{1}{2}\mathcal{A}_{\alpha _{1}}^{R_{l}}\nabla
_{R_{l}}\varepsilon _{\alpha _{1}}+\nabla _{R_{l}}\varepsilon _{\alpha _{1}}%
\mathcal{A}_{\alpha _{1}}^{R_{l}}+\mathcal{A}_{\alpha _{1}}^{P_{l}}\nabla
_{P_{l}}\varepsilon _{\alpha _{1}}+\nabla _{P_{l}}\varepsilon _{\alpha _{1}}%
\mathcal{A}_{\alpha _{1}}^{P_{l}}\right\}  \notag \\
& +\frac{i}{2}\mathcal{P}_{+}\left\{ \mathcal{A}_{\alpha
_{1}}^{P_{l}}\varepsilon _{\alpha _{1}}\mathcal{A}_{\alpha _{1}}^{R_{l}}-%
\mathcal{A}_{\alpha _{1}}^{R_{l}}\varepsilon _{\alpha _{1}}\mathcal{A}%
_{\alpha _{1}}^{P_{l}}+\varepsilon _{\alpha _{1}}\left[ \mathcal{A}_{\alpha
_{1}}^{R_{l}},\mathcal{A}_{\alpha _{1}}^{P_{l}}\right] +\left[ \mathcal{A}%
_{\alpha _{1}}^{R_{l}},\mathcal{A}_{\alpha _{1}}^{P_{l}}\right] \varepsilon
_{\alpha _{1}}\right\} .  \notag
\end{align}

Let us remark that the operator $O_{\alpha _{1}}$ is well defined since, as
explained before, the operation $D_{\alpha }\left( \mathbf{X}_{\alpha
_{1}},\alpha _{1}\right) =\left( \frac{\partial }{\partial \alpha _{1}}%
+\left\langle .\right\rangle \right) $ is independent from any
symmetrization scheme for $\varepsilon _{\alpha _{1}}\left( \mathbf{X}%
_{\alpha _{1}}\right) $ and $H\left( \mathbf{X}_{\alpha _{1}}\right) $, and
so is $O_{\alpha _{1}}.\varepsilon _{\alpha _{1}}\left( \mathbf{X}_{\alpha
_{1}}\right) =D_{\alpha }\left( \mathbf{X}_{\alpha _{1}},\alpha _{1}\right)
\varepsilon _{\alpha _{1}}\left( \mathbf{X}_{\alpha _{1}}\right) -\mathcal{P}%
_{+}\left\{ U_{\alpha _{1}}\left( D_{\alpha }\left( \mathbf{X}_{\alpha
_{1}},\alpha _{1}\right) H\left( \mathbf{X}_{\alpha _{1}}\right) \right)
U_{\alpha _{1}}^{+}\right\} $.

Since in the applications of practical interest (Bloch electrons, Dirac
Hamiltonian...), $\left( D_{\alpha }\left( \mathbf{X}_{\alpha _{1}},\alpha
_{1}\right) H\left( \mathbf{X}_{\alpha _{1}}\right) \right) $ cancels, we
will set this term to $0$ for the sake of the exposition and will consider
later its contribution. As a consequence, one has : 
\begin{equation}
\varepsilon \left( \mathbf{X}\right) =\mathcal{E}\left( \varepsilon
_{0}\left( \mathbf{X}_{0}\right) +\int_{0}^{\hbar }O_{\alpha
_{1}}.\varepsilon _{\alpha _{1}}\left( \mathbf{X}_{\alpha _{1}}\right)
d\alpha _{1}\right) .
\end{equation}
Now, similarly to Eq. $\left( \ref{deltaepsilon}\right) $ we can write : 
\begin{equation}
\varepsilon _{\alpha _{1}}\left( \mathbf{X}_{\alpha _{1}}\right)
=\varepsilon _{0}\left( \mathbf{X}_{0}\right) +\int_{0}^{\alpha
_{1}}d\varepsilon _{\alpha _{2}}\left( \mathbf{X}_{\alpha _{2}}\right)
=\varepsilon _{0}\left( \mathbf{X}_{0}\right) +\int_{0}^{\alpha _{1}}d\left(
U\left( \mathbf{X}_{\alpha _{2}}\right) H\left( \mathbf{X}_{\alpha
_{2}}\right) U^{+}\left( \mathbf{X}_{\alpha _{2}}\right) \right)
\end{equation}
which can be inserted in the expectation Eq. $\left( \ref{deltaepsilon}%
\right) $ to get the full quantum diagonal representation $\varepsilon
\left( \mathbf{X}\right) \equiv \varepsilon _{\hbar }\left( \mathbf{X}%
_{\hbar }\right) $ as : 
\begin{align}
\varepsilon _{\hbar }\left( \mathbf{X}_{\hbar }\right) & =\mathcal{E}\left(
\varepsilon _{0}\left( \mathbf{X}_{0}\right) +\int_{0}^{\hbar }d\varepsilon
_{\alpha _{1}}\left( \mathbf{X}_{\alpha _{1}}\right) d\alpha _{1}\right) 
\notag \\
& =\mathcal{E}\left( \varepsilon _{0}\left( \mathbf{X}_{0}\right)
+\int_{0}^{\hbar }O_{\alpha _{1}}.\varepsilon _{\alpha _{1}}\left( \mathbf{X}%
_{\alpha _{1}}\right) d\alpha _{1}\right)  \notag \\
& =\mathcal{E}\left( \varepsilon _{0}\left( \mathbf{X}_{0}\right)
+\int_{0}^{\hbar }O_{\alpha _{1}}.\left[ \varepsilon _{0}\left( \mathbf{X}%
_{0}\right) +\int_{0}^{\alpha _{1}}d\left[ U\left( \mathbf{X}_{\alpha
_{2}}\right) H\left( \mathbf{X}_{\alpha _{2}}\right) U^{+}\left( \mathbf{X}%
_{\alpha _{2}}\right) \right] \right] d\alpha _{1}\right)  \label{eqAA}
\end{align}
Now remark that since $\alpha _{2}<\alpha _{1}$, the terms $\mathbf{\nabla }%
_{\mathbf{X}_{\alpha _{2}}}\varepsilon _{\alpha _{2}}\left( \mathbf{X}%
_{\alpha _{2}}\right) d\mathbf{X}_{\alpha _{2}}$ and $\mathbf{\nabla }%
_{X_{\alpha _{2}}^{j}}\mathbf{\nabla }_{X_{\alpha _{2}}^{i}}\varepsilon
_{\alpha _{2}}\left( \mathbf{X}_{\alpha _{2}}\right) \left( dX_{\alpha
_{2}}^{i}dX_{\alpha _{2}}^{j}+dX_{\alpha _{2}}^{i}dX_{\alpha
_{2}}^{j}\right) $ in $d\varepsilon \left( \mathbf{X}_{\alpha _{2}}\right) =d%
\left[ U\left( \mathbf{X}_{\alpha _{2}}\right) H\left( \mathbf{X}_{\alpha
_{2}}\right) U^{+}\left( \mathbf{X}_{\alpha _{2}}\right) \right] $ do not
recombine with anything coming from $O_{\alpha _{1}}\left( \mathbf{X}%
_{\alpha _{1}}\right) $ to give a product of the type $d\mathbf{R}_{\alpha
_{2}}d\mathbf{P}_{\alpha _{2}}$that would induce a\textbf{\ }$d\alpha _{2}$
contribution to the expectation. They will thus cancel in the expectation.
As a consequence, Eq. $\left( \ref{eqAA}\right) $ can be written: 
\begin{equation*}
\varepsilon _{\hbar }\left( \mathbf{X}_{\hbar }\right) =\mathcal{E}\left(
\varepsilon _{0}\left( \mathbf{X}_{0}\right) +\int_{0}^{\hbar }O_{\alpha
_{1}}.\left[ \varepsilon _{0}\left( \mathbf{X}_{0}\right) +\int_{0}^{\alpha
_{1}}O_{\alpha _{2}}.\varepsilon _{\alpha _{2}}\left( \mathbf{X}_{\alpha
_{2}}\right) d\alpha _{2}\right] d\alpha _{1}\right)
\end{equation*}
Repeating the procedure one can then show by iteration that : 
\begin{equation}
\varepsilon _{\hbar }\left( \mathbf{X}_{\hbar }\right) =\mathcal{E}\left( %
\left[ 1+\sum_{n=1}^{\infty }\int_{0<\alpha _{n}<...<\alpha _{1}<\hbar
}O_{\alpha _{1}}...O_{\alpha _{n}}d\alpha _{1}...d\alpha _{n}\right]
\varepsilon _{0}\left( \mathbf{X}_{0}\right) \right)  \label{eqBB}
\end{equation}
At this point, two important comments have to be made, both related to the
fact that the operators $O_{\alpha _{i}}$ have been designed to depend on $%
\mathbf{X}_{\alpha _{i}}$. First, notice that the gradient appearing in the
definition of $O_{\alpha _{i}}$ have to be taken with respect to $\mathbf{X}%
_{\alpha _{i}}$, but that these operators act on functions of $\mathbf{X}%
_{\alpha _{j}}$ with $\alpha _{j}<\alpha _{i}$, such as $\varepsilon
_{0}\left( \mathbf{X}_{0}\right) $. This is not a problem however, since $%
\mathbf{X}_{\alpha _{j}}=\mathbf{X}_{\alpha _{i}}-\int_{\alpha _{j}}^{\alpha
_{i}}d\mathbf{X}_{\lambda }$ and thus the derivative of a function $F\left( 
\mathbf{X}_{\alpha _{j}}\right) $ with respect to $\mathbf{X}_{\alpha _{i}}$
is just the same as the derivative with respect to $\mathbf{X}_{\alpha _{j}}$%
, that is the gradient of the function. This the reason why, in the
definition of $O_{\alpha _{i}}$ we discarded any reference to the $\alpha
_{i}$'s in the gradients.

The second remark is that we can use the EBS operation defined in Eq. $%
\left( \ref{EBS}\right) $ and formula $\left( \ref{CXION4}\right) $to put
the series appearing in the previous formula Eq. $\left( \ref{eqBB}\right) $
at the same point. Once again, since $0<\alpha _{n}<...<\alpha _{1}$, we can
write, inside the expectation : 
\begin{align}
& \left[ \int_{0<\alpha _{n}<...<\alpha _{1}<\hbar }O_{\alpha
_{1}}...O_{\alpha _{n}}d\alpha _{1}...d\alpha _{n}\right] \varepsilon
_{0}\left( \mathbf{X}_{0}\right)  \notag \\
& =\left[ \int_{0<\alpha _{n}<...<\alpha _{1}<\hbar }e^{-\left\langle
.\right\rangle _{\alpha _{1}\rightarrow \hbar }^{S}}O_{\alpha
_{1}}...e^{-\left\langle .\right\rangle _{\alpha _{n-1}\rightarrow \alpha
_{n-2}}^{S}}O_{\alpha _{n-1}}e^{-\left\langle .\right\rangle _{\alpha
_{n}\rightarrow \alpha _{n-1}}^{S}}O_{\alpha _{n}}e^{-\left\langle
.\right\rangle _{0\rightarrow \alpha _{n}}^{S}}\right] \varepsilon
_{0}\left( \mathbf{X}_{0}\right)  \label{REMP}
\end{align}
Note that this formula is a generalization of Eq. $\left( \ref{CXION4}%
\right) $ since the $O_{\alpha _{k}}$ are not functions, but rather
operators including derivatives with respect to the canonical variables. The
part of $O_{\alpha _{k}}$ acting through some multiplication on the left or
on the right or both on the left and on the right is not problematic and Eq. 
$\left( \ref{CXION4}\right) $ applies (with the slight modification of an
irrelevant change of order in the multiplication of the functions with
respect to Eq. $\left( \ref{CXION4}\right) $). Likewise, the differential
part of $O_{\alpha _{k}}\left( \mathbf{X}_{\alpha _{k}}\right) $, such as $%
\frac{1}{2}\mathcal{A}_{\alpha _{k}}^{R_{l}}\left( \mathbf{X}_{\alpha
_{k}}\right) \nabla _{R_{\alpha _{k}}^{l}}$ acting on some function say $%
F\left( \mathbf{X}_{\alpha _{k+1}}\right) $, is not problematic since the
operator $\exp \left( -\left\langle .\right\rangle _{\alpha
_{k+1}\rightarrow \alpha _{k}}^{S}\right) $ is a (complicate) function of
the derivative with respect to the canonical variables and thus commute with 
$\nabla _{R_{\alpha _{k+1}}^{l}}$. As a consequence setting $\frac{1}{2}%
\mathcal{A}_{\alpha _{k}}^{R_{l}}\left( \mathbf{X}_{\alpha _{k}}\right)
\nabla _{R_{\alpha _{k}}^{l}}F\left( \mathbf{X}_{\alpha _{k+1}}\right) $ at
the same point through the EBS operation , that is replacing it by $\left[ 
\frac{1}{2}\mathcal{A}_{\alpha _{k}}^{R_{l}}\left( \mathbf{X}_{\alpha
_{k}}\right) \exp \left( -\left\langle .\right\rangle _{\alpha
_{k+1}\rightarrow \alpha _{k}}^{S}\right) \nabla _{R_{\alpha
_{k}}^{l}}F\left( \mathbf{X}_{\alpha _{k+1}}\right) \right] $ in the
expectation really amounts to compute $\left[ \frac{1}{2}\mathcal{A}_{\alpha
_{k}}^{R_{l}}\left( \mathbf{X}_{\alpha _{k}}\right) \nabla _{R_{\alpha
_{k}}^{l}}\exp \left( -\left\langle .\right\rangle _{\alpha
_{k+1}\rightarrow \alpha _{k}}^{S}\right) F\left( \mathbf{X}_{\alpha
_{k+1}}\right) \right] $, that is to let the EBS operation act before the
action of $O_{\alpha _{k}}\left( \mathbf{X}_{\alpha _{k+1}}\right) $. This
thus justifies formula Eq. $\left( \ref{REMP}\right) $.

The computation of Eq. $\left( \ref{REMP}\right) $ derives directly from the
definitions of the various operators involved and proceeds as follows : for $%
n>0$ , starting from $\varepsilon _{0}\left( \mathbf{X}_{0}\right) $, first
apply $e^{-\left\langle .\right\rangle _{0\rightarrow \alpha _{n}}^{S}}$ on $%
\varepsilon _{0}\left( \mathbf{X}_{0}\right) $ that will shift $\mathbf{X}%
_{0}$ by $\mathbf{X}_{\alpha _{n}}$. Then make $O_{\alpha _{n}}$ acts on the
result, apply $e^{-\left\langle .\right\rangle _{\alpha _{n}\rightarrow
\alpha _{n-1}}^{S}}$ which replaces $\mathbf{X}_{\alpha _{n}}$ by $\mathbf{X}%
_{\alpha _{n-1}}$ and so on. Ultimately at the end of the process, $\mathbf{X%
}_{0}$ is replaced by $\mathbf{X}_{\hbar }$, this last variable being
independent from the integration variables, the various integrals can be
computed easily, as the integral of a series expansion. This process of
integration will be of course the same for all similar expressions in the
sequel.

The case $n=0$ corresponding to the first term in the series expansion : $%
\mathcal{E(}\varepsilon _{0}\left( \mathbf{X}_{0}\right) )$ is of course
easily handled by replacing it with $e^{-\left\langle .\right\rangle
_{0\rightarrow \hbar }^{S}}\varepsilon _{0}\left( \mathbf{X}_{0}\right) $.

We can now deduce that (with the notation $\mathbf{X}\equiv \mathbf{X}%
_{\hbar }$): 
\begin{equation}
\varepsilon \left( \mathbf{X}\right) \equiv \varepsilon _{\hbar }\left( 
\mathbf{X}_{\hbar }\right) =\mathcal{E}\left( \left[ \mathcal{T}\exp \left[
\int_{0<\alpha <\hbar }e^{-\left\langle .\right\rangle _{\alpha \rightarrow
\hbar }^{S}}O_{\alpha }e^{-\left\langle .\right\rangle _{0\rightarrow \alpha
}^{S}}d\alpha \right] \right] \varepsilon _{0}\left( \mathbf{X}_{0}\right)
\right)  \label{Esolution}
\end{equation}
which is a compact expression for the diagonal Hamiltonian $\varepsilon
\left( \mathbf{R,P}\right) $ in terms of the ''classical''\ diagonal
Hamiltonian $\varepsilon _{0}\left( \mathbf{R,P}\right) $ in which the
classical variables $\mathbf{R}_{0}\mathbf{,P}_{0}$ have been now replaced
by the quantum ones $\mathbf{R,}$ $\mathbf{P}$ due to the $EBS$ action. Here 
$\mathcal{T}$ is the usual notation for the ''time ordered product''. Eq. $%
\left( \ref{Esolution}\right) $ is the required expression and constitutes
the main result of this paper.

We can now consider the additional contributions that would appear if we had
considered a case such that $\left( D_{\alpha }\left( \mathbf{X}_{\alpha
_{1}},\alpha _{1}\right) H\left( \mathbf{X}_{\alpha _{1}}\right) \right)
\neq 0$. Let $C\left( \mathbf{X}_{\alpha _{1}}\right) =\mathcal{P}%
_{+}\left\{ U_{\alpha _{1}}\left( D_{\alpha }\left( \mathbf{X}_{\alpha
_{1}},\alpha _{1}\right) H\left( \mathbf{X}_{\alpha _{1}}\right) \right)
U_{\alpha _{1}}^{+}\right\} $. In such a case, the repeated application of $%
O_{\alpha _{1}}+C\left( \mathbf{X}_{\alpha _{1}}\right) $ yields rather : 
\begin{eqnarray*}
\varepsilon _{\hbar }\left( \mathbf{X}_{\hbar }\right) &=&\mathcal{E}\left( %
\left[ 1+\sum_{n=1}^{\infty }\int_{0<\alpha _{n}<...<\alpha _{1}<\alpha
}O_{\alpha _{1}}...O_{\alpha _{n}}d\alpha _{1}...d\alpha _{n}\right]
\varepsilon _{0}\left( \mathbf{X}_{0}\right) \right) \\
&&+\mathcal{E}\left( \left[ 1+\sum_{n=1}^{\infty }\int_{0<\alpha
_{n}<...<\alpha _{1}<\alpha }O_{\alpha _{1}}...O_{\alpha _{n-1}}d\alpha
_{1}...d\alpha _{n}\right] C\left( \mathbf{X}_{\alpha _{n}}\right) \right)
\end{eqnarray*}
Using the same tricks as before it leads directly to the following
expression for the diagonalized energy operator : 
\begin{eqnarray*}
&&\mathcal{E}\left( \left[ \mathcal{T}\exp \left[ \int_{0<\alpha <\hbar
}e^{-\left\langle .\right\rangle _{\alpha \rightarrow \hbar }^{S}}O_{\alpha
}e^{-\left\langle .\right\rangle _{0\rightarrow \alpha }^{S}}d\alpha \right] %
\right] \varepsilon _{0}\left( \mathbf{X}_{0}\right) \right) \\
&&+\mathcal{E}\int_{\alpha _{0}}^{\hbar }\left( \left[ \mathcal{T}\exp \left[
\int_{\alpha _{0}<\alpha <\hbar }e^{-\left\langle .\right\rangle _{\alpha
\rightarrow \hbar }^{S}}O_{\alpha }e^{-\left\langle .\right\rangle _{\alpha
_{0}\rightarrow \alpha }^{S}}d\alpha \right] \right] C\left( \mathbf{X}%
_{\alpha _{0}}\right) d\alpha _{0}\right)
\end{eqnarray*}
\ \ 

Clearly the practical application of Eq. $\left( \ref{Esolution}\right) $
requires the knowledge of the transformation matrices $U_{\alpha }$ which
enter into the definition of the operators $O_{\alpha }$.

\subsection{The transformation matrix \ $U$}

Note first that there is a certain arbitrariness in the choice of the
unitary matrix $U_{\alpha}\left( \mathbf{X}\right) $ as explained in \cite%
{SERIESPIERRE} which reflects a kind of gauge invariance. Actually,
multiplying the transformation matrix $U_{\alpha}\left( \mathbf{X}\right) $\
on the right by a diagonal unitary matrix yields an other diagonalization,
equivalent to the previous one. In particular it allows to choose $n$\
conditions for the diagonal entries of $U_{\alpha}\left( \mathbf{X}\right) $
($n\times n$ being the size of $U_{\alpha}\left( \mathbf{X}\right) $). An
explicit choice will be done below to simplify our expressions.

To find the transformation matrix $U$ we use the same approach as for the
diagonalization of the Hamiltonian, by writing : 
\begin{align}
U_{\hbar }\left( \mathbf{X}_{\hbar }\right) & =\mathcal{E}\left( U\left( 
\mathbf{X}_{0}\right) +\int_{0}^{\hbar }dU_{\alpha }\left( \mathbf{X}%
_{\alpha }\right) \right)  \notag \\
& =\mathcal{E}\left( \int_{0}^{\hbar }\left( \frac{\partial }{\partial
\alpha }+\left\langle .\right\rangle \right) U_{\alpha }\left( \mathbf{X}%
_{\alpha }\right) d\alpha \right)
\end{align}
we can find $\left( \partial _{\alpha }+\left\langle .\right\rangle \right)
U_{\alpha }\left( \mathbf{X}_{\alpha }\right) $ by using again the
diagonalization process leading to Eq. $\left( \ref{diago}\right) $. Indeed
we had 
\begin{align}
\left( \frac{\partial }{\partial \alpha _{1}}+\left\langle .\right\rangle
\right) \varepsilon _{\alpha _{1}}\left( \mathbf{X}_{\alpha _{1}}\right) &
=\left( \left( \frac{\partial }{\partial \alpha _{1}}+\left\langle
.\right\rangle \right) U_{\alpha _{1}}\right) U_{\alpha _{1}}^{+}\varepsilon
_{\alpha _{1}}+\varepsilon _{\alpha _{1}}U_{\alpha _{1}}\left( \frac{%
\partial }{\partial \alpha _{1}}+\left\langle .\right\rangle \right)
U_{\alpha _{1}}^{+}  \notag \\
& +U_{\alpha _{1}}\left( \frac{\partial }{\partial \alpha _{1}}+\left\langle
.\right\rangle \right) H\left( \mathbf{X}_{\alpha _{1}}\right) U_{\alpha
_{1}}^{+}  \notag \\
& +\frac{1}{2}\left\{ \mathcal{A}_{\alpha _{1}}^{R_{l}}\nabla
_{R_{l}}\varepsilon _{\alpha _{1}}+\nabla _{R_{l}}\varepsilon _{\alpha _{1}}%
\mathcal{A}_{\alpha _{1}}^{R_{l}}+\mathcal{A}_{\alpha _{1}}^{P_{l}}\nabla
_{P_{l}}\varepsilon _{\alpha _{1}}+\nabla _{P_{l}}\varepsilon _{\alpha _{1}}%
\mathcal{A}_{\alpha _{1}}^{P_{l}}\right\}  \notag \\
& +\frac{i}{2}\left\{ \mathcal{A}_{\alpha _{1}}^{P_{l}}\varepsilon _{\alpha
_{1}}\mathcal{A}_{\alpha _{1}}^{R_{l}}-\mathcal{A}_{\alpha
_{1}}^{R_{l}}\varepsilon _{\alpha _{1}}\mathcal{A}_{\alpha
_{1}}^{P_{l}}+\varepsilon _{\alpha _{1}}\left[ \mathcal{A}_{\alpha
_{1}}^{R_{l}},\mathcal{A}_{\alpha _{1}}^{P_{l}}\right] +\left[ \mathcal{A}%
_{\alpha _{1}}^{R_{l}},\mathcal{A}_{\alpha _{1}}^{P_{l}}\right] \varepsilon
_{\alpha _{1}}\right\}  \label{U1}
\end{align}
In addition, the unitarity condition $U_{\alpha }U_{\alpha }^{+}=1$ implies
the relation $\left( \partial _{\alpha }+\left\langle .\right\rangle \right)
\left( U_{\alpha }U_{\alpha }^{+}\right) =0$ which reads : 
\begin{equation*}
0=\left( \left( \frac{\partial }{\partial \alpha }+\left\langle
.\right\rangle \right) U_{\alpha }\right) U_{\alpha }^{+}+U_{\alpha }\left( 
\frac{\partial }{\partial \alpha }+\left\langle .\right\rangle \right)
U_{\alpha }^{+}-\frac{i}{2}\left( \nabla _{P_{i}}U_{\alpha }\nabla
_{R_{i}}U_{\alpha }^{+}-\nabla _{R_{i}}U_{\alpha }\nabla _{P_{i}}U_{\alpha
}^{+}\right)
\end{equation*}
or, in a more compact way: 
\begin{equation}
0=\left( \left( \frac{\partial }{\partial \alpha }+\left\langle
.\right\rangle \right) U_{\alpha }\right) U_{\alpha }^{+}+U_{\alpha }\left( 
\frac{\partial }{\partial \alpha }+\left\langle .\right\rangle \right)
U_{\alpha }^{+}+\frac{i}{2}\left[ \mathcal{A}_{\alpha }^{R_{l}},\mathcal{A}%
_{\alpha }^{P_{l}}\right]  \label{U2}
\end{equation}
Mixing the two equations Eqs. $\left( \ref{U1}\right) \left( \ref{U2}\right) 
$, we obtain that Eq. $\left( \ref{U1}\right) $ after projection on the non
diagonal part becomes the equality 
\begin{align}
\mathcal{P}_{-}\left[ \left( \left( \frac{\partial }{\partial \alpha }%
+\left\langle .\right\rangle \right) U_{\alpha }\right) U_{\alpha
}^{+},\varepsilon _{\alpha }\right] & =-\mathcal{P}_{-}\left( U_{\alpha
}\left( \left( \frac{\partial }{\partial \alpha }+\left\langle
.\right\rangle \right) H\left( \mathbf{X}_{\alpha }\right) \right) U_{\alpha
}^{+}\right)  \notag \\
& +\frac{i}{2}\varepsilon _{\alpha }\mathcal{P}_{-}\left[ \mathcal{A}%
_{\alpha }^{R_{l}},\mathcal{A}_{\alpha }^{P_{l}}\right]  \notag \\
& -\frac{1}{2}\mathcal{P}_{-}\left\{ \mathcal{A}_{\alpha _{1}}^{R_{l}}\nabla
_{R_{l}}\varepsilon _{\alpha _{1}}+\nabla _{R_{l}}\varepsilon _{\alpha _{1}}%
\mathcal{A}_{\alpha _{1}}^{R_{l}}+\mathcal{A}_{\alpha _{1}}^{P_{l}}\nabla
_{P_{l}}\varepsilon _{\alpha _{1}}+\nabla _{P_{l}}\varepsilon _{\alpha _{1}}%
\mathcal{A}_{\alpha _{1}}^{P_{l}}\right\}  \notag \\
& -\frac{i}{2}\mathcal{P}_{-}\left\{ \mathcal{A}_{\alpha
_{1}}^{P_{l}}\varepsilon _{\alpha _{1}}\mathcal{A}_{\alpha _{1}}^{R_{l}}-%
\mathcal{A}_{\alpha _{1}}^{R_{l}}\varepsilon _{\alpha _{1}}\mathcal{A}%
_{\alpha _{1}}^{P_{l}}+\varepsilon _{\alpha _{1}}\left[ \mathcal{A}_{\alpha
_{1}}^{R_{l}},\mathcal{A}_{\alpha _{1}}^{P_{l}}\right] +\left[ \mathcal{A}%
_{\alpha _{1}}^{R_{l}},\mathcal{A}_{\alpha _{1}}^{P_{l}}\right] \varepsilon
_{\alpha _{1}}\right\}  \label{U3}
\end{align}
Both conditions Eqs. $\left( \ref{U2}\right) \left( \ref{U3}\right) $ can be
solved for $\left( \partial _{\alpha }+\left\langle .\right\rangle \right)
U_{\alpha }$. Actually, decomposing $\left[ \left( \partial _{\alpha
}+\left\langle .\right\rangle \right) U_{\alpha }\right] U_{\alpha
}^{+}=hr+ahr$ in Hermitian $hr$ and anti-Hermitian $ahr$ part, Eq. $\left( %
\ref{U3}\right) $ reads 
\begin{equation}
\mathcal{P}_{-}\left[ hr+ahr,\varepsilon _{\alpha }\right] =M\left(
U_{\alpha }\right)  \label{P}
\end{equation}
where $M\left( U_{\alpha }\right) $ is the r.h.s of Eq. $\left( \ref{U3}%
\right) $. We can now use a particular choice of gauge. We fix our
diagonalization process by setting $\mathcal{P}_{+}\left( ahr\right) =0$.
Practically it corresponds to multiply $U_{\alpha }$ on the right by a
unitary diagonal matrix $D_{\alpha }$ (which has $n$\ degree of freedom)
such that $\mathcal{P}_{+}\left( \left( \frac{\partial }{\partial \alpha }%
+\left\langle .\right\rangle \right) U_{\alpha }D_{\alpha }\right) D_{\alpha
}^{+}U_{\alpha }^{+}$ is Hermitian. This gives $n$ conditions that determine 
$D_{\alpha }$.

Now, from Eq. $\left( \ref{P}\right) $ and given our gauge choice for $%
U_{\alpha },$ we deduce that $\mathcal{P}_{-}\left( ahr\right) =ahr$ and $%
\left[ ahr,\varepsilon _{\alpha }\right] =\frac{1}{2}\left( M\left(
U_{\alpha }\right) +M^{+}(U_{\alpha })\right) $. To find $ahr$ we aim now at
inverting the commutator $\left[ .,\varepsilon _{\alpha }\right] $ in the
left hand side of this equation. First note that, since we have assumed the
diagonalization is possible, we can safely assume that the right hand side $%
\frac{1}{2}\left( M\left( U_{\alpha }\right) +M^{+}(U_{\alpha })\right) $
lies in the image of the operator $\left[ .,\varepsilon _{\alpha }\right] $.
It is in fact the case in all practical cases. Since both $ahr$ and $M\left(
U_{\alpha }\right) $ are anti-diagonal by construction, we just need to
invert the commutator with $\varepsilon _{\alpha }$, $\left[ .,\varepsilon
_{\alpha }\right] $ on the space of anti-diagonal matrices. To do so, we
first need to find the kernel of $\left[ .,\varepsilon _{\alpha }\right] $
on this space and check if it is null.

Assume first that the bands of the system are non degenerate, that is $%
\left( \varepsilon _{\alpha }\right) _{n}\neq \left( \varepsilon _{\alpha
}\right) _{m}$ for $m\neq n$. As a consequence an element $V$ in the kernel
of the commutator satisfies 
\begin{equation*}
V_{nm}\left[ \left( \varepsilon _{\alpha }\right) _{m}-\left( \varepsilon
_{\alpha }\right) _{n}\right] =-i\alpha \left( \nabla _{P_{i}}\left(
\varepsilon _{\alpha }\right) _{n}\nabla _{R_{i}}V_{nm}-\nabla
_{R_{i}}\left( \varepsilon _{\alpha }\right) _{n}\nabla _{P_{i}}V_{nm}\right)
\end{equation*}
that is : 
\begin{equation*}
V_{nm}=-i\alpha \left( \nabla _{P_{i}}\left( \varepsilon _{\alpha }\right)
_{n}\nabla _{R_{i}}V_{nm}-\nabla _{R_{i}}\left( \varepsilon _{\alpha
}\right) _{n}\nabla _{P_{i}}V_{nm}\right) \frac{1}{\left[ \left( \varepsilon
_{\alpha }\right) _{n}-\left( \varepsilon _{\alpha }\right) _{m}\right] }
\end{equation*}
Since the beginning, we have assumed that all the functions involved are
regular in $\alpha $ and have a series expansion in this parameter (as well
as in the canonical variables). Thus, iterating the last relation for $%
V_{nm} $ yields that, for each $V_{nm}$ expanded as a series expansion in $%
\alpha $, at each order in $\alpha $, $V_{nm}=0$. As a consequence, the
kernel of the commutator with $\varepsilon _{\alpha }$ is null and we can
thus formally write $ahr$ under the form : 
\begin{equation}
ahr=\widetilde{N}\left( \alpha \right) .U_{\alpha }  \label{N}
\end{equation}
where we have defined : 
\begin{equation}
\widetilde{N}\left( \alpha \right) .U_{\alpha }=\frac{1}{2}\left( \left[
.,\varepsilon _{\alpha }\right] ^{-1}\right) \left( M\left( U_{\alpha
}\right) +M^{+}(U_{\alpha })\right)  \label{N1}
\end{equation}
and the inverse of the commutator operation $\left[ ,\varepsilon _{\alpha }%
\right] $ satisfies obviously : \ 
\begin{equation*}
\left[ \left[ .,\varepsilon _{\alpha }\right] ^{-1}.M,\varepsilon _{\alpha }%
\right] =\left[ .,\varepsilon _{\alpha }\right] ^{-1}.\left[ M,\varepsilon
_{\alpha }\right] =M\text{\ \ \ for \ \ }M\text{ anti-diagonal }
\end{equation*}
If the Band are degenerate, we need to define the inverse of the commutator
more carefully. Actually since some bands may be degenerate, the kernel of
the commutator $\left[ .,\varepsilon _{\alpha }\right] $ has no reason to be
null now. This operator is thus no more bijective and we will not be able to
define its inverse uniquely. This non uniqueness will lead to a non unique
definition of $ahr$. This is not astonishing. Actually, when the Band are
degenerate, as in the Dirac case, we do not look for a diagonal
Hamiltonian,but rather for a Block diagonal Hamiltonian. This leaves thus
more freedom for the diagonalization process and as a consequence, for the
definition of $\left[ .,\varepsilon _{\alpha }\right] ^{-1}$. Practically,
we proceed as follows. Decompose our space of non diagonal matrices of
operators denoted $AD$ in the following way. Write $AD=\ker \left( \left[
.,\varepsilon _{\alpha }\right] \right) \oplus ad_{1}$ with $ad_{1}$ an
arbitrary supplementary space of $\ker \left( \left[ .,\varepsilon _{\alpha }%
\right] \right) $. Now, given $M$ an element of $\func{Im}\left( \left[
.,\varepsilon _{\alpha }\right] \right) $, chose a antecedent $m_{1}$ of $M$
with respect to $\left[ .,\varepsilon _{\alpha }\right] $, and decompose $%
m_{1}$ with respect to the previous decomposition, $m_{1}=m_{2}+m_{3}$ where 
$m_{2}\in \ker \left( \left[ .,\varepsilon _{\alpha }\right] \right) $, $%
m_{3}\in ad_{1}$. Now set $\left[ .,\varepsilon _{\alpha }\right]
^{-1}.M=m_{3}$. Given a chosen decomposition $\ker \left( \left[
.,\varepsilon _{\alpha }\right] \right) \oplus ad_{1}$, $m_{3}$ is unique
since if $\hat{m}_{1}$ is another antecedent of $M$, $\hat{m}_{1}-m_{1}\in
\ker \left( \left[ .,\varepsilon _{\alpha }\right] \right) $ and as a
consequence $\hat{m}_{1}=m_{3}+\left( m_{2}+\hat{m}_{1}-m_{1}\right) $ is
the decomposition of $\hat{m}_{1}$.

One can check that 
\begin{eqnarray*}
\left[ \left[ .,\varepsilon _{\alpha }\right] ^{-1}.M,\varepsilon _{\alpha }%
\right] &=&\left[ m_{3},\varepsilon _{\alpha }\right] \\
&=&\left[ m_{1}-m_{2},\varepsilon _{\alpha }\right] \\
&=&\left[ m_{1},\varepsilon _{\alpha }\right] =M
\end{eqnarray*}
as needed to solve Eq. $\left( \ref{P}\right) $. On the other hand, notice
that : $\left[ .,\varepsilon _{\alpha }\right] ^{-1}\left[ M,\varepsilon
_{\alpha }\right] $ is not equal to $M$ in general. Actually, choosing $M$
as an antecedent of $\left[ M,\varepsilon _{\alpha }\right] $ and
decomposing it gives an element $m_{3}$ that a priori depends on the
supplementary space chosen. As a consequence, the operation $\left[
.,\varepsilon _{\alpha }\right] ^{-1}$ is only a right inverse to $\left[
.,\varepsilon _{\alpha }\right] $. This is not a problem, since it allows
anyway to write our solution for Eq. $\left( \ref{P}\right) $ as : 
\begin{equation}
ahr=\widetilde{N}\left( \alpha \right) .U_{\alpha }
\end{equation}
where we have again defined : 
\begin{equation}
\widetilde{N}\left( \alpha \right) .U_{\alpha }=\frac{1}{2}\left( \left[
.,\varepsilon _{\alpha }\right] ^{-1}\right) \left( M\left( U_{\alpha
}\right) +M^{+}(U_{\alpha })\right)
\end{equation}
The non unicity in the definition of $\left[ .,\varepsilon _{\alpha }\right]
^{-1}$ is reflected in the choice of decomposition for the space $AD$. Let
us insist on the fact that this choice is itself the consequence of the
Bands degeneracy that allows for a larger freedom of gauge choice for a
block diagonal representation.

To complete the determination of $U_{\alpha }$ we still need to deduce $hr$
which is readily obtained from the unitarity condition Eq. $\left( \ref{U2}%
\right) $ as: 
\begin{equation}
hr=-\frac{i}{4}\left[ \mathcal{A}_{\alpha }^{R_{l}},\mathcal{A}_{\alpha
}^{P_{l}}\right]  \label{hr}
\end{equation}
Indeed, Eq. $\left( \ref{hr}\right) $ follows easily from Eq. $\left( \ref%
{U2}\right) $, when noting that $\left( \left( \frac{\partial }{\partial
\alpha }+\left\langle .\right\rangle \right) U_{\alpha }\right) U_{\alpha
}^{+}+U_{\alpha }\left( \frac{\partial }{\partial \alpha }+\left\langle
.\right\rangle \right) U_{\alpha }^{+}$ is hermitian due to the fact that
for any operator $A$, $\left\langle .\right\rangle A^{+}=\left( \left\langle
.\right\rangle A\right) ^{+}$ (and this is true in particular for $U_{\alpha
}$). This assertion can be checked on an arbitrary expansion in $\mathbf{R}$
and $\mathbf{P}$ of $A$, since for an arbitrary monomial $%
R_{l}^{n}P_{l}^{k}A_{n,k}$ with $A_{n,k}$ an arbitrary coefficient matrix, $%
\left\langle .\right\rangle \left( R_{l}^{n}P_{l}^{k}\right) A_{n,k}^{+}=%
\frac{i}{2}\left( nkR_{l}^{n-1}P_{l}^{k-1}\right) A_{n,k}^{+}=\left( -\frac{i%
}{2}nkP_{l}^{k-1}R_{l}^{n-1}A_{n,k}\right) ^{+}$ and this last quantity is
equal to $\left[ \left\langle .\right\rangle \left(
P_{l}^{k}R_{l}^{n}\right) A_{n,k}\right] ^{+}$.

Gathering the results Eqs.$\left( \ref{N1}\right) \left( \ref{hr}\right) $
allows us to introduce the operator $N_{\alpha }$ operating on $U_{\alpha }$
in the following manner $N_{\alpha }.U_{\alpha }=\left( hr+ahr\right)
U_{\alpha },$ so that we can write 
\begin{equation*}
\left[ \left( \frac{\partial }{\partial \alpha }+\left\langle .\right\rangle
\right) U_{\alpha }\right] =N_{\alpha }.U_{\alpha }
\end{equation*}
with $N_{\alpha }$ given explicitly by the following expression 
\begin{equation*}
N_{\alpha }.U_{\alpha }=\frac{1}{2}\left( \left[ .,\varepsilon _{\alpha }%
\right] ^{-1}\right) \left( M\left( U_{\alpha }\right) +M^{+}(U_{\alpha
})\right) U_{\alpha }-\frac{i}{4}\left[ \mathcal{A}_{\alpha }^{R_{l}},%
\mathcal{A}_{\alpha }^{P_{l}}\right] U_{\alpha }
\end{equation*}
As for the energy diagonalization, this expression can be rewritten in terms
of physical quantities. Recall that $M\left( U_{\alpha }\right) $ is the
r.h.s of Eq. $\left( \ref{U3}\right) $, so that 
\begin{eqnarray}
\frac{1}{2}\left( M\left( U_{\alpha }\right) +M^{+}(U_{\alpha })\right) &=&-%
\frac{1}{2}\mathcal{P}_{-}\left\{ \mathcal{A}_{\alpha _{1}}^{R_{l}}\nabla
_{R_{l}}\varepsilon _{\alpha _{1}}+\nabla _{R_{l}}\varepsilon _{\alpha _{1}}%
\mathcal{A}_{\alpha _{1}}^{R_{l}}+\mathcal{A}_{\alpha _{1}}^{P_{l}}\nabla
_{P_{l}}\varepsilon _{\alpha _{1}}+\nabla _{P_{l}}\varepsilon _{\alpha _{1}}%
\mathcal{A}_{\alpha _{1}}^{P_{l}}\right\}  \notag \\
&&-\frac{i}{4}\mathcal{P}_{-}\left\{ \mathcal{A}_{\alpha
_{1}}^{P_{l}}\varepsilon _{\alpha _{1}}\mathcal{A}_{\alpha _{1}}^{R_{l}}-%
\mathcal{A}_{\alpha _{1}}^{R_{l}}\varepsilon _{\alpha _{1}}\mathcal{A}%
_{\alpha _{1}}^{P_{l}}+\varepsilon _{\alpha _{1}}\left[ \mathcal{A}_{\alpha
_{1}}^{R_{l}},\mathcal{A}_{\alpha _{1}}^{P_{l}}\right] +\left[ \mathcal{A}%
_{\alpha _{1}}^{R_{l}},\mathcal{A}_{\alpha _{1}}^{P_{l}}\right] \varepsilon
_{\alpha _{1}}+H.C.\right\}  \notag \\
&&+\frac{i}{4}\mathcal{P}_{-}\varepsilon _{\alpha }\left[ \mathcal{A}%
_{\alpha }^{R_{l}},\mathcal{A}_{\alpha }^{P_{l}}\right] +\frac{i}{4}\mathcal{%
P}_{-}\left[ \mathcal{A}_{\alpha }^{R_{l}},\mathcal{A}_{\alpha }^{P_{l}}%
\right] \varepsilon _{\alpha }  \notag \\
&=&-\frac{1}{2}\mathcal{P}_{-}\left\{ \mathcal{A}_{\alpha
_{1}}^{R_{l}}\nabla _{R_{l}}\varepsilon _{\alpha _{1}}+\nabla
_{R_{l}}\varepsilon _{\alpha _{1}}\mathcal{A}_{\alpha _{1}}^{R_{l}}+\mathcal{%
A}_{\alpha _{1}}^{P_{l}}\nabla _{P_{l}}\varepsilon _{\alpha _{1}}+\nabla
_{P_{l}}\varepsilon _{\alpha _{1}}\mathcal{A}_{\alpha _{1}}^{P_{l}}\right\} 
\notag \\
&&-\frac{i}{4}\mathcal{P}_{-}\left\{ \left[ \varepsilon _{\alpha },\mathcal{A%
}_{\alpha }^{R_{l}}\right] \mathcal{A}_{\alpha }^{P_{l}}-\left[ \varepsilon
_{\alpha },\mathcal{A}_{\alpha }^{P_{l}}\right] \mathcal{A}_{\alpha
}^{R_{l}}+H.C.\right\}  \label{mucalc}
\end{eqnarray}
where $H.C.$\ stands for the Hermitian conjugate. Note also that $U_{\alpha
}\left( \left( \frac{\partial }{\partial \alpha }+\left\langle
.\right\rangle \right) H\left( \mathbf{X}_{\alpha }\right) \right) U_{\alpha
}^{+}$ is hermitian, as a consequence both of the hermiticity of $H\left( 
\mathbf{X}_{\alpha }\right) $ and the fact that $\left( \left\langle
.\right\rangle H\left( \mathbf{X}_{\alpha }\right) \right) ^{+}=\left\langle
.\right\rangle H\left( \mathbf{X}_{\alpha }\right) ^{+}=\left\langle
.\right\rangle H\left( \mathbf{X}_{\alpha }\right) $ (see the assertion
below Eq. $\left( \ref{hr}\right) $). Ultimately, we are led to the
following expression for $N_{\alpha }.U_{\alpha }$:

\begin{align}
N_{\alpha }.U_{\alpha }& =-\left[ .,\varepsilon _{\alpha }\right] ^{-1}.%
\left[ \mathcal{P}_{-}\left\{ \frac{1}{2}\mathcal{A}_{\alpha }^{R_{l}}\nabla
_{R_{l}}\varepsilon _{\alpha }+\nabla _{R_{l}}\varepsilon _{\alpha }\mathcal{%
A}_{\alpha }^{R_{l}}+\mathcal{A}_{\alpha }^{P_{l}}\nabla _{P_{l}}\varepsilon
_{\alpha }+\nabla _{P_{l}}\varepsilon _{\alpha }\mathcal{A}_{\alpha
}^{P_{l}}\right\} \right.  \notag \\
& +\mathcal{P}_{-}\left\{ U_{\alpha }\left( \left( \frac{\partial }{\partial
\alpha }+\left\langle .\right\rangle \right) H\left( \mathbf{X}_{\alpha
}\right) \right) U_{\alpha }^{+}\right\} \left. +\frac{i}{4}\mathcal{P}%
_{-}\left\{ \left[ \varepsilon _{\alpha },\mathcal{A}_{\alpha }^{R_{l}}%
\right] \mathcal{A}_{\alpha }^{P_{l}}-\left[ \varepsilon _{\alpha },\mathcal{%
A}_{\alpha }^{P_{l}}\right] \mathcal{A}_{\alpha }^{R_{l}}\right\} +H.C.%
\right] U_{\alpha }  \notag \\
& -\frac{i}{4}\mathcal{P}_{-}\left\{ \left[ \mathcal{A}_{\alpha }^{R_{l}},%
\mathcal{A}_{\alpha }^{P_{l}}\right] U_{\alpha }\right\}  \label{Nnew}
\end{align}
where $\varepsilon _{\alpha }\left( \mathbf{X}_{\alpha }\right) $ is
computed recursively as explained before.

As for $\varepsilon\left( \mathbf{X}\right) $, we can therefore write for $%
U\left( \mathbf{X}\right) \equiv U_{\hbar}\left( \mathbf{X}_{\hbar }\right) $
:

\begin{equation}
U\left( \mathbf{X}\right) =\mathcal{E}\left( \left[ \mathcal{T}\exp \left[
\int_{0<\alpha <\hbar }e^{-\left\langle .\right\rangle _{\alpha \rightarrow
\hbar }^{S}}N_{\alpha }e^{-\left\langle .\right\rangle _{0\rightarrow \alpha
}^{S}}d\alpha \right] \right] U_{0}\left( \mathbf{X}_{0}\right) \right)
\label{Usolution}
\end{equation}
This expression has a very similar structure as the solution for the energy
Eq. $\left( \ref{Esolution}\right) $ except that the operator $O_{\alpha }$
has to be replaced by $N_{\alpha }.$

To end up this section, and given the solution derived for $U_{\alpha }$, we
can rewrite the operator $O_{\alpha }$ in an simpler form. Indeed starting
again from Eq. $\left( \ref{diago}\right) $ using the unitarity condition
for $U_{\alpha }$\ yields : 
\begin{align}
\left( \frac{\partial }{\partial \alpha _{1}}+\left\langle .\right\rangle
\right) \varepsilon _{\alpha _{1}}\left( \mathbf{X}_{\alpha _{1}}\right) & =%
\mathcal{P}_{+}\left[ \left( \left( \frac{\partial }{\partial \alpha _{1}}%
+\left\langle .\right\rangle \right) U_{\alpha _{1}}\right) U_{\alpha
_{1}}^{+},\varepsilon _{\alpha _{1}}\right] -\frac{i}{2}\varepsilon _{\alpha
}\mathcal{P}_{+}\left[ \mathcal{A}_{\alpha }^{R_{l}},\mathcal{A}_{\alpha
}^{P_{l}}\right]  \notag \\
& +U_{\alpha _{1}}\left( \frac{\partial }{\partial \alpha _{1}}+\left\langle
.\right\rangle \right) H\left( \mathbf{X}_{\alpha _{1}}\right) U_{\alpha
_{1}}^{+}  \notag \\
& +\frac{1}{2}\mathcal{A}_{\alpha _{1}}^{R_{l}}\nabla _{R_{l}}\varepsilon
_{\alpha _{1}}+\nabla _{R_{l}}\varepsilon _{\alpha _{1}}\mathcal{A}_{\alpha
_{1}}^{R_{l}}+\mathcal{A}_{\alpha _{1}}^{P_{l}}\nabla _{P_{l}}\varepsilon
_{\alpha _{1}}+\nabla _{P_{l}}\varepsilon _{\alpha _{1}}\mathcal{A}_{\alpha
_{1}}^{P_{l}}  \notag \\
& +\frac{i}{2}\left\{ \mathcal{A}_{\alpha _{1}}^{P_{l}}\varepsilon _{\alpha
_{1}}\mathcal{A}_{\alpha _{1}}^{R_{l}}-\mathcal{A}_{\alpha
_{1}}^{R_{l}}\varepsilon _{\alpha _{1}}\mathcal{A}_{\alpha
_{1}}^{P_{l}}+\varepsilon _{\alpha _{1}}\left[ \mathcal{A}_{\alpha
_{1}}^{R_{l}},\mathcal{A}_{\alpha _{1}}^{P_{l}}\right] +\left[ \mathcal{A}%
_{\alpha _{1}}^{R_{l}},\mathcal{A}_{\alpha _{1}}^{P_{l}}\right] \varepsilon
_{\alpha _{1}}\right\}
\end{align}
Now, recall our gauge condition which states that $\mathcal{P}_{+}\left(
\left( \frac{\partial }{\partial \alpha }+\left\langle .\right\rangle
\right) U_{\alpha }\right) U_{\alpha }^{+}$\ is Hermitian, so that $\mathcal{%
P}_{+}\left[ \left( \left( \frac{\partial }{\partial \alpha _{1}}%
+\left\langle .\right\rangle \right) U_{\alpha _{1}}\right) U_{\alpha
_{1}}^{+},\varepsilon _{\alpha _{1}}\right] $ is antihermitian, as the
commutator of two hermitian quantities. As a consequence, $\mathcal{P}_{+}%
\left[ \left( \left( \frac{\partial }{\partial \alpha _{1}}+\left\langle
.\right\rangle \right) U_{\alpha _{1}}\right) U_{\alpha
_{1}}^{+},\varepsilon _{\alpha _{1}}\right] +H.C.=0.$ Then, use the fact
that $\varepsilon _{\alpha }\left( \mathbf{X}_{\alpha }\right) $\ is assumed
to be an Hermitian operator, so that $\left( \frac{\partial }{\partial
\alpha _{1}}+\left\langle .\right\rangle \right) \varepsilon _{\alpha
_{1}}\left( \mathbf{X}_{\alpha _{1}}\right) $ is equal to half its sum with
its hermitic conjugate. A computation similar to Eq. $\left( \ref{mucalc}%
\right) $, leads us directly to the expression : 
\begin{align*}
\left( \frac{\partial }{\partial \alpha _{1}}+\left\langle .\right\rangle
\right) \varepsilon _{\alpha _{1}}\left( \mathbf{X}_{\alpha _{1}}\right) & =%
\frac{1}{2}\mathcal{P}_{+}\left\{ \mathcal{A}_{\alpha }^{R_{l}}\nabla
_{R_{l}}\varepsilon _{\alpha }+\nabla _{R_{l}}\varepsilon _{\alpha }\mathcal{%
A}_{\alpha }^{R_{l}}+\mathcal{A}_{\alpha }^{P_{l}}\nabla _{P_{l}}\varepsilon
_{\alpha }+\nabla _{P_{l}}\varepsilon _{\alpha }\mathcal{A}_{\alpha
}^{P_{l}}\right\} \\
& +\left[ \frac{i}{4}\mathcal{P}_{+}\left\{ \left[ \varepsilon _{\alpha },%
\mathcal{A}_{\alpha }^{R_{l}}\right] \mathcal{A}_{\alpha }^{P_{l}}-\left[
\varepsilon _{\alpha },\mathcal{A}_{\alpha }^{P_{l}}\right] \mathcal{A}%
_{\alpha }^{R_{l}}\right\} +H.C.\right] \\
& +U_{\alpha _{1}}\left( \frac{\partial }{\partial \alpha _{1}}+\left\langle
.\right\rangle \right) H\left( \mathbf{X}_{\alpha _{1}}\right) U_{\alpha
_{1}}^{+}
\end{align*}
The action of $O_{\alpha }$ is obtained by skipping $U_{\alpha _{1}}\left( 
\frac{\partial }{\partial \alpha _{1}}+\left\langle .\right\rangle \right)
H\left( \mathbf{X}_{\alpha _{1}}\right) U_{\alpha _{1}}^{+}$\ from the last
term, so that : 
\begin{align}
O_{\alpha }\varepsilon _{\alpha }\left( \mathbf{X}_{\alpha }\right) & =\frac{%
1}{2}\mathcal{P}_{+}\left\{ \mathcal{A}_{\alpha }^{R_{l}}\nabla
_{R_{l}}\varepsilon _{\alpha }+\nabla _{R_{l}}\varepsilon _{\alpha }\mathcal{%
A}_{\alpha }^{R_{l}}+\mathcal{A}_{\alpha }^{P_{l}}\nabla _{P_{l}}\varepsilon
_{\alpha }+\nabla _{P_{l}}\varepsilon _{\alpha }\mathcal{A}_{\alpha
}^{P_{l}}\right\}  \notag \\
& +\left[ \frac{i}{4}\mathcal{P}_{+}\left\{ \left[ \varepsilon _{\alpha },%
\mathcal{A}_{\alpha }^{R_{l}}\right] \mathcal{A}_{\alpha }^{P_{l}}-\left[
\varepsilon _{\alpha },\mathcal{A}_{\alpha }^{P_{l}}\right] \mathcal{A}%
_{\alpha }^{R_{l}}\right\} +H.C.\right]  \label{Onew}
\end{align}
Let us ultimately recall, that in most of the physical applications of
interest for us (in particular for Dirac and Bloch electrons) the term $%
\left( \frac{\partial }{\partial \alpha }+\left\langle .\right\rangle
\right) H\left( \mathbf{X}_{\alpha }\right) $\ will cancel.\ This
simplification will be assumed for the rest of the paper.

\subsection{The full system H and U}

We can now write the solution of our diagonalization procedure for a general
matrix valued Hamiltonian through an unitary transformation $U$ as the
solution of the following system of differential equations 
\begin{align}
\varepsilon\left( \mathbf{X}\right) & =\mathcal{E}\left( \left[ \mathcal{T}%
\exp\left[ \int_{0<\alpha<\hbar}e^{-\left\langle .\right\rangle
_{\alpha\rightarrow\hbar}^{S}}O_{\alpha}e^{-\left\langle .\right\rangle
_{0\rightarrow\alpha}^{S}}d\alpha\right] \right] \varepsilon_{0}\left( 
\mathbf{X}_{0}\right) \right)  \label{E} \\
U\left( \mathbf{X}\right) & =\mathcal{E}\left( \left[ \mathcal{T}\exp\left[
\int_{0<\alpha<\hbar}e^{-\left\langle .\right\rangle
_{\alpha\rightarrow\hbar}^{S}}N_{\alpha}e^{-\left\langle .\right\rangle
_{0\rightarrow\alpha}^{S}}d\alpha\right] \right] U_{0}\left( \mathbf{X}%
_{0}\right) \right)  \label{U}
\end{align}
where $O_{\alpha}$ and $N_{\alpha}$ are given respectively by Eqs.$\left( %
\ref{Onew}\right) $ and $\left( \ref{Nnew}\right) $.\ The only
pre-requirement is that the diagonal form at $\alpha=0$, $\varepsilon
_{0}\left( \mathbf{R}_{0},\mathbf{P}_{0}\right) $ is known, i.e., when $%
\mathbf{R}$ and $\mathbf{P}$ are\ considered as classical commuting
variables (which means that $U_{0}\left( \mathbf{X}_{0}\right) $ is known).
Of course these equations do not allow to find directly $\varepsilon\left( 
\mathbf{X}\right) $, $U\left( \mathbf{X}\right) $ since those quantities are
involved on the R.H.S. of these relations. However, they allow us to produce
solutions for $\varepsilon\left( \mathbf{X}\right) $ and $U\left( \mathbf{X}%
\right) $ recursively in a series expansion in $\hbar$. Moreover, and as
needed, the results of our previous sections show that the matrices $%
\varepsilon\left( \mathbf{X}\right) $ and $U\left( \mathbf{X}\right) $
obtained through this process are independent of any choice of
symmetrization. A particular choice in the way of arranging the variables
for $U\left( \mathbf{X}\right) =0$ will lead to differently symmetrized, but
identical, operators.

As it will appear clearly later on, having both $\varepsilon \left( \mathbf{X%
}\right) $ and $U\left( \mathbf{X}\right) $ at order $n$ in $\hbar $, and
reinserting in the exponential of Eqs. $\left( \ref{E}\right) \left( \ref{U}%
\right) $ allows us to find $\varepsilon \left( \mathbf{X}\right) $ and $%
U\left( \mathbf{X}\right) $ at order $n+1$ in $\hbar $. But before solving
recursively the set of equations Eqs.$\left( \ref{E}\right) \left( \ref{U}%
\right) $, we first compare the present approach with the one developed in
article \cite{SERIESPIERRE}.

\section{Link with the differential equation of ref. \protect\cite%
{SERIESPIERRE}}

In \cite{SERIESPIERRE} we developed a different less general approach which
led to the differential equation 
\begin{align}
\frac{\partial }{\partial \alpha }\varepsilon _{\alpha }\left( \mathbf{X}%
_{\alpha },\alpha \right) & =\left[ \partial _{\alpha }U_{\alpha }\left( 
\mathbf{X}_{\alpha }\right) U_{\alpha }^{+}\left( \mathbf{X}_{\alpha
}\right) ,\varepsilon _{\alpha }\left( \mathbf{X}_{\alpha }\right) \right] +%
\frac{1}{2}\left\{ \mathcal{A}_{\alpha }^{R_{l}}\nabla _{R_{l}}\varepsilon
_{\alpha }+\nabla _{R_{l}}\varepsilon _{\alpha }\mathcal{A}_{\alpha
}^{R_{l}}+\mathcal{A}_{\alpha }^{P_{l}}\nabla _{P_{l}}\varepsilon _{\alpha
}+\nabla _{P_{l}}\varepsilon _{\alpha }\mathcal{A}_{\alpha }^{P_{l}}\right\}
\notag \\
& +\frac{i}{2}\left\{ \left[ \varepsilon _{\alpha },\mathcal{A}_{\alpha
}^{R_{l}}\right] \mathcal{A}_{\alpha }^{P_{l}}-\left[ \varepsilon _{\alpha },%
\mathcal{A}_{\alpha }^{P_{l}}\right] \mathcal{A}_{\alpha }^{R_{l}}\right\} -%
\frac{i}{2}\left[ \varepsilon _{\alpha },\left[ \mathcal{A}_{\alpha
}^{R_{l}},\mathcal{A}_{\alpha }^{P_{l}}\right] \right]  \notag \\
& +\left\{ U_{\alpha }\left\langle H\left( \mathbf{X}_{\alpha }\right)
\right\rangle U_{\alpha }^{+}-\frac{i}{2}\left[ \mathcal{B}_{\alpha
}\varepsilon _{\alpha }-\varepsilon _{\alpha }\mathcal{B}_{\alpha }\right]
-\left\langle \varepsilon _{\alpha }\right\rangle \right\}  \label{equalast}
\end{align}
(in the right hand side we have skipped the explicit dependence in $\alpha $
for the sake of simplicity) which was coupled to the evolution of the
transformation matrix $U_{\alpha }\left( \mathbf{X}_{\alpha }\right) $ as a
function of $\alpha $ \cite{SERIESPIERRE} : 
\begin{equation}
0=\partial _{\alpha }U_{\alpha }(\mathbf{X}_{\alpha }\mathbf{)}U_{\alpha
}^{+}(\mathbf{X}_{\alpha }\mathbf{)+}U\mathbf{_{\alpha }(X_{\alpha }\mathbf{)%
}}\partial _{\alpha }U\mathbf{_{\alpha }^{+}(X_{\alpha }\mathbf{)}}-\frac{i}{%
2}\left( \mathcal{B}_{\alpha }-\mathcal{B}_{\alpha }^{+}\right) +\frac{i}{2}%
\left[ \mathcal{A}_{\alpha }^{R_{l}},\mathcal{A}_{\alpha }^{P_{l}}\right]
\label{Uevolution}
\end{equation}
where $-\frac{i}{2}\mathcal{B}_{\alpha }=\left( Asym\left[ \nabla
_{R^{l}}\nabla _{P_{l}}U_{\alpha }\left( \mathbf{X}_{\alpha }\right) \right]
\right) U_{\alpha }^{+}\left( \mathbf{X}_{\alpha }\right) =\left\langle
U_{\alpha }\right\rangle U_{\alpha }^{+}$.With these two equations Eqs. $%
\left( \ref{equalast}\right) $ and $\left( \ref{Uevolution}\right) $ at
hand, the diagonalization process can be performed. Actually, since all
quantities are matrix valued and since $\varepsilon _{\alpha }\left( \mathbf{%
X}_{\alpha }\right) $ is by definition a diagonal matrix, we can separate
the energy equation Eq. $\left( \ref{equalast}\right) $ in a diagonal and a
off-diagonal part such that we are led to the following two equations 
\begin{align}
\frac{\partial }{\partial \alpha }\varepsilon _{\alpha }\left( \mathbf{X}%
_{\alpha }\right) & =\mathcal{P}_{+}[\text{R.H.S}.\text{ of Eq. \ref%
{equalast}}]  \label{eq1} \\
0& =\mathcal{P}_{-}[\text{R.H.S}.\text{ of Eq. \ref{equalast}}]  \label{eq2}
\end{align}
In \cite{SERIESPIERRE} it was claimed that those three Eqs. $\left( \ref%
{Uevolution}\right) \left( \ref{eq1}\right) \left( \ref{eq2}\right) $ allow
us to determine recursively in powers of $\alpha $ the energy of the quantum
system in question. Actually, the integration over $\alpha $ of Eq. $\left( %
\ref{eq1}\right) $ gives $\varepsilon _{\alpha }\left( \mathbf{X}_{\alpha
}\right) $ at order $n$ in $\alpha $ when knowing all quantities at order $%
n-1$. By the same token, Eqs. $\left( \ref{eq2}\right) $ and $\left( \ref%
{Uevolution}\right) $ (whose meaning is that $U_{\alpha }\left( \mathbf{X}%
_{\alpha }\right) $ is unitary at each order in $\alpha $) involve $\partial
_{\alpha }U_{\alpha }\left( \mathbf{X}_{\alpha }\right) $, and allow to
recover $U_{\alpha }\left( \mathbf{X}_{\alpha }\right) $ at order $n$ by
integration over $\alpha $. As a consequence, the diagonalization process is
perfectly controlled order by order in the series expansion in $\alpha $. In 
\cite{SERIESPIERRE} we also provided two physical examples at the order $%
\hbar ^{2}$.

Now we want to show that our solution Eq. $\left( \ref{Esolution}\right) $
satisfies the differential equation Eq. $\left( \ref{equalast}\right) $.

To do so, we first rewrite $\left( \ref{equalast}\right) $ and $\left( \ref%
{Uevolution}\right) $ with the notation of this paper. First, $\left( \ref%
{Uevolution}\right) $, is in fact $\left( \ref{U2}\right) $. Actually, $-%
\frac{i}{2}\left( \mathcal{B}_{\alpha }-\mathcal{B}_{\alpha }^{+}\right)
=\left\langle U_{\alpha }\right\rangle U_{\alpha }^{+}+U_{\alpha
}\left\langle U_{\alpha }^{+}\right\rangle $.

Second, $\left( \ref{eq2}\right) $ is identical to $\left( \ref{U3}\right) $%
. It can be seen by starting with $\left( \ref{eq2}\right) $, and noting
that in $\left( \ref{eq2}\right) $ 
\begin{eqnarray*}
&&\left[ \partial _{\alpha }U_{\alpha }\left( \mathbf{X}_{\alpha }\right)
U_{\alpha }^{+}\left( \mathbf{X}_{\alpha }\right) ,\varepsilon _{\alpha
}\left( \mathbf{X}_{\alpha }\right) \right] -\frac{i}{2}\left[ \mathcal{B}%
_{\alpha }\varepsilon _{\alpha }-\varepsilon _{\alpha }\mathcal{B}_{\alpha }%
\right] \\
&=&\left[ \partial _{\alpha }U_{\alpha }\left( \mathbf{X}_{\alpha }\right)
U_{\alpha }^{+}\left( \mathbf{X}_{\alpha }\right) ,\varepsilon _{\alpha
}\left( \mathbf{X}_{\alpha }\right) \right] +\left\langle U_{\alpha
}\right\rangle U_{\alpha }^{+}\varepsilon _{\alpha }-\varepsilon _{\alpha
}\left\langle U_{\alpha }\right\rangle U_{\alpha }^{+} \\
&=&\left[ \left( \left( \frac{\partial }{\partial \alpha }+\left\langle
.\right\rangle \right) U_{\alpha }\right) U_{\alpha }^{+},\varepsilon
_{\alpha }\right]
\end{eqnarray*}
which, reinserted in $\left( \ref{eq2}\right) $ and projected on the non
diagonal subspace through $\mathcal{P}_{-}$\ leads to $\left( \ref{U3}%
\right) $ after some rearrangements. Note that the term proportionnal to $%
\frac{\partial }{\partial \alpha }H\left( \mathbf{X}_{\alpha }\right) $ is
missing in $\left( \ref{eq2}\right) $ compared to $\left( \ref{U3}\right) $%
,since it was assumed to be nul in \cite{SERIESPIERRE}.

Now, we show that $\varepsilon _{\alpha }\left( \mathbf{X}_{\alpha }\right) $
as defined in $\left( \ref{E}\right) $, satisfies $\left( \ref{eq1}\right) $%
. We will use the hermiticity of $\varepsilon _{\alpha }\left( \mathbf{X}%
_{\alpha },\alpha \right) $ as well as the previous gauge condition stating
that $\mathcal{P}_{+}\left( \left( \frac{\partial }{\partial \alpha }%
+\left\langle .\right\rangle \right) U_{\alpha }\right) U_{\alpha }^{+}$ is
hermitian. As proved in the previous section, it implies that $\left[ \left(
\left( \frac{\partial }{\partial \alpha }+\left\langle .\right\rangle
\right) U_{\alpha }\right) U_{\alpha }^{+},\varepsilon _{\alpha }\right] $
is antihermitian. We also assume $\left\langle H\left( \mathbf{X}_{\alpha
}\right) \right\rangle =0$, to be consistent with the previous sections, but
including it would not harm. We are led to: 
\begin{eqnarray*}
\frac{\partial }{\partial \alpha }\varepsilon _{\alpha }\left( \mathbf{X}%
_{\alpha }\right) &=&\frac{1}{2}\mathcal{P}_{+}[\text{R.H.S}.\text{ of Eq. %
\ref{equalast}}+H.C.] \\
&=&\frac{1}{2}\mathcal{P}_{+}\left\{ \mathcal{A}_{\alpha }^{R_{l}}\nabla
_{R_{l}}\varepsilon _{\alpha }+\nabla _{R_{l}}\varepsilon _{\alpha }\mathcal{%
A}_{\alpha }^{R_{l}}+\mathcal{A}_{\alpha }^{P_{l}}\nabla _{P_{l}}\varepsilon
_{\alpha }+\nabla _{P_{l}}\varepsilon _{\alpha }\mathcal{A}_{\alpha
}^{P_{l}}\right\} \\
&&+\frac{i}{4}\left\{ \left[ \varepsilon _{\alpha },\mathcal{A}_{\alpha
}^{R_{l}}\right] \mathcal{A}_{\alpha }^{P_{l}}-\left[ \varepsilon _{\alpha },%
\mathcal{A}_{\alpha }^{P_{l}}\right] \mathcal{A}_{\alpha }^{R_{l}}-\left[
\varepsilon _{\alpha },\left[ \mathcal{A}_{\alpha }^{R_{l}},\mathcal{A}%
_{\alpha }^{P_{l}}\right] \right] +H.C.\right\} -\left\langle \varepsilon
_{\alpha }\right\rangle \\
&=&O_{\alpha }\varepsilon _{\alpha }\left( \mathbf{X}_{\alpha }\right)
-\left\langle \varepsilon _{\alpha }\right\rangle
\end{eqnarray*}

Now, from Eq. $\left( \ref{E}\right) $ we can write the partial differential
with respect to $\hbar $ :

\begin{align}
\frac{\partial }{\partial \hbar }\varepsilon \left( \mathbf{X}_{\hbar
}\right) & =\frac{\partial }{\partial \hbar }\mathcal{E}\left( \left[ 
\mathcal{T}\exp \left[ \int_{0<\alpha _{1}<\hbar }e^{-\left\langle
.\right\rangle _{\alpha _{1}\rightarrow \hbar }^{S}}O_{\alpha
_{1}}e^{-\left\langle .\right\rangle _{0\rightarrow \alpha _{1}}^{S}}d\alpha
_{1}\right] \right] \varepsilon _{0}\left( \mathbf{X}_{0}\right) \right) 
\notag \\
& =-\left\langle .\right\rangle \varepsilon _{\hbar }\left( \mathbf{X}%
_{\hbar }\right)  \notag \\
+& \mathcal{E}\left( \sum_{n}\int\limits_{0<\alpha _{n}<...<\alpha
_{2}<\hbar }O_{\hbar }e^{-\left\langle .\right\rangle _{\alpha
_{2}\rightarrow \hbar }^{S}}O_{\alpha _{2}}..e^{-\left\langle .\right\rangle
_{\alpha _{n-1}\rightarrow \alpha _{n-2}}^{S}}O_{\alpha
_{n-1}}e^{-\left\langle .\right\rangle _{\alpha _{n}\rightarrow \alpha
_{n-1}}^{S}}O_{\alpha _{n}}e^{-\left\langle .\right\rangle _{0\rightarrow
\alpha _{n}}^{S}}\varepsilon _{0}\left( \mathbf{X}_{0}\right) \right)  \notag
\\
& =-\left\langle .\right\rangle \varepsilon _{\hbar }\left( \mathbf{X}%
_{\hbar }\right) +  \notag \\
& O_{\hbar }\mathcal{E}\left( \sum_{n}\int\limits_{0<\alpha _{n}<..<\alpha
_{2}<\hbar }e^{-\left\langle .\right\rangle _{\alpha _{2}\rightarrow \hbar
}^{S}}O_{\alpha _{2}}..e^{-\left\langle .\right\rangle _{\alpha
_{n-1}\rightarrow \alpha _{n-2}}^{S}}O_{\alpha _{n-1}}e^{-\left\langle
.\right\rangle _{\alpha _{n}\rightarrow \alpha _{n-1}}^{S}}O_{\alpha
_{n}}e^{-\left\langle .\right\rangle _{0\rightarrow \alpha
_{n}}^{S}}\varepsilon _{0}\left( \mathbf{X}_{0}\right) \right)  \notag \\
& =-\left\langle .\right\rangle \varepsilon _{\hbar }\left( \mathbf{X}%
_{\hbar }\right) +O_{\hbar }\varepsilon _{\hbar }\left( \mathbf{X}_{\hbar
}\right)
\end{align}
The first equality has been obtained by using that by construction, $\frac{%
\partial }{\partial \hbar }e^{-\left\langle .\right\rangle _{\alpha
_{1}\rightarrow \hbar }^{S}}=-S_{\mathbf{X}_{\hbar }}\left\langle
.\right\rangle _{\hbar }S_{\mathbf{X}_{\hbar }}$ and by remarking that in
computation of the partial derivative $\frac{\partial }{\partial \hbar }$,
the canonical variables and the expectation operator remain unchanged (that
is $\mathbf{X}_{\hbar }$ is seen as a constant). We have given previously
the expression for $O_{\hbar }\varepsilon _{\hbar }\left( \mathbf{Y}_{\hbar
}\right) $. Using moreover Eq. $\left( \ref{Uevolution}\right) $ one gets
directly that $\left( \varepsilon \left( \mathbf{R},\mathbf{P}\right)
-\varepsilon _{0}\left( \mathbf{R},\mathbf{P}\right) \right) $ is a solution
of the differential $\left( \ref{equalast}\right) $.

\section{Dynamical operators and commutation algebra}

In this section we will see that new non-commuting position and momentum
operators which have contributions from Berry connections emerge during the
diagonalization and are more suitably to correspond to physical operators
(for the physical discussion of this point see \cite{PIERRE1}\cite%
{PIERREBLOCH}\cite{SERIESPIERRE}\cite{ALAIN}). This discussion may be
skipped, and the reader can directly move to the next section dealing with
the general expressions for the diagonalized Hamiltonian to the first and
second order.

From Eq.$\left( \ref{diago}\right) $ one sees that the operator $O_{\alpha
}=\left( \partial _{\alpha }+\left\langle .\right\rangle \right) $ acting on 
$\varepsilon _{0}\left( \mathbf{X}_{\alpha }\right) $ can be decomposed as a
sum of a \textquotedblright translation\textquotedblright\ operator $T$ and
a \textquotedblright magnetization\textquotedblright\ $M$ operator (this
terminology is explained in \cite{SERIESPIERRE}) 
\begin{equation}
O_{\alpha }\varepsilon _{0}\left( \mathbf{X}_{\alpha }\right) =\left(
T_{\alpha }+M_{\alpha }\right) \varepsilon _{0}\left( \mathbf{X}_{\alpha
}\right) .  \label{equadiff}
\end{equation}
where the \textquotedblright magnetization\textquotedblright\ operator acts
as 
\begin{equation}
M_{\alpha }\varepsilon _{0}\left( \mathbf{X}_{\alpha }\right) =\frac{i}{2}%
\mathcal{P}_{+}\left\{ \left[ \varepsilon _{\alpha },\mathcal{A}_{\alpha
}^{R_{l}}\right] \mathcal{A}_{\alpha }^{P_{l}}-\left[ \varepsilon _{\alpha },%
\mathcal{A}_{\alpha }^{P_{l}}\right] \mathcal{A}_{\alpha }^{R_{l}}\right\} +%
\mathcal{P}_{+}\left[ U_{\alpha }\left( \left( \frac{\partial }{\partial
\alpha }+\left\langle .\right\rangle \right) H\left( \mathbf{X}_{\alpha
}\right) \right) U_{\alpha }^{+}\right]  \label{M}
\end{equation}
and 
\begin{equation}
T_{\alpha }\varepsilon _{0}\left( \mathbf{X}_{\alpha }\right) =\frac{1}{2}%
\mathcal{P}_{+}\left\{ \mathcal{A}_{\alpha }^{R_{l}}\nabla
_{R_{l}}\varepsilon _{0}\left( \mathbf{X}_{\alpha }\right) +\nabla
_{R_{l}}\varepsilon _{0}\mathcal{A}_{\alpha }^{R_{l}}\left( \mathbf{X}%
_{\alpha }\right) +\mathcal{A}_{\alpha }^{P_{l}}\nabla _{P_{l}}\varepsilon
_{0}\left( \mathbf{X}_{\alpha }\right) +\nabla _{P_{l}}\varepsilon _{0}%
\mathcal{A}_{\alpha }^{P_{l}}\left( \mathbf{X}_{\alpha }\right) \right\}
\label{TT}
\end{equation}
To inspect the action of $T_{\alpha }$ , let us skip the magnetization
contribution $M_{\alpha }$ for the sake of clarity, and consider the
following relevant contribution for the computation of the diagonalized
Hamiltonian : 
\begin{equation}
\mathcal{ET}\exp \left[ \int_{0<\alpha <\hbar }e^{-\left\langle
.\right\rangle _{\alpha \rightarrow \hbar }^{S}}T_{\alpha }e^{-\left\langle
.\right\rangle _{0\rightarrow \alpha }^{S}}d\alpha \right] \varepsilon
_{0}\left( \mathbf{X}_{0}\right)  \notag
\end{equation}
When developed in series, as explained in the previous sections, it is given
by 
\begin{equation}
\mathcal{E}\sum_{n=0}^{n}\left[ \int_{0<\alpha _{n}<...<\alpha _{1}<\hbar
}e^{-\left\langle .\right\rangle _{\alpha _{1}\rightarrow \hbar
}^{S}}T_{\alpha _{1}}...e^{-\left\langle .\right\rangle _{\alpha
_{n-1}\rightarrow \alpha _{n-2}}^{S}}T_{\alpha _{n-1}}e^{-\left\langle
.\right\rangle _{\alpha _{n}\rightarrow \alpha _{n-1}}^{S}}T_{\alpha
_{n}}e^{-\left\langle .\right\rangle _{0\rightarrow \alpha _{n}}^{S}}\right]
\varepsilon _{0}\left( \mathbf{X}_{0}\right)  \label{tran}
\end{equation}
Recall that each operation of the kind $e^{-\left\langle .\right\rangle
_{\alpha i\rightarrow \alpha _{i-1}}^{S}}$ is going along with a change of
variable $\mathbf{X}_{\alpha _{i}}\rightarrow \mathbf{X}_{\alpha _{i-1}}$.

We find $\mathcal{ET}\exp \left[ \int_{0<\alpha <\hbar }e^{-\left\langle
.\right\rangle _{\alpha \rightarrow \hbar }^{S}}T_{\alpha }e^{-\left\langle
.\right\rangle _{0\rightarrow \alpha }^{S}}d\alpha \right] \varepsilon
_{0}\left( \mathbf{X}\right) $ by the following ansatz. We assume that $%
\mathcal{ET}\exp \left[ \int_{0<\alpha <\hbar }e^{-\left\langle
.\right\rangle _{\alpha \rightarrow \hbar }^{S}}T_{\alpha }e^{-\left\langle
.\right\rangle _{0\rightarrow \alpha }^{S}}d\alpha \right] \varepsilon
_{0}\left( \mathbf{X}\right) =\varepsilon _{0}\left( \mathbf{X}_{\hbar }%
\mathbf{+\emph{A}_{\hbar }^{\mathbf{X}_{\hbar }}}\right) $ where $\mathbf{%
\emph{A}_{\hbar }^{\mathbf{X}}\equiv \emph{A}_{\hbar }^{\mathbf{X}}}\left( 
\mathbf{X}_{\hbar }\mathbf{,}\hbar \right) $ has to be determined and a
particular (arbitrary) choice of symmetrization of the variables $\mathbf{X}$
has been made. Due to the form of the left hand side in the previous
relation Eq. $\left( \ref{tran}\right) $, $\mathbf{\emph{A}_{\hbar }^{%
\mathbf{X}}}$ is of order $\hbar $. We choose to write $\varepsilon _{0}$
and $\mathbf{\emph{A}_{\hbar }^{\mathbf{X}}}$ such that the powers of $%
\mathbf{R}$ and $\mathbf{P}$ are in a completely symmetrized form, that is
we sum over all equally weighted permutations of the canonical variables in
the series expansion of $\varepsilon _{0}\left( \mathbf{X}\right) $ (and $%
\mathbf{\emph{A}_{\hbar }^{\mathbf{X}}}$) . Then, $\varepsilon _{0}\left( 
\mathbf{X+\emph{A}_{\hbar }^{\mathbf{X}}}\right) $ is obtained by replacing $%
\mathbf{X}$ by $\mathbf{X+\emph{A}_{\hbar }^{\mathbf{X}}}$ in the series
expansion of $\varepsilon _{0}$. Note that, for later purposes $\left\langle
.\right\rangle _{\hbar }\varepsilon _{0}\left( \mathbf{X+\emph{A}_{\hbar }^{%
\mathbf{X}}}\right) $ is of order $\hbar ^{2}$. To differentiate the right
hand side of Eq.\ $\left( \ref{tran}\right) $ with respect to $\hbar $ we
proceed in the following way.

On one hand, assuming that $\varepsilon _{0}$ does not depends explicitly on 
$\hbar $, $\frac{\partial }{\partial \hbar }\varepsilon _{0}\left( \mathbf{X+%
\emph{A}_{\hbar }^{\mathbf{X}}}\right) $ is equal to : 
\begin{equation*}
\frac{\partial }{\partial \hbar }\varepsilon _{0}\left( \mathbf{X+\emph{A}%
_{\hbar }^{\mathbf{X}}}\right) =\frac{\partial \mathbf{\emph{A}_{\hbar }^{%
\mathbf{X}}}}{\partial \hbar }\mathbf{\nabla }_{\mathbf{X}}\varepsilon
_{0}\left( \mathbf{X+\emph{A}_{\hbar }^{\mathbf{X}}}\right)
\end{equation*}
Note that in the last expression, the product $\frac{\partial \mathbf{\emph{A%
}_{\hbar }^{\mathbf{X}}}}{\partial \hbar }\mathbf{\nabla }_{\mathbf{X}%
}\varepsilon _{0}\left( \mathbf{X+\emph{A}_{\hbar }^{\mathbf{X}}}\right) $
has to be understood in the sense that $\frac{\partial \mathbf{\emph{A}%
_{\hbar }^{\mathbf{X}}}}{\partial \hbar }$ replaces $\mathbf{X+\emph{A}%
_{\hbar }^{\mathbf{X}}}$ at each place the gradient is acting on the series
expansion of $\varepsilon _{0}$.

On the other hand, the same infinitesimal variation can computed through the
left hand side of Eq. $\left( \ref{tran}\right) $ : 
\begin{align*}
& \frac{\partial }{\partial \hbar }\varepsilon _{0}\left( \mathbf{X}_{\hbar }%
\mathbf{+\emph{A}_{\hbar }^{\mathbf{X}_{\hbar }}}\right) \\
& =\frac{\partial }{\partial \hbar }\mathcal{E}\sum_{n=0}^{n}\left[
\int\limits_{0<\alpha _{n}<...<\alpha _{1}<\hbar }e^{-\left\langle
.\right\rangle _{\alpha _{1}\rightarrow \hbar }^{S}}T_{\alpha
_{1}}..e^{-\left\langle .\right\rangle _{\alpha _{n-1}\rightarrow \alpha
_{n-2}}^{S}}T_{\alpha _{n-1}}e^{-\left\langle .\right\rangle _{\alpha
_{n}\rightarrow \alpha _{n-1}}^{S}}T_{\alpha _{n}}e^{-\left\langle
.\right\rangle _{0\rightarrow \alpha _{n}}^{S}}\right] \varepsilon
_{0}\left( \mathbf{X}_{0}\right) \\
& =-\left\langle .\right\rangle _{\hbar }\mathcal{E}\sum_{n=0}^{n}\left[
\int\limits_{0<\alpha _{n}<...<\alpha _{1}<\hbar }e^{-\left\langle
.\right\rangle _{\alpha _{1}\rightarrow \hbar }^{S}}T_{\alpha
_{1}}..e^{-\left\langle .\right\rangle _{\alpha _{n-1}\rightarrow \alpha
_{n-2}}^{S}}T_{\alpha _{n-1}}e^{-\left\langle .\right\rangle _{\alpha
_{n}\rightarrow \alpha _{n-1}}^{S}}T_{\alpha _{n}}e^{-\left\langle
.\right\rangle _{0\rightarrow \alpha _{n}}^{S}}\right] \varepsilon
_{0}\left( \mathbf{X}_{0}\right) \\
& +\mathcal{E}\left( \sum_{n}\left[ \int\limits_{0<\alpha _{n}<...<\alpha
_{2}<\hbar }T_{\hbar }e^{-\left\langle .\right\rangle _{\alpha
_{2}\rightarrow \hbar }^{S}}T_{\alpha _{2}}..e^{-\left\langle .\right\rangle
_{\alpha _{n-1}\rightarrow \alpha _{n-2}}^{S}}T_{\alpha
_{n-1}}e^{-\left\langle .\right\rangle _{\alpha _{n}\rightarrow \alpha
_{n-1}}^{S}}T_{\alpha _{n}}e^{-\left\langle .\right\rangle _{0\rightarrow
\alpha _{n}}^{S}}\right] \varepsilon _{0}\left( \mathbf{X}_{0}\right) \right)
\\
& \left( -\left\langle .\right\rangle _{\hbar }+T_{\hbar }\right)
\varepsilon _{0}\left( \mathbf{X}_{\hbar }\mathbf{+\emph{A}_{\hbar }^{%
\mathbf{X}_{\hbar }}}\right)
\end{align*}
As a consequence, one is left ultimately with : 
\begin{equation}
\left[ -\left\langle .\right\rangle _{\hbar }+T_{\hbar }\right] \varepsilon
_{0}\left( \mathbf{X+\emph{A}_{\hbar }^{\mathbf{X}}}\right) =\frac{\partial 
\mathbf{\emph{A}_{\hbar }^{\mathbf{X}}}}{\partial \hbar }\mathbf{\nabla }_{%
\mathbf{X}}\varepsilon _{0}\left( \mathbf{X+\emph{A}_{\hbar }^{\mathbf{X}}}%
\right)  \label{TH}
\end{equation}
Now consider $\mathcal{P}_{+}\left[ \mathcal{A}_{\alpha }^{\mathbf{R}}\right]
$ and $\mathcal{P}_{+}\left[ \mathcal{A}_{\alpha }^{\mathbf{P}}\right] $,
with the following definitions for the \textquotedblright
non-projected\textquotedblright\ Berry connections $\mathcal{A}_{\alpha }^{%
\mathbf{R}}=i\left[ U_{\alpha }\mathbf{\nabla }_{\mathbf{P}}U_{\alpha }^{+}%
\right] $ and $\mathcal{A}_{\alpha }^{\mathbf{P}}=-i\left[ U_{\alpha }%
\mathbf{\nabla }_{\mathbf{R}}U_{\alpha }^{+}\right] $. Define also for
convenience $\mathcal{P}_{+}\left[ \mathcal{A}_{\alpha }^{\mathbf{X}}\right]
=\left( \mathcal{P}_{+}\left[ \mathcal{A}_{\alpha }^{\mathbf{R}}\right] ,%
\mathcal{P}_{+}\left[ \mathcal{A}_{\alpha }^{\mathbf{P}}\right] \right) $.
Our last equation Eq. $\left( \ref{TH}\right) $\ is thus equivalent to : 
\begin{align*}
& \frac{1}{2}\mathcal{P}_{+}\left[ \mathcal{A}_{\hbar }^{\mathbf{X}}\right] .%
\left[ \mathbf{\nabla }_{\mathbf{X}}\varepsilon _{0}\left( \mathbf{X+\emph{A}%
_{\hbar }^{\mathbf{X}}}\right) +\mathbf{\nabla }_{\mathbf{X}}\left( \mathbf{%
\emph{A}_{\hbar }^{\mathbf{X}}}\right) ^{i}\left( \mathbf{X}\right) \nabla _{%
\mathbf{X}_{i}}\varepsilon _{0}\left( \mathbf{X+\emph{A}_{\hbar }^{\mathbf{X}%
}}\right) \right] +H.C.-\left\langle .\right\rangle _{\hbar }\varepsilon
_{0}\left( \mathbf{X+\emph{A}_{\hbar }^{\mathbf{X}}}\right) \\
& =\frac{\partial \mathbf{\emph{A}_{\hbar }^{\mathbf{X}}}}{\partial \hbar }%
\mathbf{\nabla }_{\mathbf{X}\varepsilon _{0}0}\left( \mathbf{X+\emph{A}%
_{\hbar }^{\mathbf{X}}}\right)
\end{align*}
where $H.C.$ stands for the Hermitian conjugate. The term\textbf{\ }$\mathbf{%
\nabla }_{\mathbf{X}}\varepsilon _{0}\left( \mathbf{X+\emph{A}_{\hbar }^{%
\mathbf{X}}}\right) $ stems for the gradient of $\varepsilon _{0}\left( 
\mathbf{X}\right) $ evaluated at $\left( \mathbf{X+\emph{A}_{\hbar }^{%
\mathbf{X}}}\right) $. As before $\mathbf{\nabla }_{\mathbf{X}}\left( 
\mathbf{\emph{A}_{\hbar }^{\mathbf{X}}}\right) ^{i}\left( \mathbf{X}\right)
\nabla _{\mathbf{X}_{i}}\varepsilon _{0}\left( \mathbf{X+\emph{A}_{\hbar }^{%
\mathbf{X}}}\right) $ has to be understood in the sense that $\mathbf{\nabla 
}_{\mathbf{X}}\left( \mathbf{\emph{A}_{\hbar }^{\mathbf{X}}}\right) ^{i}$
replaces $\mathbf{X+\emph{A}_{\hbar }^{\mathbf{X}}}$ at each place the
gradient is acting on the series expansion of $\varepsilon _{0}$. To be able
to compare both sides of this equation, one has to symmetrize both
expressions in the same way. As a consequence the left hand side has to be
rewritten in the same symmetrized form as the right hand side : 
\begin{align*}
& \frac{1}{2}\mathcal{P}_{+}\left[ \mathcal{A}_{\hbar }^{\mathbf{X}}\right] .%
\left[ \mathbf{\nabla }_{\mathbf{X}}\varepsilon _{0}\left( \mathbf{X+\emph{A}%
_{\hbar }^{\mathbf{X}}}\right) +\mathbf{\nabla }_{\mathbf{X}}\left( \mathbf{%
\emph{A}_{\hbar }^{\mathbf{X}}}\right) ^{i}\left( \mathbf{X}\right) \nabla _{%
\mathbf{X}_{i}}\varepsilon _{0}\left( \mathbf{X+\emph{A}_{\hbar }^{\mathbf{X}%
}}\right) \right] +H.C. \\
& =\frac{1}{2}\left\{ \mathcal{P}_{+}\left[ \mathcal{A}_{\hbar }^{\mathbf{X}}%
\right] .\left[ \mathbf{\nabla }_{\mathbf{X}}\varepsilon _{0}\left( \mathbf{%
X+\emph{A}_{\hbar }^{\mathbf{X}}}\right) +\mathbf{\nabla }_{\mathbf{X}%
}\left( \mathbf{\emph{A}_{\hbar }^{\mathbf{X}}}\right) ^{i}\left( \mathbf{X}%
\right) \nabla _{\mathbf{X}_{i}}\varepsilon _{0}\left( \mathbf{X+\emph{A}%
_{\hbar }^{\mathbf{X}}}\right) \right] \right\} _{S}+H.C.+C\left( \mathbf{X}%
\right)
\end{align*}
This formula requires some explanation. In the left hand side, the
multiplication by $\mathcal{P}_{+}\left[ \mathcal{A}_{\alpha }^{\mathbf{X}}%
\right] $ is performed half on the left and half on the right, as implied by
definition of the translation operator. On the right hand side, all
expressions are seen as symmetrized in a way that $\mathcal{P}_{+}\left[ 
\mathcal{A}_{\alpha }^{\mathbf{X}}\right] $ has been inserted at each place
where the $\mathbf{\nabla }_{\mathbf{X}}$ is acting (exactly as for $\frac{%
\partial \mathbf{\emph{A}_{\hbar }^{\mathbf{X}}}\left( \mathbf{X}\right) }{%
\partial \hbar }\mathbf{\nabla }_{\mathbf{X}}\varepsilon _{0}\left( \mathbf{%
X+\emph{A}_{\hbar }^{\mathbf{X}}}\right) $). The $\left\{ {}\right\} _{S}$
is there to recall this full symmetrization in the variables. The term $%
C\left( \mathbf{X}\right) $ is the correction due to this change of
symmetrization while moving the $\mathcal{P}_{+}\left[ \mathcal{A}_{\alpha
}^{\mathbf{X}}\right] $ inside the series expansion of $\varepsilon _{0}$.
By construction, it involves the powers of gradients of $\mathcal{P}_{+}%
\left[ \mathcal{A}_{\alpha }^{\mathbf{X}}\right] $ and $\varepsilon _{0}$
and is of order $\hbar ^{2}$. In practice, this term depends specifically on
the problem at hand (that is on the form of the Hamiltonian) and can be
computed order by order in $\hbar $. Since later on, we will consider only
the order $\hbar ^{2}$, it will turn out that this term will be negligible
due to an integration.

We can now write the differential equation : 
\begin{align*}
\frac{\partial \mathbf{\emph{A}_{\hbar }^{\mathbf{X}}}\left( \mathbf{X}%
\right) }{\partial \hbar }\mathbf{\nabla }_{\mathbf{X}}\varepsilon
_{0}\left( \mathbf{X+\emph{A}_{\hbar }^{\mathbf{X}}}\right) & =\frac{1}{2}%
\left\{ \mathcal{P}_{+}\left[ \mathcal{A}_{\alpha }^{\mathbf{X}}\right] .%
\left[ \mathbf{\nabla }_{\mathbf{X}}\varepsilon _{0}\left( \mathbf{X+\emph{A}%
_{\hbar }^{\mathbf{X}}}\right) +\mathbf{\nabla }_{\mathbf{X}}\left( \mathbf{%
\emph{A}_{\hbar }^{\mathbf{X}}}\right) ^{i}\left( \mathbf{X}\right) \nabla _{%
\mathbf{X}_{i}}\varepsilon _{0}\left( \mathbf{X+\emph{A}_{\hbar }^{\mathbf{X}%
}}\right) \right] \right\} _{S} \\
& +H.C.+C\left( \mathbf{X}\right) -\left\langle .\right\rangle _{\hbar
}\varepsilon _{0}\left( \mathbf{X+\emph{A}_{\hbar }^{\mathbf{X}}}\right)
\end{align*}
where all expressions are now symmetrized in the same way. Now, remark that,
given our previous remarks, $C\left( \mathbf{X}\right) -\left\langle
.\right\rangle _{\hbar }\varepsilon _{0}\left( \mathbf{X+\emph{A}_{\hbar }^{%
\mathbf{X}}}\right) $ is of order $\hbar ^{2}$. It implies that, after
integration, it will contribute only to the third order in $\hbar $ to $%
\mathbf{\emph{A}_{\hbar }^{\mathbf{X}}}$. As a consequence, neglecting this
term in first approximation (that will be indeed the case in our
applications) we thus deduce that $\mathbf{\emph{A}_{\hbar }^{\mathbf{X}}}$
satisfies the following differential equation : 
\begin{equation}
\mathcal{P}_{+}\left[ \mathcal{A}_{\hbar }^{\mathbf{X}}\right] +\frac{1}{2}%
\left( \mathcal{P}_{+}\left[ \mathcal{A}_{\hbar }^{\mathbf{X}}\right] .%
\mathbf{\nabla }_{\mathbf{X}}\right) \mathbf{\emph{A}_{\hbar }^{\mathbf{X}}}%
+H.C.=\frac{\partial \mathbf{\emph{A}_{\hbar }^{\mathbf{X}}}}{\partial \hbar 
}  \label{phase}
\end{equation}
Given that for $\hbar =0$, one has $\mathbf{\emph{A}_{\hbar }^{\mathbf{X}}=0}
$, the solution of this equation can be written recursively as: 
\begin{align}
\mathbf{\emph{A}_{\hbar }^{\mathbf{X}}}& =\int_{0<\alpha <\hbar }S_{\mathbf{X%
}_{\hbar }}\left[ \mathcal{P}_{+}\left[ \mathcal{A}_{\alpha }^{\mathbf{X}}%
\right] +\frac{1}{2}\left( \left( \mathcal{P}_{+}\left[ \mathcal{A}_{\alpha
}^{\mathbf{X}}\right] .\mathbf{\nabla }_{\mathbf{X}}\right) \mathbf{\emph{A}%
_{\alpha }^{\mathbf{X}}}+H.C.\right) \right] d\alpha  \label{phaseproj} \\
& =\int_{0<\alpha <\hbar }S_{\mathbf{X}_{\hbar }}\mathcal{P}_{+}\left[ 
\mathcal{A}_{\alpha }^{\mathbf{X}}\right] d\alpha +\int_{0<\alpha <\hbar }S_{%
\mathbf{X}_{\hbar }}\frac{1}{2}\left[ \left[ \mathcal{P}_{+}\left[ \mathcal{A%
}_{\alpha }^{\mathbf{X}}\right] .\mathbf{\nabla }_{\mathbf{X}}\int_{0<\alpha
_{1}<\alpha }\mathcal{P}_{+}\left[ \mathcal{A}_{\alpha _{1}}^{\mathbf{X}}%
\right] \right] +H.C.\right] d\alpha _{1}d\alpha  \notag \\
& +...
\end{align}
Recall that $S_{\mathbf{X}_{\hbar }}$ is the shift of variable $\mathbf{X}%
_{\alpha }\rightarrow \mathbf{X}_{\hbar }=\mathbf{X}$, so that in the
integrals, the variables $\mathbf{X}$ are inert, only the explicit
dependence in $\alpha $ is integrated on. Let us also stress that in the
previous integrals, the variables $\mathbf{X}_{\alpha }$ in $\emph{A}%
_{\alpha }^{\mathbf{R}}$ and $\emph{A}_{\alpha }^{\mathbf{P}}$ have been
replaced by $\mathbf{X}$ and are thus constant with respect to the $\alpha $
integrations. The reason is that in the differential equation Eq. $\left( %
\ref{phase}\right) $, $\mathbf{X}$ is seen as constant, only $\hbar $ is
running. Actually, the equation involves only the partial derivative $\frac{%
\partial \mathbf{\emph{A}_{\hbar }^{\mathbf{X}}}}{\partial \hbar }$. Having
now the solution for $\mathbf{\emph{A}_{\hbar }^{\mathbf{X}}}$ as a function
of the Berry phases we can write our solution for the exponentiated action
of the translation operator. Dividing $\mathbf{\emph{A}_{\hbar }^{\mathbf{X}}%
}$ in two components with respect to $\mathbf{R}$ and $\mathbf{P}$, $\mathbf{%
\emph{A}_{\hbar }^{\mathbf{X}}\equiv }\left( \emph{A}_{\mathbf{R}},\emph{A}_{%
\mathbf{P}}\right) $, one has : 
\begin{equation}
\mathcal{ET}\exp \left[ \int_{0<\alpha <\hbar }e^{-\left\langle
.\right\rangle _{\alpha \rightarrow \hbar }^{S}}T_{\alpha }e^{-\left\langle
.\right\rangle _{0\rightarrow \alpha }^{S}}d\alpha \right] \varepsilon
_{0}\left( \mathbf{X}\right) =\varepsilon _{0}\left( \mathbf{x}\right)
\label{epsrp}
\end{equation}
with $\mathbf{x=}\left( \mathbf{r},\mathbf{p}\right) $ and $\mathbf{r}$ and $%
\mathbf{p}$ are new coordinate and momentum operators corrected by Berry
connections terms in the following way : 
\begin{align}
\mathbf{r}& \equiv \mathbf{R+}\emph{A}_{\mathbf{R}}  \notag \\
\mathbf{p}& \equiv \mathbf{P+}\emph{A}_{\mathbf{P}}  \label{rp}
\end{align}
justifying the name translation operator for $T_{\alpha }$. The inclusion of
the corrections due to $C\left( \mathbf{X}\right) $ can be performed in the
following way. Shifting $\mathbf{\emph{A}_{\hbar }^{\mathbf{X}}}$ by a
correction $\mathbf{\emph{A}_{\hbar }^{\mathbf{X}}+}\delta \mathbf{\emph{A}%
_{\hbar }^{\mathbf{X}}}$ with $\mathbf{\emph{A}_{\hbar }^{\mathbf{X}}}$ the
solution previously found for Eq. $\left( \ref{phase}\right) $, gives the
following equation for $\delta \mathbf{\emph{A}_{\hbar }^{\mathbf{X}}}$ : 
\begin{align*}
& \frac{1}{2}\mathcal{P}_{+}\left[ \mathcal{A}_{\hbar }^{\mathbf{X}}\right] .%
\left[ \mathbf{\nabla }_{\mathbf{X}}\left( \mathbf{\delta \emph{A}_{\hbar }^{%
\mathbf{X}}}\right) ^{i}\left( \mathbf{X}\right) \nabla _{\mathbf{X}%
_{i}}\varepsilon _{0}\left( \mathbf{X+\emph{A}_{\hbar }^{\mathbf{X}}+\delta 
\emph{A}_{\hbar }^{\mathbf{X}}}\right) \right] +H.C.+C\left( \mathbf{X}%
\right) -\left\langle .\right\rangle _{\hbar }\varepsilon _{0}\left( \mathbf{%
X+\emph{A}_{\hbar }^{\mathbf{X}}}\right) \\
& =\frac{\partial \delta \mathbf{\emph{A}_{\hbar }^{\mathbf{X}}}\left( 
\mathbf{X}\right) }{\partial \hbar }\mathbf{\nabla }_{\mathbf{X}}\varepsilon
_{0}\left( \mathbf{X+\emph{A}_{\hbar }^{\mathbf{X}}+\delta \emph{A}_{\hbar
}^{\mathbf{X}}}\right)
\end{align*}
Note that $C\left( \mathbf{X}\right) $ is computed using the derivatives of $%
\mathbf{\emph{A}_{\hbar }^{\mathbf{X}}+}\delta \mathbf{\emph{A}_{\hbar }^{%
\mathbf{X}}}$. Expanding $\delta \mathbf{\emph{A}_{\hbar }^{\mathbf{X}}}%
\left( \mathbf{X}\right) $ as an $\hbar $ series expansion of completely
symmetrized function $\delta \mathbf{\emph{A}_{\hbar }^{\mathbf{X}\left(
n\right) }}$ of $\mathbf{R}$ and $\mathbf{P}$ (starting with $n=3)$\ allows,
at least theoretically to find the $\delta \mathbf{\emph{A}_{\hbar }^{%
\mathbf{X}\left( n\right) }}$ recursively by solving an equation of the kind
: 
\begin{equation*}
\frac{\partial \delta \mathbf{\emph{A}_{\hbar }^{\mathbf{X}\left( n+1\right)
}}\left( \mathbf{X}\right) }{\partial \hbar }=F\left( \mathbf{X},\delta 
\mathbf{\emph{A}_{\hbar }^{\mathbf{X}\left( n\right) }}\left( \mathbf{X}%
\right) ,...,\delta \mathbf{\emph{A}_{\hbar }^{\mathbf{X}\left( n\right) }}%
\left( \mathbf{X}\right) \right)
\end{equation*}
where $F\left( \mathbf{X},\delta \mathbf{\emph{A}_{\hbar }^{\mathbf{X}\left(
n\right) }}\left( \mathbf{X}\right) ,...,\delta \mathbf{\emph{A}_{\hbar }^{%
\mathbf{X}\left( n\right) }}\left( \mathbf{X}\right) \right) $ is determined
by replacing $\varepsilon _{0}$, $\mathbf{\emph{A}_{\hbar }^{\mathbf{X}}}$, $%
\mathcal{P}_{+}\left[ \mathcal{A}_{\hbar }^{\mathbf{X}}\right] $ by their
expressions at the $n$-th order. As said before, in our practical
considerations the term $\delta \mathbf{\emph{A}_{\hbar }^{\mathbf{X}}}%
\left( \mathbf{X}\right) $ will always be neglected.

The variables $\mathbf{x}=\mathbf{X}+\mathbf{\emph{A}_{\hbar }^{\mathbf{X}}}$
we have defined seem thus to be the natural ones arising in the
diagonalization process.

Note that at the lowest order we have : 
\begin{align}
\mathbf{r} & =\mathbf{R}+\hbar\mathcal{P}_{+}\left[ \mathcal{A}_{0}^{\mathbf{%
R}}\right] \equiv\mathbf{R}+\emph{A}_{0}^{\mathbf{R}}  \notag \\
\mathbf{p} & =\mathbf{P}+\hbar\mathcal{P}_{+}\left[ \mathcal{A}_{0}^{\mathbf{%
P}}\right] \equiv\mathbf{P}+\emph{A}_{0}^{\mathbf{P}}
\end{align}
where $\emph{A}_{0}^{\mathbf{R}}$ and $\emph{A}_{0}^{\mathbf{P}}$ are the
usual ''semiclassical Berry connections''\ defined previously \cite%
{SERIESPIERRE}.

From equations Eq. $\left( \ref{rp}\right) $ we readily deduce the following
non trivial algebra 
\begin{align}
\left[ r_{i},r_{j}\right] & =i\hbar ^{2}\Theta _{ij}^{rr}=i\hbar ^{2}\left(
\nabla _{P_{i}}\emph{A}_{R_{j}}-\nabla _{P_{j}}\emph{A}_{R_{i}}\right)
+\hbar ^{2}\left[ \emph{A}_{R_{j}},\emph{A}_{R_{i}}\right]  \notag \\
\left[ p_{i},p_{j}\right] & =i\hbar ^{2}\Theta _{ij}^{pp}=-i\hbar ^{2}\left(
\nabla _{R_{i}}\emph{A}_{P_{j}}-\nabla _{R_{j}}\emph{A}_{P_{i}}\right)
+\hbar ^{2}\left[ \emph{A}_{P_{i}},\emph{A}_{P_{j}}\right]  \notag \\
\left[ p_{i},r_{j}\right] & =-i\hbar \delta _{ij}+i\hbar ^{2}\Theta
_{ij}^{pr}=-i\hbar \delta _{ij}-i\hbar ^{2}\left( \nabla _{R_{i}}\emph{A}%
_{R_{j}}+\nabla _{P_{j}}\emph{A}_{P_{i}}\right) +\hbar ^{2}\left[ \emph{A}%
_{P_{i}},\emph{A}_{R_{j}}\right]  \label{commalgebra}
\end{align}
where the terms $\Theta _{ij}$ are the definitions of Berry curvatures. Of
course these non trivial commutation relations also give new contributions
to the equations of motion and thus lead to new phenomena \cite{SERIESPIERRE}
\cite{PIERREBLOCH}\cite{PIERRE1}. The commutation relations are valid to any
order in $\hbar ,$ but in practice we can compute them as well as the energy 
$\varepsilon \left( \mathbf{X}\right) $ in a series expansion in $\hbar .$
Relations Eqs. $\left( \ref{M}\right) \left( \ref{epsrp}\right) $ will be
helpful when writing the explicit expression of $\varepsilon \left( \mathbf{X%
}\right) $ in a series expansion in $\hbar $ in the following section.

\section{Series expansion in $\hbar$}

The exact expression Eq. $\left( \ref{Esolution}\right) $, can now be
expanded in a series expansion in $\hbar $. Note that we will always
identify $\varepsilon _{\hbar }\left( \mathbf{R}_{\hbar },\mathbf{P}_{\hbar
}\right) $ with $\varepsilon \left( \mathbf{R},\mathbf{P}\right) $. We also
implicitly assume for convenience that all expressions are symmetrized in $%
\mathbf{R}$ and $\mathbf{P}$, in such a way that for all expression
depending on $\mathbf{R}$ and $\mathbf{P}$, in the series expansion of this
expression, all powers of the momentum are put half on the left and half on
the right. Any other choice of symmetrization would be of course suitable.
Recall at this point some previous notations : $\mathcal{A}_{\alpha }^{%
\mathbf{R}}=i\left[ U_{\alpha }\left( \mathbf{X}_{\alpha }\right) \mathbf{%
\nabla }_{\mathbf{P}_{\alpha }}U_{\alpha }^{+}\left( \mathbf{X}_{\alpha
}\right) \right] $ and $\mathcal{A}_{\alpha }^{\mathbf{P}}=-i\left[
U_{\alpha }\left( \mathbf{X}_{\alpha }\right) \mathbf{\nabla }_{\mathbf{R}%
_{\alpha }}U_{\alpha }^{+}\left( \mathbf{X}_{\alpha }\right) \right] $.as
well as $\mathcal{A}_{\alpha }^{\mathbf{X}}=\left( \mathcal{A}_{\alpha }^{%
\mathbf{R}},\mathcal{A}_{\alpha }^{\mathbf{P}}\right) $ and $\mathcal{A}%
_{\alpha }^{\mathbf{X}}=\mathcal{A}_{\hbar }^{\mathbf{X}}$. We will also
denote $\mathcal{A}_{0}^{\mathbf{R}_{\alpha }}$, and $\mathcal{A}_{0}^{%
\mathbf{P}_{\alpha }}$ the zeroth \ order Berry connections evaluated at $%
\mathbf{X}_{\alpha }$, that is $\mathcal{A}_{0}^{\mathbf{R}_{\alpha }}\equiv 
\mathcal{A}_{0}^{\mathbf{R}_{\alpha }}\left( \mathbf{X}_{\alpha }\right) =i%
\left[ U_{0}\left( \mathbf{X}_{\alpha }\right) \mathbf{\nabla }_{\mathbf{P}%
_{\alpha }}U_{0}^{+}\left( \mathbf{X}_{\alpha }\right) \right] $, $\mathcal{A%
}_{0}^{\mathbf{P}_{\alpha }}\equiv \mathcal{A}_{0}^{\mathbf{P}_{\alpha
}}\left( \mathbf{X}_{\alpha }\right) =-i\left[ U_{0}\left( \mathbf{X}%
_{\alpha }\right) \mathbf{\nabla }_{\mathbf{P}_{\alpha }}U_{0}^{+}\left( 
\mathbf{X}_{\alpha }\right) \right] $, and $\mathcal{A}_{0}^{\mathbf{X}%
_{\alpha }}=\left( \mathcal{A}_{0}^{\mathbf{R}_{\alpha }},\mathcal{A}_{0}^{%
\mathbf{P}_{\alpha }}\right) $.

\subsection{First order in $\hbar$}

At the first order we obviously get the following expression :

\begin{equation*}
\varepsilon \left( \mathbf{X}\right) =\varepsilon _{0}\left( \mathbf{X}%
\right) +\int_{0}^{\hbar }S_{\mathbf{X}_{\hbar }}\left[ O_{\alpha
}\varepsilon _{0}\left( \mathbf{X}_{\alpha }\right) \right] d\alpha
\end{equation*}
This expression requires some explanations. The action of $O_{\alpha }$ on $%
\varepsilon _{0}\left( \mathbf{X}_{\alpha }\right) $ leads to an expression $%
F(\mathbf{X}_{\alpha },\alpha )$ which depends both on $\mathbf{X}_{\alpha }$
and $\alpha $. The shift operation replaces $\mathbf{X}_{\alpha }$ by $%
\mathbf{X}_{\hbar }\equiv \mathbf{X}$ so that the integration is (trivially)
performed on the variable $\alpha $ only. To stress this point, let us first
introduce a convenient notation. Recall that at the lowest order : 
\begin{eqnarray*}
O_{\alpha }\varepsilon _{0}\left( \mathbf{X}_{\alpha }\right) &=&\mathcal{P}%
_{+}\left\{ \frac{1}{2}\left( \mathcal{A}_{\alpha }^{R_{l}}\nabla
_{R_{l}}\varepsilon _{0}\left( \mathbf{X}_{\alpha }\right) +\nabla
_{R_{l}}\varepsilon _{0}\left( \mathbf{X}_{\alpha }\right) \mathcal{A}%
_{\alpha }^{R_{l}}\left( \mathbf{X}_{\alpha }\right) +\mathcal{A}_{\alpha
}^{P_{l}}\nabla _{P_{l}}\varepsilon _{0}\left( \mathbf{X}_{\alpha }\right)
+\nabla _{P_{l}}\varepsilon _{0}\left( \mathbf{X}_{\alpha }\right) \mathcal{A%
}_{\alpha }^{P_{l}}\right) \right\} \\
&&\mathcal{P}_{+}\left\{ \frac{i}{4}\left[ \varepsilon _{0}\left( \mathbf{X}%
_{\alpha }\right) ,\mathcal{A}_{\alpha }^{R_{l}}\right] \mathcal{A}_{\alpha
}^{P_{l}}-\frac{i}{4}\left[ \varepsilon _{0}\left( \mathbf{X}_{\alpha
}\right) ,\mathcal{A}_{\alpha }^{P_{l}}\right] \mathcal{A}_{\alpha
}^{R_{l}}+H.C.\right\}
\end{eqnarray*}
where the Berry phases can be replaced by their value at zeroth order
(evaluated at $\mathbf{X}_{\alpha }$), that is : 
\begin{eqnarray*}
O_{\alpha }\varepsilon _{0}\left( \mathbf{X}_{\alpha }\right) &=&\mathcal{P}%
_{+}\left\{ \frac{1}{2}\left( \mathcal{A}_{0}^{R_{\alpha l}}\nabla _{R\alpha
_{l}}\varepsilon _{0}\left( \mathbf{X}_{\alpha }\right) +\nabla _{R_{\alpha
l}}\varepsilon _{0}\left( \mathbf{X}_{\alpha }\right) \mathcal{A}%
_{0}^{R_{\alpha l}}+\mathcal{A}_{0}^{P_{\alpha l}}\nabla _{P_{l}}\varepsilon
_{0}\left( \mathbf{X}_{\alpha }\right) +\nabla _{P_{l}}\varepsilon
_{0}\left( \mathbf{X}_{\alpha }\right) \mathcal{A}_{0}^{P_{\alpha l}}\right)
\right\} \\
&&\mathcal{P}_{+}\left\{ \frac{i}{4}\left[ \varepsilon _{0}\left( \mathbf{X}%
_{\alpha }\right) ,\mathcal{A}_{0}^{R_{\alpha l}}\right] \mathcal{A}%
_{0}^{P_{\alpha l}}-\frac{i}{4}\left[ \varepsilon _{0}\left( \mathbf{X}%
_{\alpha }\right) ,\mathcal{A}_{0}^{P_{\alpha l}}\right] \mathcal{A}%
_{0}^{R_{\alpha l}}+H.C.\right\}
\end{eqnarray*}
since $\left[ U_{\alpha }\left( \left( \frac{\partial }{\partial \alpha }%
+\left\langle .\right\rangle \right) H\left( \mathbf{X}_{\alpha }\right)
\right) U_{\alpha }^{+}\right] =0$, at that order.

Now, let define the "covariant derivative" operator :$\mathcal{D}_{\mathbf{X}%
_{\alpha }}=\left( \mathcal{D}_{\mathbf{R}_{\alpha }},\mathcal{D}_{\mathbf{P}%
_{\alpha }}\right) $%
\begin{eqnarray*}
\mathcal{D}_{\mathbf{R}_{\alpha }} &=&\left[ \nabla _{\mathbf{R}_{\alpha }}+%
\frac{i}{2}\mathcal{A}_{0}^{\mathbf{P}_{\alpha }},.\right] \\
\mathcal{D}_{\mathbf{P}_{\alpha }} &=&\left[ \nabla _{\mathbf{P}_{\alpha }}-%
\frac{i}{2}\mathcal{A}_{0}^{\mathbf{R}_{\alpha }},.\right]
\end{eqnarray*}
and as usual $\mathcal{D}_{\mathbf{X}}\equiv \mathcal{D}_{\mathbf{X}_{\hbar
}}$, so that we can write directly : 
\begin{equation}
O_{\alpha }\varepsilon _{0}\left( \mathbf{X}_{\alpha }\right) =\frac{1}{2}%
\mathcal{P}_{+}\left[ \left( \mathcal{D}_{\mathbf{X}_{\alpha }}\varepsilon
_{0}\left( \mathbf{X}_{\alpha }\right) \right) \mathcal{A}_{0}^{\mathbf{X}%
_{\alpha }}+H.C.\right]  \notag
\end{equation}

Therefore, at this order we have 
\begin{align}
\varepsilon \left( \mathbf{X}\right) & =\varepsilon _{0}\left( \mathbf{X}%
\right) +\int_{0}^{\hbar }S_{\mathbf{X}_{\hbar }}\left[ \frac{1}{2}\mathcal{P%
}_{+}\left[ \left( \mathcal{D}_{\mathbf{X}_{\alpha }}\varepsilon _{0}\left( 
\mathbf{X}_{\alpha }\right) \right) \mathcal{A}_{0}^{\mathbf{X}_{\alpha
}}+H.C.\right] \right] d\alpha +O(\hbar ^{2})  \notag \\
& =\varepsilon _{0}\left( \mathbf{X}\right) +\frac{1}{2}\int_{0}^{\hbar }%
\mathcal{P}_{+}\left[ \left( \mathcal{D}_{\mathbf{X}}\varepsilon _{0}\left( 
\mathbf{X}\right) \right) \mathcal{A}_{0}^{\mathbf{X}}+H.C.\right] d\alpha
+O(\hbar ^{2})  \notag \\
& =\varepsilon _{0}\left( \mathbf{X}\right) +\frac{1}{2}\mathcal{P}_{+}\left[
\left( \mathcal{D}_{\mathbf{X}}\varepsilon _{0}\left( \mathbf{X}\right)
\right) \mathcal{A}_{0}^{\mathbf{X}}+H.C.\right] \int_{0}^{\hbar }d\alpha
+O(\hbar ^{2})  \notag \\
& =\varepsilon _{0}\left( \mathbf{X}\right) +\frac{\hbar }{2}\mathcal{P}_{+}%
\left[ \left( \mathcal{D}_{\mathbf{X}}\varepsilon _{0}\left( \mathbf{X}%
\right) \right) \mathcal{A}_{0}^{\mathbf{X}}+H.C.\right] +O(\hbar ^{2})
\end{align}
In the three last lines, we have introduced a slight abuse of notation.
Consistently with our conventions, we should have written $\mathcal{A}_{0}^{%
\mathbf{X}_{\hbar }}$ rather than $\mathcal{A}_{0}^{\mathbf{X}}$. However,
since in the final expression, all terms have to be expressed at $\mathbf{X}%
_{\hbar }=\mathbf{X}$ and not at $\mathbf{X}_{0}$, no confusion should arise
and $\mathcal{A}_{0}^{\mathbf{X}}$ will, from now, always stand for $%
\mathcal{A}_{0}^{\mathbf{X}_{\hbar }}$ in the final results. Note that, as
stated before, the integral over $\alpha $ in the previous computations is
trivial once the shift of variable has been performed.

The previous formula can be expanded in several ways. one has, 
\begin{align}
\varepsilon \left( \mathbf{X}\right) & =\varepsilon _{0}\left( \mathbf{X}%
\right) +\mathcal{P}_{+}\left\{ \frac{\hbar }{2}\left( \mathcal{A}%
_{0}^{X_{l}}\nabla _{X_{l}}\varepsilon _{0}\left( \mathbf{X}\right) +\nabla
_{X_{l}}\varepsilon _{0}\left( \mathbf{X}\right) \mathcal{A}%
_{0}^{X_{l}}\right) \right\}  \notag \\
& +\frac{i\hbar }{2}\mathcal{P}_{+}\left\{ \left[ \varepsilon _{0}\left( 
\mathbf{X}\right) ,\mathcal{A}_{0}^{R_{l}}\right] \mathcal{A}_{0}^{P_{l}}-%
\left[ \varepsilon _{0}\left( \mathbf{X}\right) ,\mathcal{A}_{0}^{P_{l}}%
\right] \mathcal{A}_{0}^{R_{l}}\right\} +O(\hbar ^{2})
\end{align}
or, in terms of covariant variables: 
\begin{equation*}
\varepsilon \left( \mathbf{X}\right) =\varepsilon _{0}\left( \mathbf{x}%
\right) +\frac{i\hbar }{2}\mathcal{P}_{+}\left\{ \left[ \varepsilon
_{0}\left( \mathbf{X}\right) ,\mathcal{A}_{0}^{R_{l}}\right] \mathcal{A}%
_{0}^{P_{l}}-\left[ \varepsilon _{0}\left( \mathbf{X}\right) ,\mathcal{A}%
_{0}^{P_{l}}\right] \mathcal{A}_{0}^{R_{l}}\right\} +O(\hbar ^{2})
\end{equation*}
One then recover the formula first derived in ref. \cite{SERIESPIERRE} where 
$x=\left( \mathbf{r},\mathbf{p}\right) $ are given by the expression Eq. $%
\left( \ref{rp}\right) $. As previously mentioned, we only need to know $%
U_{0}\left( \mathbf{X}\right) $ and thus the zeroth order Berry phases to
get the energy expansion at the first order.

We end up this paragraph by providing the first order expansion for $U$ that
will be used for the second order diagonalization: 
\begin{align}
U\left( \mathbf{X}\right) & =U_{0}\left( \mathbf{X}\right) +\int_{0}^{\hbar
}S_{\mathbf{X}_{\hbar }}\left[ N_{\alpha }.U_{0}\left( \mathbf{X}_{\alpha
}\right) \right] d\alpha \equiv U_{0}\left( \mathbf{X}\right) +\hbar
U_{1}\left( \mathbf{X}\right) U_{0}\left( \mathbf{X}\right)  \notag \\
& =\left( 1-\hbar \left[ .,\varepsilon _{0}\right] ^{-1}.\left[ \mathcal{P}%
_{-}\left\{ \frac{1}{2}\left( \mathcal{A}_{0}^{R_{l}}\nabla
_{R_{l}}\varepsilon _{0}\left( \mathbf{X}\right) +\nabla _{R_{l}}\varepsilon
_{0}\left( \mathbf{X}\right) \mathcal{A}_{0}^{R_{l}}+\mathcal{A}%
_{0}^{P_{l}}\nabla _{P_{l}}\varepsilon _{0}\left( \mathbf{X}\right) +\nabla
_{P_{l}}\varepsilon _{0}\left( \mathbf{X}\right) \mathcal{A}%
_{0}^{P_{l}}\right) \right\} \right. \right.  \notag \\
& \left. \left. -\frac{i}{2}\left\{ \left[ \varepsilon _{0}\left( \mathbf{X}%
\right) ,\mathcal{A}_{0}^{R_{l}}\right] \mathcal{A}_{0}^{P_{l}}-\left[
\varepsilon _{0}\left( \mathbf{X}\right) ,\mathcal{A}_{0}^{P_{l}}\right] 
\mathcal{A}_{0}^{R_{l}}\right\} \right] -\frac{i}{4}\left[ \mathcal{A}%
_{0}^{R_{l}},\mathcal{A}_{0}^{P_{l}}\right] \right) U_{0}\left( \mathbf{X}%
\right)  \label{UU1}
\end{align}

\subsection{Second order}

Given our choice of symmetrization $\left\langle \varepsilon _{0}\left( 
\mathbf{X}_{\alpha }\right) \right\rangle _{\alpha }$ is of order $1$ in $%
\alpha $, so that. $\left\langle .\right\rangle _{0\rightarrow \hbar
}^{S}.\varepsilon _{0}\left( \mathbf{X}\right) =\int_{0}^{\hbar }S_{\mathbf{X%
}_{\hbar }}\left\langle \varepsilon _{0}\left( \mathbf{X}_{\alpha }\right)
\right\rangle _{\alpha }d\alpha $ is of order $2$. As a consequence,
expanding the compact form of the energy operator to the second order, leads
to :

\begin{eqnarray*}
\varepsilon \left( \mathbf{X}\right) &=&\varepsilon _{0}\left( \mathbf{X}%
\right) +\int_{0}^{\hbar }S_{\mathbf{X}_{\hbar }}\left[ O_{\alpha
}\varepsilon _{0}\left( \mathbf{X}_{\alpha }\right) \right] d\alpha
+\int_{0}^{\hbar }\int_{0}^{\alpha _{1}}S_{\mathbf{X}_{\hbar }}\left[
O_{\alpha _{1}}S_{\mathbf{X}_{\alpha _{1}}}\left[ O_{\alpha _{2}}\varepsilon
_{0}\left( \mathbf{X}_{\alpha _{2}}\right) \right] \right] d\alpha
_{2}d\alpha _{1} \\
&&-\int_{0}^{\hbar }S_{\mathbf{X}_{\hbar }}\left\langle \varepsilon
_{0}\left( \mathbf{X}_{\alpha }\right) \right\rangle _{\alpha }d\alpha
\end{eqnarray*}
The last contribution can be computed easily. At the lowest order in $\alpha 
$, $S_{\mathbf{X}_{\hbar }}\left\langle \varepsilon _{0}\left( \mathbf{X}%
_{\alpha }\right) \right\rangle _{\alpha }=\frac{\alpha }{\hbar }%
\left\langle \varepsilon _{0}\left( \mathbf{X}\right) \right\rangle _{\hbar
} $ (actually, $S_{\mathbf{X}_{\hbar }}\left\langle \varepsilon _{0}\left( 
\mathbf{X}_{\alpha }\right) \right\rangle _{\alpha }$ is a function of $%
\mathbf{X}$ times $\alpha $ and $\left\langle \varepsilon _{0}\left( \mathbf{%
X}\right) \right\rangle _{\hbar }$ is the same function times $\hbar $). As
a consequence, $\int_{0}^{\hbar }S_{\mathbf{X}_{\hbar }}\left\langle
\varepsilon _{0}\left( \mathbf{X}_{\alpha }\right) \right\rangle _{\alpha
}d\alpha =\int_{0}^{\hbar }\frac{\alpha }{\hbar }\left\langle \varepsilon
_{0}\left( \mathbf{X}\right) \right\rangle _{\hbar }d\alpha =\frac{\hbar }{2}%
\left\langle \varepsilon _{0}\left( \mathbf{X}\right) \right\rangle $ (with
the convention $\left\langle \varepsilon _{0}\left( \mathbf{X}\right)
\right\rangle =\left\langle \varepsilon _{0}\left( \mathbf{X}\right)
\right\rangle _{\hbar }$).

The first contribution $\int_{0}^{\hbar }O_{\alpha }d\alpha \varepsilon
_{0}\left( \mathbf{X}_{\alpha }\right) $ can be expanded as before as: 
\begin{align*}
& \int_{0}^{\hbar }S_{\mathbf{X}_{\hbar }}\left[ O_{\alpha }\varepsilon
_{0}\left( \mathbf{X}_{\alpha }\right) \right] d\alpha \\
& =\int_{0}^{\hbar }S_{\mathbf{X}_{\hbar }}\mathcal{P}_{+}\left\{ \frac{1}{2}%
\left( \mathcal{A}_{\alpha }^{R_{l}}\nabla _{R_{l}}\varepsilon _{0}\left( 
\mathbf{X}_{\alpha }\right) +\nabla _{R_{l}}\varepsilon _{0}\left( \mathbf{X}%
_{\alpha }\right) \mathcal{A}_{\alpha }^{R_{l}}+\mathcal{A}_{\alpha
}^{P_{l}}\nabla _{P_{l}}\varepsilon _{0}\left( \mathbf{X}_{\alpha }\right)
+\nabla _{P_{l}}\varepsilon _{0}\left( \mathbf{X}_{\alpha }\right) \mathcal{A%
}_{\alpha }^{P_{l}}\right) \right\} d\alpha \\
& +\int_{0}^{\hbar }S_{\mathbf{X}_{\hbar }}\mathcal{P}_{+}\left\{ \frac{i}{4}%
\left[ \varepsilon _{0}\left( \mathbf{X}_{\alpha }\right) ,\mathcal{A}%
_{\alpha }^{R_{l}}\right] \mathcal{A}_{\alpha }^{P_{l}}-\frac{i}{4}\left[
\varepsilon _{0}\left( \mathbf{X}_{\alpha }\right) ,\mathcal{A}_{\alpha
}^{P_{l}}\right] \mathcal{A}_{\alpha }^{R_{l}}+H.C.\right. \\
& \left. +\left[ U_{\alpha }\left( \left( \frac{\partial }{\partial \alpha }%
+\left\langle .\right\rangle \right) H\left( \mathbf{X}_{\alpha }\right)
\right) U_{\alpha }^{+}\right] \right\} d\alpha
\end{align*}
but now, the Berry connections have to be expanded to the first order in $%
\alpha $. Note that in the second order contribution, due to the double
integral, 
\begin{equation*}
\int_{0}^{\hbar }\int_{0}^{\alpha _{1}}S_{\mathbf{X}_{\hbar }}\left[
O_{\alpha _{1}}S_{\mathbf{X}_{\alpha _{1}}}\left[ O_{\alpha _{2}}\varepsilon
_{0}\left( \mathbf{X}_{\alpha _{2}}\right) \right] \right] d\alpha
_{2}d\alpha _{1}-\frac{\hbar }{2}\left\langle \varepsilon _{0}\left( \mathbf{%
X}\right) \right\rangle
\end{equation*}
the Berry connections need only to be expanded to the zeroth order in $%
\alpha $.

To go further, we thus need to give first some expanded formula at the first
order in $\alpha $ for the Berry phases $\mathcal{A}_{\alpha }^{\mathbf{X}}$
intervening in the definition of the diagonalized Hamiltonian and the
dynamical variables.

\subsubsection{Berry connections at the first order}

To obtain $\mathcal{A}_{\alpha }^{\mathbf{X}}$ we only need the matrix $U$
at the first order as computed before : $U\left( \mathbf{X}\right) \equiv
U_{\hbar }\left( \mathbf{X}\right) =U_{0}\left( \mathbf{X}\right) +\hbar
U_{1}\left( \mathbf{X}\right) U_{0}\left( \mathbf{X}\right) $ with $%
U_{1}\left( \mathbf{X}\right) $ given by (Eq. $\left( \ref{UU1}\right) $)

\begin{align}
U_{1}\left( \mathbf{X}\right) & =-\left[ .,\varepsilon _{0}\right] ^{-1}.%
\left[ \mathcal{P}_{-}\left\{ \frac{1}{2}\left( \mathcal{A}%
_{0}^{R_{l}}\nabla _{R_{l}}\varepsilon _{0}\left( \mathbf{X}\right) +\nabla
_{R_{l}}\varepsilon _{0}\left( \mathbf{X}\right) \mathcal{A}_{0}^{R_{l}}+%
\mathcal{A}_{0}^{P_{l}}\nabla _{P_{l}}\varepsilon _{0}\left( \mathbf{X}%
\right) +\nabla _{P_{l}}\varepsilon _{0}\left( \mathbf{X}\right) \mathcal{A}%
_{0}^{P_{l}}\right) \right\} \right.  \notag \\
& \left. +\frac{i}{2}\left\{ \left[ \varepsilon _{0}\left( \mathbf{X}\right)
,\mathcal{A}_{0}^{R_{l}}\right] \mathcal{A}_{0}^{P_{l}}-\left[ \varepsilon
_{0}\left( \mathbf{X}\right) ,\mathcal{A}_{0}^{P_{l}}\right] \mathcal{A}%
_{0}^{R_{l}}\right\} \right] -\frac{i}{4}\left[ \mathcal{A}_{0}^{R_{l}},%
\mathcal{A}_{0}^{P_{l}}\right]
\end{align}
At the same order the (non-diagonal) Berry connections $\mathcal{A}_{\alpha
}^{\mathbf{X}}=\left( \mathcal{A}_{\alpha }^{\mathbf{R}},\mathcal{A}_{\alpha
}^{\mathbf{P}}\right) $ are again given by 
\begin{align*}
\mathcal{A}_{\alpha }^{\mathbf{R}}\left( \mathbf{X}_{\alpha }\right) & =i%
\left[ U_{\alpha }\left( \mathbf{X}_{\alpha }\right) \mathbf{\nabla }_{%
\mathbf{P}}U_{\alpha }^{+}\left( \mathbf{X}_{\alpha }\right) \right] =\frac{1%
}{\alpha }U_{\alpha }\left( \mathbf{X}_{\alpha }\right) \mathbf{R}_{\alpha
}U_{\alpha }^{+}\left( \mathbf{X}_{\alpha }\right) \\
\text{and }\mathcal{A}_{\alpha }^{\mathbf{P}}\left( \mathbf{X}_{\alpha
}\right) & =-i\left[ U_{\alpha }\left( \mathbf{X}_{\alpha }\right) \mathbf{%
\nabla }_{\mathbf{R}}U_{\alpha }^{+}\left( \mathbf{X}_{\alpha }\right) %
\right] =\frac{1}{\alpha }U_{\alpha }\left( \mathbf{X}_{\alpha }\right) 
\mathbf{P}_{\alpha }U_{\alpha }^{+}\left( \mathbf{X}_{\alpha }\right)
\end{align*}
where now $U_{\alpha }\left( \mathbf{X}_{\alpha }\right) $ is the
transformation to the first order in $\alpha $ i.e. $U_{0}\left( \mathbf{X}%
_{\alpha }\right) +\alpha U_{1}\left( \mathbf{X}_{\alpha }\right) ,$ in
which $\mathbf{X}$ is replaced by the running operator $\mathbf{X}_{\alpha }$%
. Using the hermiticity of $\mathcal{A}_{\alpha }^{\mathbf{X}}$, so that one
has $\mathcal{A}_{\alpha }^{\mathbf{X}}=\left( \mathcal{A}_{\alpha }^{%
\mathbf{X}}+\left( \mathcal{A}_{\alpha }^{\mathbf{X}}\right) ^{+}\right) /2$
we can expand $\mathcal{A}_{\alpha }^{\mathbf{X}}$ as : 
\begin{equation*}
\mathcal{A}_{\alpha }^{\mathbf{X}}=\left( \frac{1}{2}\left( \left[ 1+\alpha
U_{1}\left( \mathbf{X}_{\alpha }\right) \right] U_{0}\left( \mathbf{X}%
_{\alpha }\right) \right) \frac{\mathbf{X}_{\alpha }}{\alpha }\left(
U_{0}^{+}\left( \mathbf{X}_{\alpha }\right) \left[ 1+\alpha U_{1}^{+}\left( 
\mathbf{X}_{\alpha }\right) \right] \right) +\frac{1}{2}H.C.\right) -\frac{%
\mathbf{X}_{\alpha }}{\alpha }
\end{equation*}
(the $\frac{1}{\alpha }$ factor reminds that in our definition of $\mathcal{A%
}_{\alpha }^{\mathbf{X}}$ the gradient with respect to $\mathbf{X}_{\alpha }$
is normalized, i.e. divided by $\alpha $). After some recombinations, the
previous expression can be written in a more convenient form : 
\begin{align*}
\mathcal{A}_{\alpha }^{\mathbf{X}}& =\frac{1}{2\alpha }U_{0}\left( \mathbf{X}%
_{\alpha }\right) \left[ \mathbf{X}_{\alpha }\mathbf{,}U_{0}^{+}\left( 
\mathbf{X}_{\alpha }\right) \right] +H.C. \\
& +\frac{1}{2}\left[ U_{1}\left( \mathbf{X}_{\alpha }\right) \left[
U_{0}\left( \mathbf{X}_{\alpha }\right) \mathbf{X}_{\alpha }U_{0}^{+}\left( 
\mathbf{X}_{\alpha }\right) \right] +\left[ U_{0}\left( \mathbf{X}_{\alpha
}\right) \mathbf{X}_{\alpha }U_{0}^{+}\left( \mathbf{X}_{\alpha }\right) %
\right] U_{1}^{+}\left( \mathbf{X}_{\alpha }\right) \right] +H.C.-\mathbf{X}%
_{\alpha }
\end{align*}
using now the fact that at the lowest order $U_{0}\left( \mathbf{X}_{\alpha
}\right) \left[ \frac{\mathbf{X}_{\alpha }}{\alpha },U_{0}^{+}\left( \mathbf{%
X}_{\alpha }\right) \right] =\mathcal{A}_{0}^{\mathbf{X}}\left( \mathbf{X}%
_{\alpha }\right) $, one has : 
\begin{align*}
\mathcal{A}_{\alpha }^{\mathbf{X}}& =\frac{1}{2\alpha }\left[ U_{0}\left( 
\mathbf{X}_{\alpha }\right) \left[ \mathbf{X}_{\alpha }\mathbf{,}%
U_{0}^{+}\left( \mathbf{X}_{\alpha }\right) \right] +H.C.\right] \\
& +\frac{1}{2}\left[ \alpha U_{1}\left( \mathbf{X}_{\alpha }\right) \mathcal{%
A}_{0}^{\mathbf{X}}\left( \mathbf{X}_{\alpha }\right) +\alpha \mathcal{A}%
_{0}^{\mathbf{X}}\left( \mathbf{X}_{\alpha }\right) U_{1}^{+}\left( \mathbf{X%
}_{\alpha }\right) +\frac{1}{2}\left[ \mathbf{X}_{\alpha },U_{1}^{+}\left( 
\mathbf{X}_{\alpha }\right) \right] +H.C.\right]
\end{align*}
decomposing $U_{1}^{+}\left( \mathbf{X}_{\alpha }\right) $ into Hermitian
and anti-Hermitian part we are thus led to : 
\begin{align}
\mathcal{A}_{\alpha }^{\mathbf{X}}& =\frac{1}{2\alpha }\left( U_{0}\left( 
\mathbf{X}_{\alpha }\right) \left[ \mathbf{X}_{\alpha }\mathbf{,}%
U_{0}^{+}\left( \mathbf{X}_{\alpha }\right) \right] +H.C.\right) +\left[ 
\mathbf{X}_{\alpha }\mathbf{+}\alpha \mathcal{A}_{0}^{\mathbf{X}}\mathbf{,}%
\mathit{ahr}\left( U_{1}^{+}\left( \mathbf{X}_{\alpha }\right) \right) %
\right]  \notag \\
& +\frac{\alpha }{2}\left( \mathcal{A}_{0}^{\mathbf{X}}\mathit{hr}\left(
U_{1}^{+}\left( \mathbf{X}_{\alpha }\right) \right) +\mathit{hr}\left(
U_{1}^{+}\left( \mathbf{X}_{\alpha }\right) \right) \mathcal{A}_{0}^{\mathbf{%
X}}\right)  \label{interm}
\end{align}
where $\mathit{ahr}\left( Z\right) $ and $\mathit{hr}\left( Z\right) $
denote the anti-Hermitian and Hermitian part of an operator $Z$
respectively. Now using Eq. $\left( \ref{U1}\right) $, we are led to the
following expressions :

\begin{equation*}
\left[ \mathbf{X}_{\alpha }\mathbf{+}\alpha \mathcal{A}_{0}^{\mathbf{X}}%
\mathbf{,}\mathit{ahr}\left( U_{1}^{+}\left( \mathbf{X}_{\alpha }\right)
\right) \right] =\left[ B_{\alpha },\mathbf{X}_{\alpha }+\alpha \mathcal{A}%
_{0}^{\mathbf{X}_{\alpha }}\right]
\end{equation*}
and 
\begin{eqnarray*}
&&\frac{\alpha }{2}\left( \mathcal{A}_{0}^{\mathbf{X}}\mathit{hr}\left(
U_{1}^{+}\left( \mathbf{X}_{\alpha }\right) \right) +\mathit{hr}\left(
U_{1}^{+}\left( \mathbf{X}_{\alpha }\right) \right) \mathcal{A}_{0}^{\mathbf{%
X}}\right) +H.C. \\
&=&\alpha \left( \mathcal{A}_{0}^{\mathbf{X}}\mathit{hr}\left(
U_{1}^{+}\left( \mathbf{X}_{\alpha }\right) \right) +\mathit{hr}\left(
U_{1}^{+}\left( \mathbf{X}_{\alpha }\right) \right) \mathcal{A}_{0}^{\mathbf{%
X}}\right) \\
&=&-\frac{i}{4}\mathcal{A}_{0}^{\mathbf{X}}\left[ \mathcal{A}_{0}^{R_{l}},%
\mathcal{A}_{0}^{P_{l}}\right] +H.C.
\end{eqnarray*}
with $B_{\alpha }$ given by the following relation 
\begin{align}
B_{\alpha }& =-\left[ .,\varepsilon _{0}\left( \mathbf{X}_{\alpha }\right) %
\right] ^{-1}.\left( \mathcal{P}_{-}\left\{ \frac{1}{2}\mathcal{A}%
_{0}^{R\alpha _{l}}\nabla _{R_{\alpha l}}\varepsilon _{0}\left( \mathbf{X}%
_{\alpha }\right) +\frac{1}{2}\mathcal{A}_{0}^{P_{\alpha l}}\nabla
_{P_{\alpha l}}\varepsilon _{0}\left( \mathbf{X}_{\alpha }\right)
+H.C.\right\} \right.  \notag \\
& \left. -\frac{i}{4}\left\{ \left[ \varepsilon _{0}\left( \mathbf{X}%
_{\alpha }\right) ,\mathcal{A}_{0}^{R_{\alpha l}}\right] \mathcal{A}%
_{0}^{P_{\alpha l}}-\left[ \varepsilon _{0}\left( \mathbf{X}_{\alpha
}\right) ,\mathcal{A}_{0}^{P_{\alpha l}}\right] \mathcal{A}_{0}^{R_{\alpha
l}}+H.C.\right\} \right)  \notag \\
& =-\left[ .,\varepsilon _{0}\left( \mathbf{X}_{\alpha }\right) \right]
^{-1}.\left( \mathcal{P}_{-}\left\{ \frac{1}{2}\mathcal{A}_{0}^{R_{\alpha
l}}\nabla _{R_{\alpha l}}\varepsilon _{0}\left( \mathbf{X}_{\alpha }\right) +%
\frac{1}{2}\mathcal{A}_{0}^{P_{\alpha l}}\nabla _{P_{\alpha l}}\varepsilon
_{0}\left( \mathbf{X}_{\alpha }\right) +H.C.\right\} \right)  \notag \\
& +\frac{i}{4}\left\{ \mathcal{P}_{-}\mathcal{A}_{0}^{R_{\alpha l}}\mathcal{P%
}_{+}\mathcal{A}_{0}^{P_{\alpha l}}-\mathcal{P}_{-}\mathcal{A}%
_{0}^{P_{\alpha l}}\mathcal{P}_{+}\mathcal{A}_{0}^{R_{\alpha l}}+H.C.\right\}
\label{Bform}
\end{align}
Note that given our previous notations, we can write: 
\begin{equation*}
B_{\alpha }=-\left[ .,\varepsilon _{0}\left( \mathbf{X}_{\alpha }\right) %
\right] ^{-1}.\left\{ \left( \mathcal{D}_{\mathbf{X}_{\alpha }}\varepsilon
_{0}\left( \mathbf{X}_{\alpha }\right) \right) \mathcal{A}_{0}^{\mathbf{X}%
_{\alpha }}+H.C.\right\}
\end{equation*}

The two last expressions will allow to rewrite $\mathcal{A}_{\alpha }^{%
\mathbf{X}}$: To do so, we need to express carefully $\frac{1}{2\alpha }%
\left[ U_{0}\left( \mathbf{X}_{\alpha }\right) \left[ \mathbf{X}_{\alpha }%
\mathbf{,}U_{0}^{+}\left( \mathbf{X}_{\alpha }\right) \right] +H.C.\right] $.

Actually, given our choice of symmetrization, at the second order this term
is not equal to $\mathcal{A}_{0}^{\mathbf{X}}$ but also includes some second
order corrections to symmetrize properly the product $U_{0}\left( \mathbf{X}%
_{\alpha }\right) \times \left[ \mathbf{X}_{\alpha }\mathbf{,}%
U_{0}^{+}\left( \mathbf{X}_{\alpha }\right) \right] $ (see \cite{SEMIDIAG}
for this fact). Moving powers of $\mathbf{P}_{\alpha }$ and $\mathbf{R}%
_{\alpha }$ on the left or on the right yields: 
\begin{eqnarray*}
&&\frac{1}{2\alpha }\left( U_{0}\left( \mathbf{X}_{\alpha }\right) \left[ 
\mathbf{X}_{\alpha }\mathbf{,}U_{0}^{+}\left( \mathbf{X}_{\alpha }\right) %
\right] +H.C.\right) \\
&=&\mathcal{A}_{0}^{\mathbf{X}}+\frac{i}{4\alpha }\nabla _{R_{\alpha
l}}U_{0}\left( \mathbf{X}_{\alpha }\right) \nabla _{P_{\alpha l}}\left( %
\left[ \mathbf{X}_{\alpha }\mathbf{,}U_{0}^{+}\left( \mathbf{X}_{\alpha
}\right) \right] \right) -\frac{i}{4\alpha }\nabla _{P_{\alpha
l}}U_{0}\left( \mathbf{X}_{\alpha }\right) \nabla _{R_{\alpha l}}\left( %
\left[ \mathbf{X}_{\alpha }\mathbf{,}U_{0}^{+}\left( \mathbf{X}_{\alpha
}\right) \right] \right) +H.C. \\
&=&\mathcal{A}_{0}^{\mathbf{X}}+\frac{1}{4\alpha }\mathcal{A}_{0}^{P_{\alpha
l}}U_{0}\left( \mathbf{X}_{\alpha }\right) \nabla _{P_{\alpha l}}\left( %
\left[ \mathbf{X}_{\alpha }\mathbf{,}U_{0}^{+}\left( \mathbf{X}_{\alpha
}\right) \right] \right) +\frac{1}{4\alpha }\mathcal{A}_{0}^{R_{\alpha
l}}U_{0}\left( \mathbf{X}_{\alpha }\right) \nabla _{R_{\alpha l}}\left( %
\left[ \mathbf{X}_{\alpha }\mathbf{,}U_{0}^{+}\left( \mathbf{X}_{\alpha
}\right) \right] \right) \\
&=&\mathcal{A}_{0}^{\mathbf{X}}+\frac{\alpha }{4}\left( \mathcal{A}%
_{0}^{R_{\alpha l}}U_{0}\left( \mathbf{X}_{\alpha }\right) \nabla
_{R_{\alpha l}}\mathcal{A}_{0}^{\mathbf{X}}+\mathcal{A}_{0}^{P_{\alpha
l}}U_{0}\left( \mathbf{X}_{\alpha }\right) \nabla _{P_{\alpha l}}\mathcal{A}%
_{0}^{\mathbf{X}_{\alpha }}+H.C.\right) \\
&&-\frac{1}{4\alpha }\mathcal{A}_{0}^{P_{\alpha l}}\left( \nabla _{P_{\alpha
l}}U_{0}\left( \mathbf{X}_{\alpha }\right) \right) \left( \left[ \mathbf{X}%
_{\alpha }\mathbf{,}U_{0}^{+}\left( \mathbf{X}_{\alpha }\right) \right]
\right) -\frac{1}{4\alpha }\mathcal{A}_{0}^{R_{\alpha l}}\left( \nabla
_{R_{\alpha l}}U_{0}\left( \mathbf{X}_{\alpha }\right) \right) \left( \left[ 
\mathbf{X}_{\alpha }\mathbf{,}U_{0}^{+}\left( \mathbf{X}_{\alpha }\right) %
\right] \right) +H.C. \\
&=&\mathcal{A}_{0}^{\mathbf{X}}+\frac{\alpha }{4}\left( \mathcal{A}%
_{0}^{R_{\alpha l}}\nabla _{R_{\alpha l}}\mathcal{A}_{0}^{\mathbf{X}}+%
\mathcal{A}_{0}^{P_{\alpha l}}\nabla _{P_{\alpha l}}\mathcal{A}_{0}^{\mathbf{%
X}_{\alpha }}+H.C.\right) \\
&&-\frac{i}{4}\mathcal{A}_{0}^{P_{\alpha l}}\mathcal{A}_{0}^{R_{\alpha l}}%
\mathcal{A}_{0}^{\mathbf{X}}+\frac{i}{4}\mathcal{A}_{0}^{R_{\alpha l}}%
\mathcal{A}_{0}^{P_{\alpha l}}\mathcal{A}_{0}^{\mathbf{X}}+H.C. \\
&=&\mathcal{A}_{0}^{\mathbf{X}}+\frac{\alpha }{4}\left( \mathcal{A}%
_{0}^{R_{\alpha l}}\nabla _{R_{\alpha l}}\mathcal{A}_{0}^{\mathbf{X}}+%
\mathcal{A}_{0}^{P_{\alpha l}}\nabla _{P_{\alpha l}}\mathcal{A}_{0}^{\mathbf{%
X}_{\alpha }}+H.C.\right) +\left( \frac{i}{4}\mathcal{A}_{0}^{\mathbf{X}}%
\left[ \mathcal{A}_{0}^{R_{l}},\mathcal{A}_{0}^{P_{l}}\right] +H.C.\right)
\end{eqnarray*}
And as consequence, we can regroup the various terms of Eq. $\left( \ref%
{interm}\right) $, to obtain ultimately : 
\begin{equation*}
\mathcal{A}_{\alpha }^{\mathbf{X}}=\mathcal{A}_{0}^{\mathbf{X}_{\alpha
}}\left( \mathbf{R}_{\alpha }\mathbf{+}\frac{\alpha }{2}\mathcal{A}%
_{0}^{R_{\alpha }},\mathbf{P}_{\alpha }\mathbf{+}\frac{\alpha }{2}\mathcal{A}%
_{0}^{P_{\alpha }}\right) +\left[ B_{\alpha },\mathbf{X}_{\alpha }+\alpha 
\mathcal{A}_{0}^{\mathbf{X}_{\alpha }}\right]
\end{equation*}
where we introduced the notations 
\begin{equation*}
\mathcal{A}_{0}^{\mathbf{X}_{\alpha }}\left( \mathbf{R}_{\alpha }\mathbf{+}%
\frac{\alpha }{2}\mathcal{A}_{0}^{R_{\alpha }},\mathbf{P}_{\alpha }\mathbf{+}%
\frac{\alpha }{2}\mathcal{A}_{0}^{P_{\alpha }}\right) \equiv \mathcal{A}%
_{0}^{\mathbf{X}_{\alpha }}+\frac{\alpha }{4}\left\{ \mathcal{A}%
_{0}^{R_{\alpha l}}\nabla _{R_{\alpha l}}\mathcal{A}_{0}^{\mathbf{X}_{\alpha
}}+\mathcal{A}_{0}^{P_{\alpha l}}\nabla _{P_{\alpha l}}\mathcal{A}_{0}^{%
\mathbf{X}_{\alpha }}+H.C.\right\}
\end{equation*}

\subsubsection{Hamiltonian diagonalization at the second order}

We can now turn to the expression of the diagonalized Hamiltonian. Focusing
first on the double integral term : 
\begin{eqnarray*}
&&\int_{0}^{\hbar }\int_{0}^{\alpha _{1}}S_{\mathbf{X}_{\hbar }}\left[
O_{\alpha _{1}}S_{\mathbf{X}_{\alpha _{1}}}\left[ O_{\alpha _{2}}\varepsilon
_{0}\left( \mathbf{X}_{\alpha _{2}}\right) \right] \right] d\alpha
_{2}d\alpha _{1}-\frac{\hbar }{2}\left\langle \varepsilon _{0}\left( \mathbf{%
X}\right) \right\rangle \\
&=&\int_{0}^{\hbar }\int_{0}^{\alpha _{1}}S_{\mathbf{X}_{\hbar }}\left[
O_{\alpha _{1}}S_{\mathbf{X}_{\alpha _{1}}}\left[ \frac{\alpha _{2}}{2}%
\mathcal{P}_{+}\left[ \left( \mathcal{D}_{\mathbf{X}_{\alpha
_{2}}}\varepsilon _{0}\left( \mathbf{X}_{\alpha _{2}}\right) \right) 
\mathcal{A}_{0}^{\mathbf{X}_{\alpha _{2}}}+H.C.\right] \right] \right]
d\alpha _{2}d\alpha _{1}-\frac{\hbar }{2}\left\langle \varepsilon _{0}\left( 
\mathbf{X}\right) \right\rangle \\
&=&\int_{0}^{\hbar }\int_{0}^{\alpha _{1}}\frac{S_{\mathbf{X}_{\hbar }}}{2}%
\mathcal{P}_{+}\left\{ \left( \mathcal{D}_{\mathbf{X}_{\alpha _{1}}}\left( 
\frac{1}{2}S_{\mathbf{X}_{\alpha _{1}}}\mathcal{P}_{+}\left[ \left( \mathcal{%
D}_{\mathbf{X}_{\alpha _{2}}}\varepsilon _{0}\left( \mathbf{X}_{\alpha
_{2}}\right) \right) \mathcal{A}_{0}^{\mathbf{X}_{\alpha _{2}}}+H.C.\right]
\right) \right) \mathcal{A}_{0}^{\mathbf{X}_{\alpha _{1}}}+H.C.\right\}
\alpha _{2}d\alpha _{2}d\alpha _{1} \\
&&-\frac{\hbar }{2}\left\langle \varepsilon _{0}\left( \mathbf{X}\right)
\right\rangle
\end{eqnarray*}
due to the action of the shift operators, $\mathbf{X}_{\alpha _{1}}$ and $%
\mathbf{X}_{\alpha _{2}}$ are shifted to $\mathbf{X}_{\hbar }=\mathbf{X}$ at
that order. As a consequence, 
\begin{eqnarray*}
&&\int_{0}^{\hbar }\int_{0}^{\alpha _{1}}S_{\mathbf{X}_{\hbar }}\left[
O_{\alpha _{1}}S_{\mathbf{X}_{\alpha _{1}}}\left[ O_{\alpha _{2}}\varepsilon
_{0}\left( \mathbf{X}_{\alpha _{2}}\right) \right] \right] d\alpha
_{2}d\alpha _{1}-\frac{\hbar }{2}\left\langle \varepsilon _{0}\left( \mathbf{%
X}\right) \right\rangle \\
&=&\int_{0}^{\hbar }\int_{0}^{\alpha _{1}}\frac{\alpha _{2}}{4}d\alpha
_{2}d\alpha _{1}\left[ \mathcal{P}_{+}\left\{ \left( \mathcal{D}_{\mathbf{X}}%
\left[ \mathcal{P}_{+}\left[ \left( \mathcal{D}_{\mathbf{X}}\varepsilon
_{0}\left( \mathbf{X}\right) \right) \mathcal{A}_{0}^{\mathbf{X}}+H.C.\right]
\right] \right) \mathcal{A}_{0}^{\mathbf{X}}\right\} +H.C.\right] -\frac{%
\hbar }{2}\left\langle \varepsilon _{0}\left( \mathbf{X}\right) \right\rangle
\\
&=&\frac{\hbar ^{2}}{8}\left[ \mathcal{P}_{+}\left\{ \left( \mathcal{D}_{%
\mathbf{X}}\left[ \mathcal{P}_{+}\left[ \left( \mathcal{D}_{\mathbf{X}%
}\varepsilon _{0}\left( \mathbf{X}\right) \right) \mathcal{A}_{0}^{\mathbf{X}%
}+H.C.\right] \right] \right) \mathcal{A}_{0}^{\mathbf{X}}\right\} +H.C.%
\right] -\frac{\hbar }{2}\left\langle \varepsilon _{0}\left( \mathbf{X}%
\right) \right\rangle
\end{eqnarray*}
We can now turn to the determination of the single integral term. For the
sake of simplicity, we consider the case $\left[ U_{\alpha }\left( \left( 
\frac{\partial }{\partial \alpha }+\left\langle .\right\rangle \right)
H\left( \mathbf{X}_{\alpha }\right) \right) U_{\alpha }^{+}\right] =0$,
which is in fact satisfied in the usual examples of interest.. 
\begin{align*}
& \int_{0}^{\hbar }S_{\mathbf{X}_{\hbar }}\left[ O_{\alpha }\varepsilon
_{0}\left( \mathbf{X}_{\alpha }\right) \right] d\alpha \\
& =\int_{0}^{\hbar }S_{\mathbf{X}_{\hbar }}\mathcal{P}_{+}\left\{ \frac{1}{2}%
\left( \mathcal{A}_{\alpha }^{R_{l}}\nabla _{R_{l}}\varepsilon _{0}\left( 
\mathbf{X}_{\alpha }\right) +\nabla _{R_{l}}\varepsilon _{0}\left( \mathbf{X}%
_{\alpha }\right) \mathcal{A}_{\alpha }^{R_{l}}+\mathcal{A}_{\alpha
}^{P_{l}}\nabla _{P_{l}}\varepsilon _{0}\left( \mathbf{X}_{\alpha }\right)
+\nabla _{P_{l}}\varepsilon _{0}\left( \mathbf{X}_{\alpha }\right) \mathcal{A%
}_{\alpha }^{P_{l}}\right) \right\} d\alpha \\
& +\int_{0}^{\hbar }S_{\mathbf{X}_{\hbar }}\mathcal{P}_{+}\left\{ \frac{i}{4}%
\left[ \varepsilon _{0}\left( \mathbf{X}_{\alpha }\right) ,\mathcal{A}%
_{\alpha }^{R_{l}}\right] \mathcal{A}_{\alpha }^{P_{l}}-\frac{i}{4}\left[
\varepsilon _{0}\left( \mathbf{X}_{\alpha }\right) ,\mathcal{A}_{\alpha
}^{P_{l}}\right] \mathcal{A}_{\alpha }^{R_{l}}+H.C.\right\} d\alpha
\end{align*}
Introduce now the correction to the covariant derivatives at this order: $%
\mathcal{\hat{D}}_{\mathbf{X}_{\alpha }}\equiv \left( \mathcal{\hat{D}}_{%
\mathbf{R}_{\alpha }},\mathcal{\hat{D}}_{\mathbf{P}_{\alpha }}\right) $ with 
\begin{eqnarray*}
\mathcal{\hat{D}}_{\mathbf{R}_{\alpha }} &=&\left[ \nabla _{\mathbf{R}%
_{\alpha }}+\frac{i}{2}\mathcal{A}_{\alpha }^{\mathbf{P}},.\right] \\
\mathcal{\hat{D}}_{\mathbf{P}_{\alpha }} &=&\left[ \nabla _{\mathbf{P}%
_{\alpha }}-\frac{i}{2}\mathcal{A}_{\alpha }^{\mathbf{R}},.\right]
\end{eqnarray*}
Then, one can write : 
\begin{equation*}
\int_{0}^{\hbar }S_{\mathbf{X}_{\hbar }}\left[ O_{\alpha }\varepsilon
_{0}\left( \mathbf{X}_{\alpha }\right) \right] d\alpha =\int_{0}^{\hbar }S_{%
\mathbf{X}_{\hbar }}\left[ \frac{1}{2}\mathcal{P}_{+}\left[ \left( \mathcal{%
\hat{D}}_{\mathbf{X}_{\alpha }}\varepsilon _{0}\left( \mathbf{X}_{\alpha
}\right) \right) \mathcal{A}_{\alpha }^{\mathbf{X}}+H.C.\right] \right]
d\alpha
\end{equation*}
which can be expanded as : 
\begin{eqnarray}
&&\int_{0}^{\hbar }S_{\mathbf{X}_{\hbar }}\left[ O_{\alpha }\varepsilon
_{0}\left( \mathbf{X}_{\alpha }\right) \right] d\alpha  \notag \\
&=&\frac{1}{2}\int_{0}^{\hbar }S_{\mathbf{X}_{\hbar }}\left[ \mathcal{P}_{+}%
\left[ \left( \mathcal{D}_{\mathbf{X}_{\alpha }}\varepsilon _{0}\left( 
\mathbf{X}_{\alpha }\right) \right) \mathcal{A}_{0}^{\mathbf{X}_{\alpha
}}+H.C.\right] \right] d\alpha  \notag \\
&&+\frac{1}{2}\int_{0}^{\hbar }S_{\mathbf{X}_{\hbar }}\mathcal{P}_{+}\left(
\left( \frac{\alpha }{4}\mathcal{A}_{0}^{X_{\alpha l}}\nabla _{X_{\alpha l}}%
\mathcal{A}_{0}^{X_{\alpha j}}+H.C.+\left[ B_{\alpha },X_{\alpha j}+\alpha 
\mathcal{A}_{0}^{X_{\alpha j}}\right] \right) \nabla _{X_{\alpha
j}}\varepsilon _{0}\left( \mathbf{X}_{\alpha }\right) +H.C.\right) d\alpha 
\notag \\
&&+\frac{i}{2}\int_{0}^{\hbar }S_{\mathbf{X}_{\hbar }}\mathcal{P}_{+}\left( %
\left[ \varepsilon _{0}\left( \mathbf{X}_{\alpha }\right) ,\frac{\alpha }{4}%
\mathcal{A}_{0}^{X_{\alpha l}}\nabla _{X_{\alpha l}}\mathcal{A}_{0}^{\mathbf{%
R}_{\alpha }}+H.C.+\left[ B_{\alpha },\mathbf{R}_{\alpha }+\alpha \mathcal{A}%
_{0}^{\mathbf{R}_{\alpha }}\right] \right] \mathcal{A}_{0}^{\mathbf{P}%
_{\alpha }}+H.C.\right) d\alpha  \notag \\
&&-\frac{i}{2}\int_{0}^{\hbar }S_{\mathbf{X}_{\hbar }}\mathcal{P}_{+}\left( %
\left[ \varepsilon _{0}\left( \mathbf{X}_{\alpha }\right) ,\frac{\alpha }{4}%
\mathcal{A}_{0}^{X_{\alpha l}}\nabla _{X_{\alpha l}}\mathcal{A}_{0}^{\mathbf{%
P}_{\alpha }}+H.C.+\left[ B_{\alpha },\mathbf{P}_{\alpha }+\alpha \mathcal{A}%
_{0}^{\mathbf{P}_{\alpha }}\right] \right] \mathcal{A}_{0}^{\mathbf{R}%
_{\alpha }}+H.C.\right) d\alpha  \notag \\
&=&\frac{\hbar }{2}\mathcal{P}_{+}\left[ \left( \mathcal{D}_{\mathbf{X}%
}\varepsilon _{0}\left( \mathbf{X}\right) \right) \mathcal{A}_{0}^{\mathbf{X}%
}+H.C.\right]  \notag \\
&&+\frac{\hbar ^{2}}{4}\mathcal{P}_{+}\left( \left( \frac{1}{4}\mathcal{A}%
_{0}^{X_{l}}\nabla _{X_{l}}\mathcal{A}_{0}^{X_{i}}+H.C.+\left[ B,\frac{%
X_{\alpha j}}{\hbar }+\mathcal{A}_{0}^{X_{\alpha j}}\right] \right) \nabla
_{X_{j}}\varepsilon _{0}\left( \mathbf{X}\right) +H.C.\right)  \notag \\
&&+i\frac{\hbar ^{2}}{4}\mathcal{P}_{+}\left\{ \left( \varepsilon _{0}\left( 
\mathbf{X}\right) ,\frac{1}{4}\mathcal{A}_{0}^{X_{l}}\nabla _{X_{\alpha l}}%
\mathcal{A}_{0}^{\mathbf{R}}+H.C.+\left[ B,\frac{\mathbf{R}}{\hbar }+%
\mathcal{A}_{0}^{\mathbf{R}}\right] \right) \mathcal{A}_{0}^{\mathbf{P}%
}+H.C.\right\}  \notag \\
&&-i\frac{\hbar ^{2}}{4}\mathcal{P}_{+}\left\{ \left( \varepsilon _{0}\left( 
\mathbf{X}_{\alpha }\right) ,\mathcal{A}_{0}^{X_{\alpha l}}\nabla
_{X_{\alpha l}}\mathcal{A}_{0}^{\mathbf{P}}+H.C.+\left[ B,\frac{\mathbf{P}}{%
\hbar }+\mathcal{A}_{0}^{\mathbf{P}}\right] \right) \mathcal{A}_{0}^{\mathbf{%
R}}+H.C.\right\}  \notag \\
&=&\frac{\hbar }{2}\mathcal{P}_{+}\left[ \left( \mathcal{\tilde{D}}_{\mathbf{%
X}}\varepsilon _{0}\left( \mathbf{X}\right) \right) \mathcal{A}^{\mathbf{X}%
}+H.C.\right]
\end{eqnarray}
with $B\equiv B_{_{\hbar }}$ and 
\begin{equation}
\mathcal{A}^{\mathbf{X}}=\mathcal{A}_{0}^{\mathbf{X}}+\frac{\hbar }{4}\left( 
\frac{1}{2}\mathcal{A}_{0}^{X_{l}}\nabla _{X_{l}}\mathcal{A}_{0}^{\mathbf{X}%
}+\left[ B,\frac{\mathbf{X}}{\hbar }+\mathcal{A}_{0}^{\mathbf{X}}\right]
+H.C.\right)  \label{cxord2}
\end{equation}
We also introduced the notation $\mathcal{\tilde{D}}_{\mathbf{X}}\equiv
\left( \mathcal{\tilde{D}}_{\mathbf{R}},\mathcal{\tilde{D}}_{\mathbf{P}%
}\right) $ with 
\begin{eqnarray*}
\mathcal{\tilde{D}}_{\mathbf{R}} &=&\left[ \nabla _{\mathbf{R}_{\alpha }}+%
\frac{i}{2}\mathcal{A}^{\mathbf{P}},.\right] \\
\mathcal{\hat{D}}_{\mathbf{P}} &=&\left[ \nabla _{\mathbf{P}_{\alpha }}-%
\frac{i}{2}\mathcal{A}^{\mathbf{R}},.\right]
\end{eqnarray*}
Ultimately, one can gather the various terms to obtain the energy at the
second order in $\hbar $ as a function of the canonical variables 
\begin{eqnarray*}
\varepsilon \left( \mathbf{X}\right) &=&\varepsilon _{0}\left( \mathbf{X}%
\right) +\frac{\hbar }{2}\mathcal{P}_{+}\left[ \left( \mathcal{\hat{D}}_{%
\mathbf{X}}\varepsilon _{0}\left( \mathbf{X}\right) \right) \mathcal{A}^{%
\mathbf{X}}+H.C.\right] \\
&&+\frac{\hbar ^{2}}{8}\left[ \mathcal{P}_{+}\left\{ \left( \mathcal{D}_{%
\mathbf{X}}\left[ \mathcal{P}_{+}\left[ \left( \mathcal{D}_{\mathbf{X}%
}\varepsilon _{0}\left( \mathbf{X}\right) \right) \mathcal{A}_{0}^{\mathbf{X}%
}+H.C.\right] \right] \right) \mathcal{A}_{0}^{\mathbf{X}}\right\} +H.C.%
\right] -\frac{\hbar }{2}\left\langle \varepsilon _{0}\left( \mathbf{X}%
\right) \right\rangle
\end{eqnarray*}
This formula is still compact and can be, for practical purposes, expanded
in several ways. We first start by developing $\varepsilon \left( \mathbf{X}%
\right) $ as a function of the zeroth order diagonalization data, that is $%
\varepsilon _{0}\left( \mathbf{X}\right) $ and $\mathcal{A}_{0}^{\mathbf{X}}$%
. Starting first by expanding $\mathcal{A}^{\mathbf{X}}$, we are led to : 
\begin{eqnarray}
\varepsilon \left( \mathbf{X}\right) &=&\varepsilon _{0}\left( \mathbf{X}%
\right) +\frac{\hbar }{2}\mathcal{P}_{+}\left[ \mathcal{D}_{\mathbf{X}%
}\varepsilon _{0}\mathcal{A}_{0}^{\mathbf{X}}+H.C.\right] +\frac{\hbar ^{2}}{%
8}\left[ \mathcal{P}_{+}\left\{ \left( \mathcal{D}_{\mathbf{X}}\left[ \left[ 
\mathcal{D}_{\mathbf{X}}\varepsilon _{0}\mathcal{A}_{0}^{\mathbf{X}}+H.C.%
\right] \right] \right) \mathcal{A}_{0}^{\mathbf{X}}\right\} +H.C.\right] 
\notag \\
&&+\frac{\hbar ^{2}}{4}\mathcal{P}_{+}\left( \left( \frac{1}{4}\mathcal{A}%
_{0}^{X_{l}}\nabla _{X_{l}}\mathcal{A}_{0}^{X_{i}}+H.C.+\left[ B,\mathcal{A}%
_{0}^{X_{\alpha j}}\right] \right) \nabla _{X_{j}}\varepsilon
_{0}+H.C.\right)  \notag \\
&&+i\frac{\hbar ^{2}}{4}\mathcal{P}_{+}\left\{ \left[ \varepsilon _{0},\frac{%
1}{4}\mathcal{A}_{0}^{X_{l}}\nabla _{X_{\alpha l}}\mathcal{A}_{0}^{\mathbf{R}%
}+H.C.\right] \mathcal{A}_{0}^{\mathbf{P}}-\left[ \varepsilon _{0},\frac{1}{4%
}\mathcal{A}_{0}^{X_{\alpha l}}\nabla _{X_{\alpha l}}\mathcal{A}_{0}^{%
\mathbf{P}}+H.C.\right] \mathcal{A}_{0}^{\mathbf{R}}+H.C.\right\}  \notag \\
&&+i\frac{\hbar ^{2}}{4}\mathcal{P}_{+}\left\{ \left[ \varepsilon _{0},\left[
B,\frac{\mathbf{R}}{\hbar }+\mathcal{A}_{0}^{\mathbf{R}}\right] \right] 
\mathcal{A}_{0}^{\mathbf{P}}-\left[ \varepsilon _{0},\left[ B,\frac{\mathbf{P%
}}{\hbar }+\mathcal{A}_{0}^{\mathbf{P}}\right] \right] \mathcal{A}_{0}^{%
\mathbf{R}}+H.C.\right\}
\end{eqnarray}
Or, if we expand $B$ fully, 
\begin{eqnarray}
&&\varepsilon \left( \mathbf{X}\right)  \notag \\
&=&\varepsilon _{0}\left( \mathbf{X}\right) +\frac{\hbar }{2}\mathcal{P}_{+}%
\left[ \left( \mathcal{A}_{0}^{\mathbf{X}}\mathbf{\nabla }_{\mathbf{X}%
}\varepsilon _{0}\left( \mathbf{X}\right) \right) +\frac{i}{2}\left[
\varepsilon _{0}\left( \mathbf{X}\right) ,\mathcal{A}_{0}^{R_{l}}\right] 
\mathcal{A}_{0}^{P_{l}}-\frac{i}{2}\left[ \varepsilon _{0}\left( \mathbf{X}%
\right) ,\mathcal{A}_{0}^{P_{l}}\right] \mathcal{A}_{0}^{R_{l}}+H.C.\right] 
\notag \\
&&+\frac{\hbar ^{2}}{8}\mathcal{P}_{+}\left\{ \left[ \mathcal{A}_{0}^{%
\mathbf{X}}\mathbf{\nabla }_{\mathbf{X}}\mathcal{P}_{+}\left( \mathcal{A}%
_{0}^{\mathbf{X}}\mathbf{\nabla }_{\mathbf{X}}\varepsilon _{0}\left( \mathbf{%
X}\right) +i\left[ \varepsilon _{0}\left( \mathbf{X}\right) ,\mathcal{A}%
_{0}^{R_{l}}\right] \mathcal{A}_{0}^{P_{l}}+H.C.\right) \right] +H.C.\right\}
\notag \\
&&+i\frac{\hbar ^{2}}{8}\mathcal{P}_{+}\left\{ \left[ \mathcal{P}_{+}\left(
\left( \mathcal{A}_{0}^{\mathbf{X}}\mathbf{\nabla }_{\mathbf{X}}\varepsilon
_{0}\left( \mathbf{X}\right) \right) +i\left[ \varepsilon _{0}\left( \mathbf{%
X}\right) ,\mathcal{A}_{0}^{R_{l}}\right] \mathcal{A}_{0}^{P_{l}}+H.C.%
\right) ,\mathcal{A}_{0}^{R_{l}}\right] \mathcal{A}_{0}^{P_{l}}\right\} 
\notag \\
&&-i\frac{\hbar ^{2}}{8}\mathcal{P}_{+}\left\{ \left[ \mathcal{P}_{+}\left(
\left( \mathcal{A}_{0}^{\mathbf{X}}\mathbf{\nabla }_{\mathbf{X}}\varepsilon
_{0}\left( \mathbf{X}\right) \right) +i\left[ \varepsilon _{0}\left( \mathbf{%
X}\right) ,\mathcal{A}_{0}^{R_{l}}\right] \mathcal{A}_{0}^{P_{l}}+H.C.%
\right) ,\mathcal{A}_{0}^{P_{l}}\right] \mathcal{A}_{0}^{R_{l}}\right\} 
\notag \\
&&+\frac{\hbar ^{2}}{4}\mathcal{P}_{+}\left( \left( \frac{1}{4}\mathcal{A}%
_{0}^{X_{l}}\nabla _{X_{l}}\mathcal{A}_{0}^{X_{i}}+H.C.+\left[ \varepsilon
_{0},.\right] ^{-1}\left[ \mathcal{A}_{0}^{\mathbf{X}}\mathbf{\nabla }_{%
\mathbf{X}}\varepsilon _{0}+i\left[ \varepsilon _{0},\mathcal{A}_{0}^{R_{l}}%
\right] \mathcal{A}_{0}^{P_{l}}+H.C.\right] \right) \nabla
_{X_{j}}\varepsilon _{0}+H.C.\right)  \notag \\
&&+\frac{i\hbar ^{2}}{4}\mathcal{P}_{+}\left\{ \left[ \varepsilon _{0},\frac{%
1}{4}\mathcal{A}_{0}^{X_{l}}\nabla _{X_{\alpha l}}\mathcal{A}_{0}^{\mathbf{R}%
}+H.C.\right] \mathcal{A}_{0}^{\mathbf{P}}-\left[ \varepsilon _{0},\frac{1}{4%
}\mathcal{A}_{0}^{X_{\alpha l}}\nabla _{X_{\alpha l}}\mathcal{A}_{0}^{%
\mathbf{P}}+H.C.\right] \mathcal{A}_{0}^{\mathbf{R}}+H.C.\right\}  \notag \\
&&-\frac{i\hbar ^{2}}{4}\mathcal{P}_{+}\left\{ \left[ \varepsilon _{0},\left[
i\nabla _{\mathbf{P}}+\mathcal{A}_{0}^{\mathbf{R}},\left[ \varepsilon _{0},.%
\right] ^{-1}\left[ \mathcal{A}_{0}^{\mathbf{X}}\mathbf{\nabla }_{\mathbf{X}%
}\varepsilon _{0}+i\left[ \varepsilon _{0},\mathcal{A}_{0}^{R_{l}}\right] 
\mathcal{A}_{0}^{P_{l}}+H.C.\right] \right] \right] \mathcal{A}_{0}^{\mathbf{%
P}}+H.C.\right\}  \notag \\
&&+\frac{i\hbar ^{2}}{4}\mathcal{P}_{+}\left\{ \left[ \varepsilon _{0},\left[
-i\nabla _{\mathbf{R}}+\mathcal{A}_{0}^{\mathbf{P}},\left[ \varepsilon _{0},.%
\right] ^{-1}\left[ \mathcal{A}_{0}^{\mathbf{X}}\mathbf{\nabla }_{\mathbf{X}%
}\varepsilon _{0}+i\left[ \varepsilon _{0},\mathcal{A}_{0}^{R_{l}}\right] 
\mathcal{A}_{0}^{P_{l}}+H.C.\right] \right] \right] \mathcal{A}_{0}^{\mathbf{%
R}}+H.C.\right\}  \label{Efull}
\end{eqnarray}
The last expansion is useful for practical purpose, since it yields directly
the $\hbar $ expansion for the diagonalized Hamiltonian. However, it seems
more elegant and relevant to rewrite $\varepsilon \left( \mathbf{X}\right) $
as a function of the covariant variables $\mathbf{x=}\left( \mathbf{r,p}%
\right) $ defined in the preceding section. Note first that the Berry phases
Eq. $\ref{cxord2}$, $\mathcal{A}^{\mathbf{X}}=\left( \mathcal{A}^{\mathbf{R}%
}\left( \mathbf{X}\right) ,\mathcal{A}^{\mathbf{P}}\left( \mathbf{X}\right)
\right) $ with $\mathcal{A}^{\mathbf{X}}=\mathcal{A}_{0}^{\mathbf{X}}+\frac{%
\hbar }{4}\left( \mathcal{A}_{0}^{X_{l}}\nabla _{X_{l}}\mathcal{A}_{0}^{%
\mathbf{X}}+\left[ B,\mathbf{X}+\mathcal{A}_{0}^{\mathbf{X}}\right]
+H.C.\right) $, satisfy, by construction : 
\begin{align*}
\mathcal{A}^{R_{l}}\left( \mathbf{X}\right) & =\frac{1}{\hbar }\int_{\alpha
}^{\hbar }S_{\mathbf{X}_{\hbar }}\mathcal{A}_{\alpha }^{R_{l}}d\alpha \\
\mathcal{A}^{P_{l}}\left( \mathbf{X}\right) & =\frac{1}{\hbar }\int_{\alpha
}^{\hbar }S_{\mathbf{X}_{\hbar }}\mathcal{A}_{\alpha }^{P_{l}}d\alpha
\end{align*}
So that, recalling our previous definition of the variables $\mathbf{x=X+}%
\emph{A}^{\mathbf{X}}$ (see Eq. \ref{phaseproj}), we can write for $\mathbf{r%
}$, 
\begin{align}
\mathbf{r}& =\mathbf{R+}\int_{0}^{\hbar }S_{\mathbf{X}_{\hbar }}\mathcal{P}%
_{+}\mathcal{A}_{\alpha _{1}}^{\mathbf{R}}\left( \mathbf{R},\mathbf{P}%
\right) d\alpha _{1}+\int\limits_{0<\alpha <\hbar }S_{\mathbf{X}_{\hbar }}%
\frac{1}{2}\left[ \left[ \mathcal{P}_{+}\left[ \mathcal{A}_{\alpha }^{%
\mathbf{X}}\right] .\mathbf{\nabla }_{\mathbf{X}}\int\limits_{0<\alpha
_{1}<\alpha }\mathcal{P}_{+}\left[ \mathcal{A}_{\alpha _{1}}^{\mathbf{R}}%
\right] \right] +H.C.\right] d\alpha _{1}d\alpha  \notag \\
& =\mathbf{R+}\hbar \mathcal{P}_{+}\mathcal{A}^{\mathbf{R}}\left( \mathbf{X}%
\right) +\frac{\hbar }{4}^{2}\left( \left( \mathcal{P}_{+}\mathcal{A}_{0}^{%
\mathbf{R}}.\mathbf{\nabla }_{\mathbf{R}}\right) \mathcal{P}_{+}\mathcal{A}%
_{0}^{\mathbf{R}}+\left( \mathcal{P}_{+}\mathcal{A}_{0}^{\mathbf{P}}.\mathbf{%
\nabla }_{\mathbf{P}}\right) \mathcal{P}_{+}\mathcal{A}_{0}^{\mathbf{R}%
}+H.C.\right)  \notag \\
& =\mathbf{R+}\hbar \mathcal{P}_{+}\mathcal{A}_{0}^{\mathbf{R}}+\frac{\hbar
^{2}}{4}\mathcal{P}_{+}\left( \frac{1}{2}\mathcal{A}_{0}^{X_{l}}\nabla
_{X_{l}}\mathcal{A}_{0}^{\mathbf{R}}+\left[ B,\frac{\mathbf{R}}{\hbar }+%
\mathcal{A}_{0}^{\mathbf{R}}\right] +H.C.\right)  \notag \\
& +\frac{\hbar }{4}^{2}\left( \left( \mathcal{P}_{+}\mathcal{A}_{0}^{\mathbf{%
R}}.\mathbf{\nabla }_{\mathbf{R}}\right) \mathcal{P}_{+}\mathcal{A}_{0}^{%
\mathbf{R}}+\left( \mathcal{P}_{+}\mathcal{A}_{0}^{\mathbf{P}}.\mathbf{%
\nabla }_{\mathbf{P}}\right) \mathcal{P}_{+}\mathcal{A}_{0}^{\mathbf{R}%
}+H.C.\right)  \notag \\
& \equiv \mathbf{R+}\hbar \emph{A}_{0}^{\mathbf{R}}+\frac{\hbar }{2}^{2}%
\emph{A}_{1}^{\mathbf{R}}
\end{align}
For $\mathbf{p,}$ we have in the same manner the following expansion 
\begin{align}
\mathbf{p}& =\mathbf{P}+\int_{0}^{\hbar }S_{\mathbf{X}_{\hbar }}\mathcal{P}%
_{+}\mathcal{A}_{\alpha _{1}}^{\mathbf{P}}\left( \mathbf{R},\mathbf{P}%
\right) d\alpha _{1}+\int\limits_{0<\alpha <\hbar }S_{\mathbf{X}_{\hbar }}%
\frac{1}{2}\left[ \left[ \mathcal{P}_{+}\left[ \mathcal{A}_{\alpha }^{%
\mathbf{X}}\right] .\mathbf{\nabla }_{\mathbf{X}}\int\limits_{0<\alpha
_{1}<\alpha }\mathcal{P}_{+}\left[ \mathcal{A}_{\alpha _{1}}^{\mathbf{P}}%
\right] \right] +H.C.\right] d\alpha _{1}d\alpha  \notag \\
& =\mathbf{P+}\hbar \mathcal{P}_{+}\mathcal{A}^{\mathbf{R}}\left( \mathbf{X}%
\right) +\frac{\hbar }{4}^{2}\left( \left( \mathcal{P}_{+}\mathcal{A}_{0}^{%
\mathbf{R}}.\mathbf{\nabla }_{\mathbf{R}}\right) \mathcal{P}_{+}\mathcal{A}%
_{0}^{\mathbf{P}}+\left( \mathcal{P}_{+}\mathcal{A}_{0}^{\mathbf{P}}.\mathbf{%
\nabla }_{\mathbf{P}}\right) \mathcal{P}_{+}\mathcal{A}_{0}^{\mathbf{P}%
}+H.C.\right)  \notag \\
& =\mathbf{P+}\hbar \mathcal{P}_{+}\mathcal{A}^{\mathbf{P}}+\frac{\hbar ^{2}%
}{4}\mathcal{P}_{+}\left( \frac{1}{2}\mathcal{A}_{0}^{X_{l}}\nabla _{X_{l}}%
\mathcal{A}_{0}^{\mathbf{P}}+\left[ B,\frac{\mathbf{P}}{\hbar }+\mathcal{A}%
_{0}^{\mathbf{P}}\right] +H.C.\right)  \notag \\
& +\frac{\hbar }{4}^{2}\left( \frac{1}{2}\left( \mathcal{P}_{+}\mathcal{A}%
_{0}^{\mathbf{R}}.\mathbf{\nabla }_{\mathbf{R}}\right) \mathcal{P}_{+}%
\mathcal{A}_{0}^{\mathbf{P}}+\left( \mathcal{P}_{+}\mathcal{A}_{0}^{\mathbf{P%
}}.\mathbf{\nabla }_{\mathbf{P}}\right) \mathcal{P}_{+}\mathcal{A}_{0}^{%
\mathbf{P}}+H.C.\right)  \notag \\
& \equiv \mathbf{P+}\hbar \emph{A}_{0}^{\mathbf{P}}+\frac{\hbar }{2}^{2}%
\emph{A}_{1}^{\mathbf{P}}
\end{align}
As before. $\mathbf{x=}\left( \mathbf{r,p}\right) $ satisfy of course an
algebra Eq. $\left( \ref{commalgebra}\right) $.

Now, come back to the compact form of $\varepsilon \left( \mathbf{X}\right) $
: 
\begin{eqnarray*}
\varepsilon \left( \mathbf{X}\right) &=&\varepsilon _{0}\left( \mathbf{X}%
\right) +\frac{\hbar }{2}\mathcal{P}_{+}\left[ \left( \mathcal{\hat{D}}_{%
\mathbf{X}}\varepsilon _{0}\left( \mathbf{X}\right) \right) \mathcal{A}^{%
\mathbf{X}}+H.C.\right] \\
&&+\frac{\hbar ^{2}}{8}\left[ \mathcal{P}_{+}\left\{ \left( \mathcal{D}_{%
\mathbf{X}}\left[ \mathcal{P}_{+}\left[ \left( \mathcal{D}_{\mathbf{X}%
}\varepsilon _{0}\left( \mathbf{X}\right) \right) \mathcal{A}_{0}^{\mathbf{X}%
}+H.C.\right] \right] \right) \mathcal{A}_{0}^{\mathbf{X}}\right\} +H.C.%
\right] -\frac{\hbar }{2}\left\langle \varepsilon _{0}\left( \mathbf{X}%
\right) \right\rangle
\end{eqnarray*}
expanded as : 
\begin{eqnarray*}
\varepsilon \left( \mathbf{X}\right) &=&\varepsilon _{0}\left( \mathbf{X}%
\right) +\mathcal{P}_{+}\left\{ \frac{\hbar }{2}\left( \mathcal{A}%
^{X_{l}}\nabla _{X_{l}}\varepsilon _{0}\left( \mathbf{X}\right) +\nabla
_{X_{l}}\varepsilon _{0}\left( \mathbf{X}\right) \mathcal{A}^{X_{l}}\right)
\right\} \\
&&+\frac{i\hbar }{4}\left[ \mathcal{P}_{+}\left\{ \left[ \varepsilon
_{0}\left( \mathbf{X}\right) ,\mathcal{A}^{R_{l}}\right] \mathcal{A}^{P_{l}}-%
\left[ \varepsilon _{0}\left( \mathbf{X}\right) ,\mathcal{A}^{P_{l}}\right] 
\mathcal{A}^{R_{l}}\right\} +H.C.\right] \\
&&+\frac{\hbar ^{2}}{8}\mathcal{P}_{+}\left\{ \left[ \mathcal{A}_{0}^{%
\mathbf{X}}\mathbf{\nabla }_{\mathbf{X}}\mathcal{P}_{+}\left( \mathcal{A}%
_{0}^{\mathbf{X}}\mathbf{\nabla }_{\mathbf{X}}\varepsilon _{0}\left( \mathbf{%
X}\right) +i\left[ \varepsilon _{0}\left( \mathbf{X}\right) ,\mathcal{A}%
_{0}^{R_{l}}\right] \mathcal{A}_{0}^{P_{l}}+H.C.\right) \right] +H.C.\right\}
\\
&&+i\frac{\hbar ^{2}}{8}\mathcal{P}_{+}\left\{ \left[ \mathcal{P}_{+}\left(
\left( \mathcal{A}_{0}^{\mathbf{X}}\mathbf{\nabla }_{\mathbf{X}}\varepsilon
_{0}\left( \mathbf{X}\right) \right) +i\left[ \varepsilon _{0}\left( \mathbf{%
X}\right) ,\mathcal{A}_{0}^{R_{l}}\right] \mathcal{A}_{0}^{P_{l}}+H.C.%
\right) ,\mathcal{A}_{0}^{R_{l}}\right] \mathcal{A}_{0}^{P_{l}}\right\} \\
&&-i\frac{\hbar ^{2}}{8}\mathcal{P}_{+}\left\{ \left[ \mathcal{P}_{+}\left(
\left( \mathcal{A}_{0}^{\mathbf{X}}\mathbf{\nabla }_{\mathbf{X}}\varepsilon
_{0}\left( \mathbf{X}\right) \right) +i\left[ \varepsilon _{0}\left( \mathbf{%
X}\right) ,\mathcal{A}_{0}^{R_{l}}\right] \mathcal{A}_{0}^{P_{l}}+H.C.%
\right) ,\mathcal{A}_{0}^{P_{l}}\right] \mathcal{A}_{0}^{R_{l}}\right\}
\end{eqnarray*}
To the second order, and given our choice of symmetrization, the
contributions 
\begin{equation*}
\varepsilon _{0}\left( \mathbf{X}\right) +\mathcal{P}_{+}\left\{ \frac{\hbar 
}{2}\left( \mathcal{A}^{X_{l}}\nabla _{X_{l}}\varepsilon _{0}\left( \mathbf{X%
}\right) +\nabla _{X_{l}}\varepsilon _{0}\left( \mathbf{X}\right) \mathcal{A}%
^{X_{l}}\right) \right\} +\frac{\hbar ^{2}}{4}\mathcal{P}_{+}\left\{ 
\mathcal{A}_{0}^{\mathbf{X}}\mathbf{\nabla }_{\mathbf{X}}\mathcal{P}%
_{+}\left( \mathcal{A}_{0}^{\mathbf{X}}\mathbf{\nabla }_{\mathbf{X}%
}\varepsilon _{0}\left( \mathbf{X}\right) \right) +H.C.\right\}
\end{equation*}
can be recombined as : 
\begin{equation*}
\varepsilon _{0}\left( \mathbf{X+}\hbar \mathcal{P}_{+}\mathcal{A}^{X_{l}}+%
\frac{\hbar }{4}^{2}\left( \left( \mathcal{P}_{+}\mathcal{A}_{0}^{\mathbf{R}%
}.\mathbf{\nabla }_{\mathbf{R}}\right) \mathcal{P}_{+}\mathcal{A}_{0}^{%
\mathbf{R}}+\left( \mathcal{P}_{+}\mathcal{A}_{0}^{\mathbf{P}}.\mathbf{%
\nabla }_{\mathbf{P}}\right) \mathcal{P}_{+}\mathcal{A}_{0}^{\mathbf{R}%
}+H.C.\right) \right) =\varepsilon _{0}\left( \mathbf{x}\right)
\end{equation*}
On the other hand, one has also at the same order: 
\begin{eqnarray*}
&&\frac{i\hbar }{4}\mathcal{P}_{+}\left\{ \left[ \varepsilon _{0}\left( 
\mathbf{X}\right) ,\mathcal{A}^{R_{l}}\right] \mathcal{A}^{P_{l}}-\left[
\varepsilon _{0}\left( \mathbf{X}\right) ,\mathcal{A}^{P_{l}}\right] 
\mathcal{A}^{R_{l}}+H.C.\right\} \\
&&+i\frac{\hbar ^{2}}{8}\mathcal{P}_{+}\left\{ \left[ \mathcal{A}_{0}^{%
\mathbf{X}}\mathbf{\nabla }_{\mathbf{X}}\mathcal{P}_{+}\left( \left[
\varepsilon _{0}\left( \mathbf{X}\right) ,\mathcal{A}_{0}^{R_{l}}\right] 
\mathcal{A}_{0}^{P_{l}}+H.C.\right) \right] +H.C.\right\} \\
&&+i\frac{\hbar ^{2}}{8}\mathcal{P}_{+}\left\{ \left[ \mathcal{P}_{+}\left(
\left( \mathcal{A}_{0}^{\mathbf{X}}\mathbf{\nabla }_{\mathbf{X}}\varepsilon
_{0}\left( \mathbf{X}\right) \right) +H.C.\right) ,\mathcal{A}_{0}^{R_{l}}%
\right] \mathcal{A}_{0}^{P_{l}}\right\} \\
&&-i\frac{\hbar ^{2}}{8}\mathcal{P}_{+}\left\{ \left[ \mathcal{P}_{+}\left(
\left( \mathcal{A}_{0}^{\mathbf{X}}\mathbf{\nabla }_{\mathbf{X}}\varepsilon
_{0}\left( \mathbf{X}\right) \right) +H.C.\right) ,\mathcal{A}_{0}^{P_{l}}%
\right] \mathcal{A}_{0}^{R_{l}}\right\} \\
&=&\frac{i\hbar }{2}\mathcal{P}_{+}\left\{ \left[ \varepsilon _{0}\left( 
\mathbf{X+}\mathcal{P}_{+}\mathcal{A}_{0}^{\mathbf{X}}\right) ,\mathcal{A}%
^{R_{l}}\left( X\mathbf{+}\frac{\mathcal{P}_{+}\mathcal{A}_{0}^{\mathbf{X}}}{%
2}\right) \right] \mathcal{A}^{P_{l}}\left( X\mathbf{+}\frac{\mathcal{P}_{+}%
\mathcal{A}_{0}^{\mathbf{X}}}{2}\right) \right\} \\
&&-\frac{i\hbar }{2}\mathcal{P}_{+}\left\{ \left[ \varepsilon _{0}\left( 
\mathbf{X+}\mathcal{P}_{+}\mathcal{A}_{0}^{\mathbf{X}}\right) ,\mathcal{A}%
^{P_{l}}\left( X\mathbf{+}\frac{\mathcal{P}_{+}\mathcal{A}_{0}^{\mathbf{X}}}{%
2}\right) \right] \mathcal{A}^{R_{l}}\left( X\mathbf{+}\frac{\mathcal{P}_{+}%
\mathcal{A}_{0}^{\mathbf{X}}}{2}\right) \right\} \\
&=&\frac{i}{2}\hbar \mathcal{P}_{+}\left\{ \left[ \varepsilon _{0}\left( 
\mathbf{x}\right) ,\mathcal{\hat{A}}^{R_{l}}\right] \mathcal{\hat{A}}%
^{P_{l}}-\left[ \varepsilon _{0}\left( \mathbf{x}\right) ,\mathcal{\hat{A}}%
^{P_{l}}\right] \mathcal{\hat{A}}^{R_{l}}-\left[ \varepsilon _{0}\left( 
\mathbf{x}\right) ,\left[ \mathcal{\hat{A}}^{R_{l}},\mathcal{\hat{A}}^{P_{l}}%
\right] \right] \right\}
\end{eqnarray*}
where we introduced the notations 
\begin{align*}
\mathcal{\hat{A}}^{R_{l}}& =\frac{1}{2}\left[ 1-\frac{1}{2}\left( \mathcal{P}%
_{+}\mathcal{A}_{0}^{\mathbf{X}}.\mathbf{\nabla }_{\mathbf{X}}\right) \right]
\mathcal{A}^{R_{l}}\left( \mathbf{x}\right) +H.C. \\
\mathcal{\hat{A}}^{P_{l}}& =\frac{1}{2}\left[ 1-\frac{1}{2}\left( \mathcal{P}%
_{+}\mathcal{A}_{0}^{\mathbf{X}}.\mathbf{\nabla }_{\mathbf{X}}\right) \right]
\mathcal{A}^{P_{l}}\left( \mathbf{x}\right) +H.C.
\end{align*}
Ultimately,the contributions to the Hamiltonian can be recombined to yield :

\begin{align}
\varepsilon \left( \mathbf{X}\right) & =\varepsilon _{0}\left( \mathbf{x}%
\right) +\frac{i}{4}\hbar \mathcal{P}_{+}\left\{ \left[ \varepsilon
_{0}\left( \mathbf{x}\right) ,\mathcal{\hat{A}}^{R_{l}}\right] \mathcal{\hat{%
A}}^{P_{l}}-\left[ \varepsilon _{0}\left( \mathbf{x}\right) ,\mathcal{\hat{A}%
}^{P_{l}}\right] \mathcal{\hat{A}}^{R_{l}}-\left[ \varepsilon _{0}\left( 
\mathbf{x}\right) ,\left[ \mathcal{A}_{0}^{R_{l}},\mathcal{A}_{0}^{P_{l}}%
\right] \right] +H.C.\right\}  \notag \\
& -\frac{\hbar ^{2}}{8}\mathcal{P}_{+}\left\{ \left[ \left[ \varepsilon
_{0}\left( \mathbf{x}\right) ,\mathcal{A}_{0}^{R_{l}}\right] \mathcal{A}%
_{0}^{P_{l}}-\left[ \varepsilon _{0}\left( \mathbf{x}\right) ,\mathcal{A}%
_{0}^{P_{l}}\right] \mathcal{A}_{0}^{R_{l}},\mathcal{A}_{0}^{R_{k}}\right] 
\mathcal{A}_{0}^{P_{k}}\right\}  \notag \\
& +\frac{\hbar ^{2}}{8}\mathcal{P}_{+}\left\{ \left[ \left[ \varepsilon
_{0}\left( \mathbf{x}\right) ,\mathcal{A}_{0}^{R_{l}}\right] \mathcal{A}%
_{0}^{P_{l}}-\left[ \varepsilon _{0}\left( \mathbf{x}\right) ,\mathcal{A}%
_{0}^{P_{l}}\right] \mathcal{A}_{0}^{R_{l}},\mathcal{A}_{0}^{P_{k}}\right] 
\mathcal{A}_{0}^{R_{k}}\right\} -\frac{\hbar }{2}\left\langle \varepsilon
_{0}\left( \mathbf{x}\right) \right\rangle  \label{EH22}
\end{align}
Note that in the all Hamiltonian, including the Berry connections $A^{R_{l}}$
and $A^{P_{l}}$,\ we have replaced the operators $\left( \mathbf{R,P}\right) 
$\ by $\mathbf{x=}\left( \mathbf{r,p}\right) $\ at each order of the
expansion.

\section{Physical applications}

\subsection{The Dirac electron in an electric field}

To illustrate our general theory we consider the case of a Dirac electron in
an external electric field. We will obtain the block diagonal Hamiltonian to
the second order in $\hbar $ and will compare with the FW transformation
obtained in \cite{Blount}. Note that contrary to the FW which is not an
expansion in $\hbar $, the new method is valid for strong external fields
(actually a FW transformation expanded into a power series in $\hbar $ was
also recently proposed \cite{SILENKO}).

Let consider the following Dirac Hamiltonian $(c=1)$ 
\begin{equation*}
H_{1}=\mathbf{\alpha .P}+\beta m+eV\left( \mathbf{R}\right)
\end{equation*}
To compute the diagonalized Hamiltonian to the second order in $\hbar $ we
first need the zeroth order diagonalization transformation, which is the
usual FW transformation for a free particle: 
\begin{equation*}
U_{0}=\frac{E+m+\beta \mathbf{\alpha P}}{\sqrt{2E\left( E+m\right) }}
\end{equation*}
where $E=\sqrt{\mathbf{P}^{2}+m^{2}}.$ In this case, we have (with $\mathbf{%
\Sigma =1\otimes \sigma }$) 
\begin{align*}
\mathcal{A}_{0}^{R}& =iU_{0}\mathbf{\nabla }_{\mathbf{P}}U_{0}^{-1}=i\frac{%
-\beta \mathbf{\alpha .PP}+E\left( E+m\right) \beta \mathbf{\alpha }-iE%
\mathbf{P\times \Sigma }}{2E^{2}\left( E+m\right) } \\
\mathcal{A}_{0}^{P}& =-iU_{0}^{i}\mathbf{\nabla }_{\mathbf{R}}U_{0}^{-1}=0
\end{align*}
which leads to the first order projected Berry connections: 
\begin{align*}
\emph{A}_{0}^{\mathbf{R}}& =\mathcal{P}_{+}\mathcal{A}_{0}^{R}=\frac{\mathbf{%
P\times \Sigma }}{2E\left( E+m\right) } \\
\emph{A}_{0}^{\mathbf{P}}& =\mathcal{P}_{+}\mathcal{A}_{0}^{P}=0
\end{align*}
The zeroth order diagonalized energy is $\varepsilon _{0}\left( \mathbf{R},%
\mathbf{P}\right) =E=\sqrt{\mathbf{P}^{2}+m^{2}}$

Now, to complete the diagonalization process, we need the matrix $B$ Eq. (%
\ref{Bform}) that enters in the definition of the covariant variables: 
\begin{equation*}
B=\frac{\beta e}{2E}\mathcal{P}_{-}\left( \mathcal{A}_{0}^{R}\right) .%
\mathbf{\nabla }_{R}V
\end{equation*}

From $B$ and Eq.$\left( \ref{cxord2}\right) $ we can then compute the Berry
connections to the second order in $\hbar $. Here expressions simplify
greatly. Actually, one has : 
\begin{eqnarray*}
\mathcal{A}^{\mathbf{R}} &=&\mathcal{A}_{0}^{\mathbf{R}}+\frac{1}{2}\left[ B,%
\mathbf{R}+\hbar \mathcal{A}_{0}^{\mathbf{X}}\right] \\
&=&\hbar \frac{\mathbf{P\times \Sigma }}{2E\left( E+m\right) }+\frac{\hbar e%
}{2}\beta \frac{E^{2}\mathbf{\nabla }_{R}V-\left[ \mathbf{P.\nabla }_{R}V%
\right] \mathbf{P}}{4E^{5}} \\
&&+\hbar e\mathbf{\nabla }_{P}\frac{-\mathbf{\alpha .PP}+E\left( E+m\right) 
\mathbf{\alpha }}{8E^{3}\left( E+m\right) }\mathbf{\nabla }_{R}V \\
&&+\hbar e\frac{E\left( E+m\right) \mathbf{P}\times \left( \mathbf{\nabla }%
_{R}V\times \mathbf{\alpha }\right) -\left( \mathbf{P}\times \left( \mathbf{P%
}\times \mathbf{\alpha }\right) \right) \mathbf{P.\nabla }_{R}V}{%
4E^{4}\left( E+m\right) ^{2}}
\end{eqnarray*}
and 
\begin{equation*}
\mathcal{A}^{\mathbf{P}}=\frac{1}{2}\left[ B,\mathbf{P}\right] =i\hbar \frac{%
\beta e}{4E}\left( \mathcal{P}_{-}\left( \mathcal{A}_{0}^{R}\right) .\mathbf{%
\nabla }_{R}\right) \mathbf{\nabla }_{R}V
\end{equation*}
we can also write the dynamical operators as:

\begin{equation*}
\mathbf{r}\mathbf{=R+}\hbar \mathcal{P}_{+}\mathcal{A}_{0}^{\mathbf{R}}+%
\frac{\hbar ^{2}}{2}\mathcal{P}_{+}\left[ B,\mathcal{A}_{0}^{\mathbf{R}}%
\right] =\mathbf{R+}\hbar \frac{\mathbf{P\times \Sigma }}{2E\left(
E+m\right) }+\frac{\hbar ^{2}e}{2}\beta \frac{E^{2}\mathbf{\nabla }_{R}V-%
\left[ \mathbf{P.\nabla }_{R}V\right] \mathbf{P}}{4E^{5}}
\end{equation*}
and 
\begin{equation*}
\mathbf{p}=\mathbf{P}
\end{equation*}
Now using expression Eq. $\left( \ref{EH22}\right) $, we arrive at the
following expression for the diagonal representation of the energy operator
as a function of the covariant variables $\left( \mathbf{r},\mathbf{p}%
\right) $: 
\begin{eqnarray*}
\varepsilon \left( \mathbf{X}\right) &=&\varepsilon _{0}\left( \mathbf{x}%
\right) +\frac{i}{2}\hbar \mathcal{P}_{+}\left\{ \left[ \varepsilon
_{0}\left( \mathbf{x}\right) ,\mathcal{\hat{A}}^{R_{l}}\right] \mathcal{\hat{%
A}}^{P_{l}}-\left[ \varepsilon _{0}\left( \mathbf{x}\right) ,\mathcal{\hat{A}%
}^{P_{l}}\right] \mathcal{\hat{A}}^{R_{l}}\right\} \\
&=&\varepsilon _{0}\left( \mathbf{x}\right) +\frac{i}{2}\hbar \mathcal{P}%
_{+}\left\{ \left[ \varepsilon _{0}\left( \mathbf{x}\right) ,\mathcal{A}%
_{0}^{R_{l}}\right] \mathcal{A}^{P_{l}}-\left[ \varepsilon _{0}\left( 
\mathbf{x}\right) ,\mathcal{A}^{P_{l}}\right] \mathcal{A}_{0}^{R_{l}}\right\}
\end{eqnarray*}
which, once developped, leads to: 
\begin{equation}
\varepsilon =\beta \sqrt{\mathbf{p}^{2}+m^{2}}+\frac{\hbar ^{2}e}{2}\mathbf{%
\nabla }_{\mathbf{r}}.\frac{E^{2}\mathbf{\nabla }_{\mathbf{r}}V-\left[ 
\mathbf{p.\nabla }_{\mathbf{r}}V\right] \mathbf{p}}{4E^{4}}+eV\left( \mathbf{%
r}\right)  \label{Erelat}
\end{equation}
Here we have made the choice of fully symmetrizing in $\mathbf{r}$ and $%
\mathbf{p}$, that is to weight equally all permutations in $\mathbf{r}$ and $%
\mathbf{p}$ in the series expansions of our expressions. One can now check
that developing the variables $\mathbf{r}$\ as a function of the canonical
variables $\mathbf{R}$\ and $\mathbf{P}$\ yields the same expression for the
Hamiltonian as in \cite{Blount} (apart from.a small sign mistake for the
spin-orbit coupling in that reference) 
\begin{eqnarray}
&&\varepsilon =\beta \sqrt{\mathbf{P}^{2}+m^{2}}+eV\left( \mathbf{R}\right)
+\hbar \frac{\mathbf{P\times \Sigma }}{2E\left( E+m\right) }.\mathbf{\nabla }%
_{R}V  \notag \\
&&+\hbar ^{2}e\beta \frac{E^{2}\left( \mathbf{\nabla }_{R}V\right)
^{2}-\left( \mathbf{P.\nabla }_{R}V\right) ^{2}}{8E^{5}}  \notag \\
&&+\hbar ^{2}e\left( \frac{\mathbf{\nabla }_{R}^{2}V}{4E\left( E+m\right) }-%
\frac{\left( 2E^{2}+2Em+m^{2}\right) \left( \mathbf{P.\nabla }_{R}\right)
^{2}V}{8E^{4}\left( E+m\right) ^{2}}\right)
\end{eqnarray}
Note that this result can also be found directly by using the expression Eq.
(\ref{Efull}) for the diagonalized Hamiltonian as a function of the
canonical variables. Eq. (\ref{Erelat}) being fully relativistic, to compare
with the usual FW \cite{FOLDY} approach we consider the non-relativistic
limit and expand our result to second order in $\frac{1}{mc}.$ We readily
obtain the well known diagonal representation of the positive energy
(expressed in coordinates $R$ and $P$) \cite{FOLDY} : 
\begin{equation}
\varepsilon =\frac{\mathbf{P}^{2}}{2m}-\frac{\mathbf{P}^{4}}{8m^{3}c^{2}}%
+eV\left( \mathbf{R}\right) +\frac{e\hbar }{4m^{2}c^{2}}\mathbf{\mathbf{%
\sigma .}}\left( \mathbf{\nabla }_{\mathbf{R}}V\times \mathbf{p}\right) 
\mathbf{+}\frac{e\hbar ^{2}}{8m^{2}c^{2}}\mathbf{\nabla }^{2}V
\end{equation}
with $\mathbf{\sigma }$ the Pauli matrices. Note that it is the term order $%
\hbar ^{2}$ in Eq. (\ref{Erelat}) which, in the non relativistic limit,
leads to the Darwin term.

\subsection{Massless Dirac particle in a static symmetric gravitational field%
}

We consider here the Hamiltonian of a massless neutrino propagating in an
isotropic inhomogeneous curved space of metric $g^{ij}(\mathbf{R}%
)=n^{-1}\left( \mathbf{R}\right) \delta ^{ij}$.(with the convention $c=1$) 
\begin{equation}
H_{0}=\frac{1}{2}\left( \mathbf{\alpha }.\mathbf{P}F(\mathbf{R})+F(\mathbf{R}%
)\mathbf{\alpha }.\mathbf{P}\right)
\end{equation}
with $F(\mathbf{R})=n^{-1}\left( \mathbf{R}\right) $.

As for the electron in an electric field, we start by giving the zeroth
order diagonalization matrix as well as the effective energy $\varepsilon
_{0}\left( \mathbf{R},\mathbf{P}\right) $. Since at that order, $\left( 
\mathbf{R},\mathbf{P}\right) $ can be thought as commuting variables, one
finds easily that the Hamiltonian diagonalization is performed through the
following Foldy-Wouthuysen unitary matrix : 
\begin{equation}
U_{0}\left( \mathbf{P}\right) =\frac{\sqrt{\mathbf{P}^{2}}+\beta \mathbf{%
\alpha .P}}{\sqrt{2\mathbf{P}^{2}}}
\end{equation}
and 
\begin{equation}
\varepsilon _{0}\left( \mathbf{P,R}\right) =U_{0}H_{0}\left( \mathbf{P,R}%
\right) U_{0}^{+}=\frac{1}{2}\beta \left( F(\mathbf{R)}\sqrt{\mathbf{P}^{2}}+%
\sqrt{\mathbf{P}^{2}}F(\mathbf{R)}\right)
\end{equation}
We also need the Berry phases at the lowest order: 
\begin{eqnarray*}
\mathcal{A}_{0}^{\mathbf{R}} &=&i\left[ U_{0}\nabla _{\mathbf{P}}U_{0}^{+}%
\right] =i\frac{-\beta \mathbf{\alpha .PP}+E^{2}\beta \mathbf{\alpha }-iE%
\mathbf{P\times \Sigma }}{2E^{3}} \\
\mathcal{A}_{0}^{\mathbf{P}} &=&-i\left[ U\nabla _{\mathbf{R}}U^{+}\right] =0
\end{eqnarray*}
The matrix $B$ Eq. (\ref{Bform}) needed to obtain the corrections to the
energy at the second order is even simpler than for the electron in an
electric field since here (see Eq. $\left( \ref{cxord2}\right) $): 
\begin{equation}
B=-\left[ .,\varepsilon _{0}\left( \mathbf{X}_{\alpha }\right) \right] ^{-1}.%
\mathcal{P}_{-}\left\{ \frac{1}{2}\mathcal{A}_{0}^{R\alpha _{l}}\nabla
_{R_{\alpha l}}\varepsilon _{0}\left( \mathbf{X}\right) +\frac{1}{2}\nabla
_{R_{\alpha l}}\varepsilon _{0}\left( \mathbf{X}\right) \mathcal{A}%
_{0}^{R\alpha _{l}}\right\} =0  \notag
\end{equation}
Therefore Berry connections and covariant variables to the second order
reduce to $\mathcal{A}^{\mathbf{R}}=\mathcal{A}_{0}^{\mathbf{R}}$ and $%
\mathcal{A}^{\mathbf{P}}=0$ as well as $\mathbf{r}\mathbf{=R+}\hbar \mathcal{%
P}_{+}\mathcal{A}_{0}^{\mathbf{R}}=\mathbf{R+}\hbar \frac{\mathbf{P\times
\Sigma }}{2E^{2}}$, $\mathbf{p}=\mathbf{P}$.

Now, using expression Eq. $\left( \ref{EH22}\right) $, we ultimately obtain
the energy operator: 
\begin{eqnarray}
\varepsilon \left( \mathbf{P,r}\right) &=&\frac{1}{2}\beta \left( F\left( 
\mathbf{r}\right) \sqrt{\mathbf{P}^{2}}+\sqrt{\mathbf{P}^{2}}F\left( \mathbf{%
r}\right) \right) -\frac{\hbar }{2}\left\langle \varepsilon _{0}\left( 
\mathbf{X}\right) \right\rangle  \notag \\
&=&\frac{1}{2}\beta \left( F\left( \mathbf{r}\right) \sqrt{\mathbf{P}^{2}}+%
\sqrt{\mathbf{P}^{2}}F\left( \mathbf{r}\right) \right) -\frac{1}{4}\hbar %
\left[ \sqrt{\mathbf{P}^{2}},F\left( \mathbf{R}\right) \right]  \notag \\
&=&\frac{1}{2}\beta \left( F\left( \mathbf{r}\right) \sqrt{\mathbf{P}^{2}}+%
\sqrt{\mathbf{P}^{2}}F\left( \mathbf{r}\right) \right) -\frac{1}{4}\frac{%
\hbar ^{2}}{\sqrt{\mathbf{P}^{2}}}\mathbf{P\nabla }F\left( \mathbf{r}\right)
\label{Ehbar2}
\end{eqnarray}
From the last formula, we can deduce the equations of motion to the second
order approximation. Indeed, for a particle of positive energy only $%
\varepsilon \left( \mathbf{P,r}\right) =\frac{1}{2}\left( F\left( \mathbf{r}%
\right) \sqrt{\mathbf{P}^{2}}+\sqrt{\mathbf{P}^{2}}F\left( \mathbf{r}\right)
\right) -\frac{1}{4}\frac{\hbar ^{2}}{\sqrt{\mathbf{P}^{2}}}\mathbf{P\nabla }%
F\left( \mathbf{r}\right) $ with now $\mathbf{r}=\mathbf{R+}i\frac{\mathbf{P}%
\times \mathbf{S}}{\mathbf{P}^{2}}$ a $2\times 2$ matrix (the spin matrix is 
$\mathbf{S=\hbar \sigma /2}$), the usual relations $\mathbf{\dot{r}}=-\frac{i%
}{\hbar }\left[ \mathbf{r},\varepsilon \left( \mathbf{P,r}\right) \right] $
and $\mathbf{\dot{P}}=-\frac{i}{\hbar }\left[ \mathbf{P},\varepsilon \left( 
\mathbf{P,r}\right) \right] $, lead to the equations of motion 
\begin{eqnarray}
\mathbf{\dot{r}} &=&\nabla _{\mathbf{P}}\varepsilon +\hbar \mathbf{\dot{P}%
\times }\Theta ^{rr}  \notag \\
\mathbf{\dot{P}} &=&\nabla _{\mathbf{r}}\varepsilon  \label{eqmotion}
\end{eqnarray}
Note that the anomalous velocity term $\hbar \mathbf{\dot{P}\times }\Theta
^{rr}$ in Eq. $\left( \ref{eqmotion}\right) $ has to be understood as a
symmetrized expression of $\mathbf{P}$ and $\mathbf{r}$. Here, $\left[
r_{i},r_{j}\right] =i\hbar ^{2}\Theta _{ij}^{rr}=i\hbar ^{2}\varepsilon
_{ijk}\Theta _{k}^{rr}=-i\hbar ^{2}\varepsilon _{ijk}\lambda \frac{P^{k}}{%
P^{3}}$, and $\lambda =\mathbf{\sigma .P/}P$ the helicity. In addition
helicity is conserved $d\lambda /dt=0$. Eqs $\left( \ref{eqmotion}\right) $
imply the spin Hall effect of light (for the photon one just has to replace
the Pauli matrices $\mathbf{\sigma }$ by the spin one matrices) as a
consequence of the anomalous velocity term $\hbar \mathbf{\dot{P}\times }%
\Theta ^{rr}$ (see also \cite{BLIOKH2} and \cite{BLIOKHNATURE} for the
experimental confirmation of this effect)

From the Eq. $\left( \ref{eqmotion}\right) $ we deduce the following
expression for the velocity components 
\begin{eqnarray}
v^{i} &=&\frac{1}{2}\left( \frac{c}{n(\mathbf{r})}\frac{P^{i}}{P}+\frac{P^{i}%
}{P}\frac{c}{n(\mathbf{r})}\right) +\frac{\lambda \hbar }{2P^{2}}\varepsilon
_{ijk}\left( P^{k}\frac{\partial \ln n}{\partial x^{i}}\frac{c}{n(\mathbf{r})%
}+\frac{\partial \ln n}{\partial x^{i}}\frac{c}{n(\mathbf{r})}P^{k}\right) 
\notag \\
&&+\frac{\hbar ^{2}}{4}\frac{c}{n(\mathbf{r})}\left( \frac{1}{P}\partial
_{i}\ln n-\frac{P_{i}P_{j}}{P^{3}}\partial _{j}\ln n\right)
\end{eqnarray}
from which we compute the modulus of the velocity 
\begin{equation}
v=\frac{c}{n(\mathbf{r})}\left( 1+\hbar ^{2}\frac{\lambda ^{2}}{P^{2}}\left(
\left( \nabla \ln n\right) ^{2}-\frac{1}{P^{2}}\left( \mathbf{P.}\nabla \ln
n\right) ^{2}\right) \right) ^{1/2}+O(\hbar ^{3})
\end{equation}
This expression shows a very small correction to the usual expression $v=c/n(%
\mathbf{r})$ of order $\hbar ^{2}$ due to the interaction between the
polarization state and external inhomogeneities.

\section{Conclusion}

In this paper, we presented a new diagonalization method for a generic
matrix valued Hamiltonian which leads to a diagonal representation where the
operator energy takes an elegant and compact form. This approach requires
the introduction of some new mathematical objects like non-commuting
operators evolving with the Planck constant promoted as a running variable
and thus reveals a mathematical structure reminiscent of the stochastic
calculus. It also shows once more the very important role played by Berry
phases in these systems as the energy operator is written in terms of
covariant dynamical operators containing Berry connections and satisfying a
non-commutative algebra.

It was also found that the diagonal representation of the energy is solution
of a differential equation in $\hbar $ presented previously in \cite%
{SERIESPIERRE} and which could only be solved recursively in a series
expansion in $\hbar .$ Actually the formal exact solution presented here can
also be written explicitly as a series expansion in $\hbar ,$ but it appears
that the derivation of the coefficients of this expansion is now much more
easier. Indeed we could give the expression of the energy and the dynamical
variables to the second order in $\hbar $ for a generic matrix valued
Hamiltonian. We then applied this method, first to the simple case of a
Dirac electron in an external electric field which allowed to recover the
usual Pauli Hamiltonian in the non relativistic limit including the Darwin
term of order $\hbar ^{2}$, and second to the neutrino in a gravitational
field. This is obviously a good check for the validity of the proposed
method. We leave for subsequent work its application to more complicated
systems in condensed matter or relativistic particle physics such as Bloch
and Dirac electrons in interaction.

\end{document}